\documentclass{article}
\usepackage{amsfonts}
\usepackage[fleqn]{amsmath}
\usepackage{enumitem}
\usepackage{graphicx}
\usepackage[left=2cm, right=2cm]{geometry}
\usepackage[dvipsnames, table]{xcolor}
\usepackage{multicol}
\usepackage{multirow}
\usepackage{booktabs}
\usepackage{float}
\usepackage[authoryear,round]{natbib}
\usepackage{hyperref}
\hypersetup{colorlinks=true, linkcolor=blue, citecolor=blue, filecolor=magenta, urlcolor=blue}
\usepackage{mathabx}
\usepackage{indentfirst}
\usepackage[flushleft]{threeparttable}

\usepackage{subcaption}
\usepackage{caption}

\DeclareCaptionLabelFormat{subfigfull}{Figure~#2:}
\makeatletter
\renewcommand\p@subfigure{}
\makeatother

\newcounter{FOCequation}

\title{Climate Policy and the Energy Transition}

\author{Roy Sarkis\thanks{Email: roy.sarkis@epfl.ch}\\ College of Management\\ EPFL}
\date{}

\begin{document}
\maketitle
\begin{abstract}

This paper studies the macroeconomic dynamics of climate policy in a multi-sector dynamic general equilibrium model with renewable and non-renewable energy, sector-specific capital adjustment frictions, household energy demand, and endogenous fossil resource dynamics. The central mechanism is that decarbonization requires reallocating energy use and installed capital: fossil energy demand can contract immediately, while renewable capacity and abatement adjust only gradually. The analysis delivers four results. First, gradual policy implementation sharply reduces transition costs: relative to immediate implementation, gradual emissions caps improve welfare by 2.26 percentage points under comprehensive regulation and by 5.06 percentage points under firm-only regulation. Second, renewable energy subsidies and non-renewable energy taxes support renewable capital accumulation and reduce, but do not eliminate, the welfare cost of front-loaded tightening. Third, sectoral coverage changes the welfare ranking across implementation speeds. Firm-only regulation performs better under gradual implementation because it shields utility-relevant household energy services, but becomes nearly as costly as the carbon-price-only transition under immediate implementation. Fourth, endogenous fossil exploration and stock-dependent extraction costs transmit climate policy into lower extraction, fewer discoveries, and a declining shadow value of reserves, providing a structural mechanism for stranded fossil assets. The results show that deep decarbonization can be achieved at substantially lower macroeconomic cost when policy manages the speed and incidence of energy-capital reallocation.

\end{abstract}
\section{Introduction}

How costly is the transition from fossil to renewable energy? A large body of research suggests that the answer depends primarily on the optimal design of carbon pricing and on the direction of technological change. In many integrated assessment models (IAMs) and dynamic stochastic general equilibrium (DSGE) frameworks, optimal policy can often be characterized by carbon‑pricing rules derived from steady‑state or balanced‑growth considerations, abstracting from the short‑ and medium‑run adjustment costs of decarbonization. This paper instead takes a transition‑centric perspective: it studies how the macroeconomic costs of climate policy depend on the path of the energy transition when installed energy capital and fossil reserves adjust only gradually.

Yet the global energy system is built on large stocks of installed capital that adjust only gradually over time. Fossil extraction infrastructure, power plants, and renewable generation capacity cannot be reallocated instantaneously in response to policy incentives. As a result, the speed of decarbonization is constrained not only by substitution possibilities or technological innovation, but also by the slow reallocation of energy capital across sectors. The central question of this paper is therefore not only what long-run emissions target is achieved, but how quickly the economy can reach it and at what cost. This paper asks how energy-system adjustment, sector-specific capital stocks, and fossil resource dynamics shape the macroeconomic transition toward lower-carbon energy systems, and whether alternative climate policy designs can accelerate this transition while limiting short-run welfare losses.

% A key quantitative result is that these transition costs are not primarily driven by the convex installation-cost term itself. Counterfactual transitions with near-zero capital adjustment costs deliver similar welfare losses, indicating that the main bottleneck is the predetermined sectoral structure of energy capital and the persistent contraction in effective fossil energy input, rather than installation frictions alone.

To study this question, a DSGE model with renewable and non-renewable energy sectors, capital adjustment frictions, endogenous fossil resource dynamics, and energy use by both firms and households was developed. A distinctive feature of the framework is the inclusion of exploration investment and stock-dependent extraction costs, which link fossil production to the evolution of underground reserves and allow climate policy to influence both current extraction and future resource availability. Climate policy operates through taxes and subsidies that affect energy demand as well as investment in fossil and renewable capital. This framework allows to compare price-based climate policies, implemented through carbon pricing, with quantity-based policies that regulate emissions directly through binding caps. This structure allows environmental policy to operate through energy substitution and capital reallocation while also capturing the role of fossil resource dynamics through endogenous extraction and exploration decisions.

The analysis yields four main findings. First, under the calibrated model economy, all transition policies considered in the paper generate short-run welfare losses while converging to the targeted long-run emissions reduction. Second, implementation timing is the dominant margin: holding the instrument set fixed, gradual implementation improves welfare relative to immediate implementation by 2.26 percentage points under comprehensive regulation and by 5.06 percentage points under firm-only regulation. The largest welfare loss arises under an immediate carbon-price-only transition, in which fossil energy use contracts before renewable capital has expanded sufficiently, creating a temporary shortage of effective energy capacity. Third, complementary instruments mitigate but do not eliminate the cost of immediate implementation. An immediate full policy package combining a carbon price, a renewable subsidy, and a non-renewable energy tax substantially improves welfare relative to a carbon-price-only transition, but remains markedly more costly than gradual implementation with the same broad instrument set. Fourth, sectoral coverage interacts with timing. Under gradual implementation, regulating firm emissions alone reduces welfare losses relative to comprehensive caps because firms compensate through higher abatement effort while a utility-relevant margin, household energy services, is shielded from direct compression. Under immediate implementation, however, this advantage reverses: firm-only regulation forces firms to absorb the entire adjustment at once and becomes nearly as costly as the carbon-price-only transition, while comprehensive coverage spreads part of the adjustment burden across firms and households.

A key mechanism behind this result is an internal form of carbon leakage between firms and households. When only firms are directly regulated, the decline in regulated firm fossil demand can temporarily lower the producer price of fossil energy during the transition and induce households to use more fossil energy relative to the comprehensive-cap case. In the model, this mechanism operates through relative price movements rather than through an explicit consumer price index: because the final good is the numeraire, the incidence of firm-level regulation appears through changes in energy prices, factor returns, and profits. This general equilibrium leakage shifts a larger share of the aggregate emissions reduction onto firms while leaving households less exposed to direct reductions in fossil energy use. Under gradual implementation, this can make firm-only regulation welfare-superior to a comprehensive economy-wide emissions cap, even though it reallocates emissions toward households.

The model also provides a structural mechanism for the endogenous creation of stranded fossil assets. Under a permanent climate mandate, climate policy compresses the profitability of fossil extraction and lowers the shadow value of in-ground fossil reserves. Resource firms respond by reducing both current extraction and exploration investment, leading to a persistent decline in new discoveries relative to the laissez-faire benchmark. As a result, the fossil resource sector contracts over time as both extraction and exploration adjust downward in response to the lower profitability of fossil energy production.

These results suggest that the macroeconomic costs of climate policy depend critically on transition dynamics. Relative to the existing literature, the contribution of this paper is to combine sector-specific capital adjustment frictions, endogenous fossil reserves and exploration, and household as well as firm energy demand in a single DSGE framework. Building on this unified environment, the paper makes four contributions. First, it introduces sector-specific energy capital into a DSGE framework with renewable and non-renewable energy, allowing climate policy to generate large transitional macroeconomic effects when fossil energy use is compressed before renewable capacity and abatement can fully adjust. Second, it embeds endogenous fossil exploration and stock-dependent extraction costs, linking climate policy to the evolution of underground reserves and providing a structural mechanism for the gradual creation of stranded fossil assets as climate policy compresses the shadow value of in-ground resources. Third, it highlights an internal carbon leakage channel between firms and households: emissions-based policies that regulate only firm emissions alter relative fossil energy prices during the transition, shift emissions toward households, and can reduce aggregate welfare losses relative to comprehensive economy-wide caps by redistributing the adjustment burden. Fourth, it compares transition policies that achieve the same long-run emissions reduction while varying implementation timing, sectoral coverage, and instrument set. This comparison shows that welfare rankings cannot be inferred from the emissions target alone: a narrow immediate carbon-price-only policy is highly costly, a broader immediate policy package reduces this loss substantially, and the firm-only gradual cap delivers the lowest welfare cost in the baseline calibration.

The paper contributes to the literature in three main respects. First, it relates to DSGE analyses of climate policy that study optimal carbon taxation in general equilibrium; the main contribution is on the transition dynamics generated by energy capital adjustment and fossil resource depletion. Second, the paper connects with research on directed technical change in the context of environmental policy by highlighting a complementary mechanism: while that literature emphasizes endogenous innovation responses to environmental policy, the  analysis focuses instead on the dynamics of installed capital reallocation across energy technologies. Third, it contributes to the literature on fossil resource depletion and climate policy by embedding endogenous exploration and stock-dependent extraction costs within a DSGE framework. Most existing DSGE climate models abstract from at least one of the margins that are central to the analysis: they often exclude explicit household energy demand, treat fossil energy as an elastically supplied flow input without endogenous reserves and exploration, or assume frictionless capital adjustment across energy sectors. By combining household and firm energy use, explicit fossil resource dynamics, and sector-specific energy-capital accumulation in a unified framework, the model makes it possible to study how these three margins jointly shape the transition costs of decarbonization and the incidence of climate policy across sectors.. A more detailed discussion of the related literature is provided in Section~\ref{section_literature_review}.

Taken together, the analysis highlights a central implication for climate policy design. The macroeconomic cost of decarbonization depends not only on the long-run substitutability between clean and dirty energy sources, but also on the speed at which energy capital can be reallocated across sectors and on whether the policy instrument set in place actively supports that reallocation. Understanding these transition dynamics is therefore essential for evaluating the economic consequences of climate policy. The analysis emphasizes that welfare comparisons across climate policies depend critically on transition dynamics rather than steady-state outcomes alone.

% The paper proceeds as follows. Section \ref{section_literature_review} reviews the related literature. Section \ref{section_model} presents the model environment and characterizes the decentralized equilibrium. Section \ref{section_results} analyzes the quantitative results. Section~\ref{section_results} begins by studying the propagation of energy productivity shocks, then evaluates the welfare effects of alternative climate policy instruments, and finally examines transition dynamics under both price-based and quantity-based climate policies. Section \ref{section_conclusion} concludes.
The analysis begins with a review of the related literature in Section \ref{section_literature_review}. The model and decentralized equilibrium are characterized in Section \ref{section_model}. Section~\ref{section_results} begins by studying the propagation of energy productivity shocks, then evaluates the welfare effects of alternative climate policy instruments, and finally examines transition dynamics under both price-based and quantity-based climate policies. Section \ref{section_conclusion} concludes.

\section{Literature review}
\label{section_literature_review}

This paper studies how energy-system adjustment shapes the macroeconomic costs of decarbonization policies and the speed of the transition from fossil to renewable energy. While many climate-economy models emphasize long-run substitution between clean and fossil energy inputs, the analysis focuses on the transition path: fossil energy use can contract quickly after policy tightening, while renewable capacity, fossil reserves, and sectoral capital stocks adjust only gradually. The literature relevant to this paper can be organized around four related strands: optimal carbon pricing in climate-macroeconomic models, directed technical change and energy substitution, exhaustible-resource dynamics and stranded fossil assets, and instrument choice under climate policy.

The benchmark DSGE analysis of optimal carbon taxation is provided by \citet{Golosov2014}, who derive a closed-form characterization of the marginal external damage from emissions and show that the optimal carbon tax is proportional to current output under their assumptions. Their framework provides the standard general-equilibrium benchmark for carbon pricing. Subsequent work extends it by incorporating fiscal distortions and resource dynamics. \citet{Barrage2019} shows that optimal carbon taxation must be determined jointly with the broader fiscal system when labor and capital taxes are distortionary, while \citet{Hassler2018} emphasize the role of fossil-resource extraction and the shadow value of carbon in shaping optimal climate policy. Earlier integrated assessment models, including \citet{nordhaus1994,nordhaus2008} and \citet{nordhaus_boyer_2000}, embed climate damages and emissions in Ramsey-style growth frameworks, but typically use highly aggregated production structures and largely exogenous technological progress. Relative to this literature, the present paper focuses less on the analytical characterization of the optimal carbon tax and more on the transition dynamics generated by energy-sector capital, household energy demand, and fossil-resource adjustment.

A second strand studies how environmental policy affects technology and substitution between clean and dirty inputs. \citet{Acemoglu2012} develop a model of directed technical change in which innovation responds to environmental policy incentives. They demonstrate that, when the degree of substitution across clean and dirty inputs is sufficiently large, carbon taxes together with temporary innovation subsidies can redirect technological progress toward clean sectors while preserving long-run growth. This insight is relevant here because substitution elasticities in household and firm energy bundles shape both shock propagation and policy rankings. \citet{Hassler2021} emphasize a related mechanism in which rising fossil-energy prices induce directed energy-saving technical change, making long-run substitution possibilities larger than short-run ones. The key difference is that technological change is exogenous in the present paper: the analysis studies how policy reallocates installed energy capital and fossil-resource use, rather than how it redirects innovation. Related empirical research reinforces this focus. Using OPEC announcements as a source of exogenous variation, \citet{kanzig2021} finds that energy-market shocks have economically meaningful aggregate effects.

The model also builds on the literature on exhaustible resources. \citet{Hotelling1931} establishes the canonical result that the shadow value of an exhaustible resource rises at the rate of interest along an optimal extraction path, while \citet{Dasgupta1974} analyze optimal depletion and the interaction between resource scarcity and economic growth. The broader macroeconomic implications of resource scarcity are surveyed by \citet{van_der_ploeg2011}. More recent work, such as \citet{vanderploeg2012}, shows that anticipated improvements in renewable alternatives can generate Green-Paradox incentives by encouraging resource owners to accelerate extraction before fossil rents decline. The present framework incorporates this intertemporal logic through endogenous exploration, stock-dependent extraction costs, and a reserve shadow value, but embeds it in a DSGE model with renewable capital and climate policy instruments.

A related macro-energy literature studies stranded fossil assets and transition costs under decarbonization. \citet{Mercure2018SFFA} quantify the macroeconomic impact of stranded fossil fuel assets under alternative technology and policy trajectories, while \citet{ChenLandryReilly2022} estimate stranded assets in fossil extraction and coal power generation using a global general equilibrium framework with detailed energy-sector capital stocks.\footnote{See also \citet{StrandedAssetsReview} for a recent survey of the stranded-assets literature.} Relative to these studies, the present paper provides a tractable DSGE mechanism linking stranded fossil assets to climate policy through declining extraction profitability, lower exploration, and a falling shadow value of underground reserves.

Recent DSGE contributions increasingly incorporate environmental policy, pollution, and energy sectors into macroeconomic models. \citet{Annicchiarico2015} study environmental policy in a New Keynesian DSGE model with pollution and abatement dynamics, while \citet{annicchiarico2021} survey the broader environmental DSGE literature. \citet{Varga2022} develop E-QUEST, a multisector framework for the European Union with dirty and clean energy sectors, exhaustible fossil resources, and substitution between fossil and clean energy inputs. \citet{Coenen2024} develop a macroeconomic framework with detailed energy-sector representation and use it to quantify the effects of decarbonization policies. Relative to these contributions, the present paper emphasizes the interaction between sector-specific energy capital, fossil-resource dynamics, household energy demand, and policy timing within a unified DSGE framework.

The paper also relates to the literature on instrument choice in environmental policy. \citet{GoulderParry2008} emphasize that environmental instruments should be evaluated not only by cost effectiveness, but also by incidence, uncertainty, and political feasibility. A related second-best taxation literature shows that carbon-tax welfare effects depend on interactions with pre-existing fiscal distortions and revenue recycling \citep{BovenbergGoulder1996,Goulder1995,Parry1997}. The present paper highlights a complementary dynamic margin: even without distortionary labor or capital taxation, instrument choice can substantially affect welfare during the transition because different instruments load the adjustment burden onto different margins. Carbon prices compress fossil use directly, renewable subsidies support clean-capacity expansion, non-renewable energy taxes alter fossil demand, and sectoral coverage shifts the incidence of adjustment between firms and households.

Finally, the analysis is related to work on carbon leakage and the incidence of climate policy. A large literature studies cross-border leakage in emissions trading systems, with mixed evidence across ex-ante and ex-post studies \citep{ETSLeakageReview}. Firm-level analyses of the European Union Emissions Trading System (EU ETS) find limited evidence of carbon leakage and industrial relocation, although some sectors and firm types appear more exposed to investment leakage risks \citep[e.g.][]{EUETSInvestmentLeakage,CarbonLeakageProspects}. The internal carbon leakage mechanism studied here is different: emissions do not shift abroad, but are reallocated between regulated firms and unregulated households within a closed economy through endogenous changes in fossil-energy prices.\footnote{Recent empirical work also studies leakage through global supply-chain reconfiguration and sourcing responses to carbon pricing and border adjustments; see, for example, \citet{SupplyChainLeakage}.} The model therefore connects leakage to household energy demand and general-equilibrium incidence rather than to international relocation.

Together, these strands show that the macroeconomic cost of decarbonization depends not only on long-run substitution possibilities, but also on the speed of energy-system adjustment, the response of fossil-resource supply, and the incidence of policy across sectors. The model uses these mechanisms to study transition dynamics, welfare rankings across policy instruments, and the interaction between fossil-resource depletion and renewable-capital accumulation.

\section{Model}
\label{section_model}
The economy is modeled as an infinite-horizon dynamic general equilibrium framework featuring environmental externalities, endogenous fossil resource depletion, and renewable energy production. All markets are perfectly competitive. The main distortion in the decentralized economy arises from the climate externality, while fossil-resource dynamics provide an additional intertemporal propagation mechanism through extraction, exploration, and stock-dependent extraction costs. The remainder of this section describes the building blocks of the decentralized equilibrium. Sections~\ref{model_households} to \ref{model_energy_sector} introduce households, final-goods firms, and the energy sectors; Sections~\ref{model_environment} and \ref{model_government} describe the environmental block and government budget. Section~\ref{model_equilibrium} defines market clearing and the main distortions; and Sections~\ref{model_exogenous_shocks} and \ref{model_calibration} specify the exogenous shock processes and the calibration strategy.

\subsection{Households}
\label{model_households}
A representative household derives utility from the consumption of final goods, direct energy services, and environmental quality. Allowing energy to enter utility captures the fact that households consume energy services such as heating, electricity, and mobility, which are not perfectly substitutable with non-energy goods. Including environmental quality in utility captures the direct welfare cost of pollution, in addition to its effect on production through climate damages, introduced in the final goods production function in Section~\ref{model_final_goods_firm}. This specification therefore allows environmental policy to affect welfare through three channels: production damages, household energy services, and environmental quality.

The inclusion of environmental quality in household utility follows the environmental macroeconomics literature, including \citet{Acemoglu2012}, where household welfare depends directly on consumption and the state of the environment. The direct utility from household energy services is best interpreted as a reduced-form representation of the services households obtain from heating, electricity, and mobility. This specification makes household exposure to energy costs explicit: policies that compress household energy services generate direct utility losses, whereas policies that operate only through production affect households indirectly through income and prices.

The model abstracts from labor supply decisions to isolate the substitution margins between capital, renewable energy, and non-renewable energy during the climate transition. As a result, the model also abstracts from the labor-leisure margin and from interactions between carbon taxation and pre-existing labor tax distortions. Following energy-services approaches in macroeconomic models such as \citet{Bergmann2018}, households derive utility from both final goods consumption and energy services. The household's period utility is given by
\begin{align}
\label{utility_function}
U_t = \frac{\psi_C C_t^{1-\rho_C}}{1-\rho_C} + \frac{\psi_E E_{HH,t}^{1-\rho_E}}{1-\rho_E} + \frac{\phi_Q Q_t^{1-\rho_Q}}{1-\rho_Q}.
\end{align}

Although environmental quality is determined in equilibrium by aggregate pollution, households take \(Q_t\) as given when choosing consumption and energy demand. Individual households therefore do not internalize how their own non-renewable energy use affects aggregate emissions, the pollution stock, and environmental quality. As a result, the private marginal cost of non-renewable energy consumption is lower than the social marginal cost, generating an environmental externality in the competitive equilibrium.

The household's energy services \((E_{HH,t})\) are modeled as a CES composite of renewable \((E_{HHR,t})\) and non-renewable \((E_{HHNR,t})\) energy, following the use of nested energy-consumption bundles in \citet{Varga2022}. The CES specification allows for imperfect substitution between energy sources; a simple additive aggregator would instead impose perfect substitutability and a constant marginal rate of substitution:
\begin{align}
E_{HH,t} = \left[ \Omega^{1/\theta} E_{HHR,t}^{(\theta-1)/\theta} + (1-\Omega)^{1/\theta} E_{HHNR,t}^{(\theta-1)/\theta} \right]^{\theta/(\theta-1)},
\end{align}
where \(\theta\) governs the degree of substitutability across renewable and non-renewable energy inputs, and \(\Omega \in (0,1)\) determines the relative weight of renewable energy in the household energy bundle.

The household owns the capital used in the final-goods ($Y$), renewable-energy ($R$), and non-renewable-energy ($NR$) sectors. They receive rental income from the three capital stocks, distributed profits from firms, and government transfers. They allocate resources between final-goods consumption, renewable and non-renewable energy use, and investment in the three capital stocks so as to maximize lifetime utility \eqref{utility_function}, subject to

\begin{align}
C_t &+ P_{ER,t}(1-\tau_{ER,t})E_{HHR,t}+ P_{ENR,t}(1+\tau_{ENR,t})E_{HHNR,t} 
+ \sum_{j \in \{Y,R,NR\}} I_{j,t} \quad + \sum_{j \in \{Y,R,NR\}} AC_{j,t} \nonumber  \\
&= r_{Y,t}K_{t-1}+ r_{R,t}K_{R,t-1}+ r_{NR,t}K_{NR,t-1} + \Pi_t + Tr_t.
\end{align}
% where $P_{ER,t}$ and $P_{ENR,t}$ represent the prices of renewable and non-renewable energy in terms of the final good. Since the final good is the numeraire, the model does not feature an explicit CPI block, so tax incidence is reflected indirectly through relative prices and household income components. The subsidy rate on renewable energy and the tax rate on non-renewable energy are defined as $\tau_{ER}$ and $\tau_{ENR}$, respectively. 
% Households own all firms and receive total profits $\Pi_t$ defined as
The residual income term \(\Pi_t\)\footnote{The term \(\Pi_t\) denotes residual dividend income remitted to households, not pure profits. Perfect competition requires only that firms take prices as given. It does not require zero accounting income when the production function exhibits decreasing returns in the explicitly modeled private inputs -- so that an unmodeled fixed factor earns a residual share -- or when scarcity rents accrue to owned resource stocks, as in the non-renewable energy sector.} is defined as

\begin{equation*}
\Pi_t = \pi_{Y,t} + \pi_{ER,t} + \pi_{ENR,t}.
\end{equation*}
The left-hand side contains household expenditures on final consumption, renewable household energy, non-renewable household energy, gross investment in the three sector-specific capital stocks, and capital adjustment costs. The right-hand side contains rental income from capital employed in the final-goods sector, the renewable-energy sector, and the non-renewable-energy sector, followed by firm dividends and government transfers. The variables \(P_{ER,t}\) and \(P_{ENR,t}\) denote renewable and non-renewable energy prices in units of the final good. Since the final good is the num\'eraire, the model does not include an explicit consumer price index; tax incidence is reflected through relative energy prices and household income. The policy instruments \(\tau_{ER,t}\) and \(\tau_{ENR,t}\) denote the renewable-energy subsidy and the non-renewable-energy tax, respectively.

For each sector $j \in \{Y, R, NR\}$, the physical capital stock evolves according to the standard law of motion:
\begin{equation}
    K_{j,t} = (1-\delta_j)K_{j,t-1} + I_{j,t}
\end{equation}
where $0 \leq \delta_j \leq 1$ is the sector-specific capital depreciation rate, and $I_{j,t}$ represents gross investment. 

Installing new capital requires real resources. Following the convex capital adjustment cost framework of \citet{Hayashi1982}, the following quadratic functional form is assumed for tractability
\begin{equation}
    AC_{j,t} = \frac{h_j}{2} \left( \frac{I_{j,t}}{K_{j,t-1}} - \delta_j \right)^2 K_{j,t-1},
\end{equation}
where $h_j$ determines the magnitude of the investment friction in sector $j$. These installation costs are borne by the household and therefore enter the aggregate resource constraint as additional uses of final output. They make rapid changes in sectoral capital stocks costly and shape the timing of capital reallocation during the transition.

\subsection{Final goods firm}
\label{model_final_goods_firm}
The final good is produced by a representative producer using capital together with an aggregate energy composite. Gross output ($Y_t$) is reduced by climate damages from the accumulated pollution stock ($M_t$) and is defined as 
\begin{align}
\label{output}
Y_t = D(M_t)\, A_{Y,t}\, K_{t-1}^{\alpha_K} E_{F,t}^{\alpha_E},
\end{align}
where climate damages ($D(M_t)$) affect total factor productivity multiplicatively. 
Following the integrated assessment literature initiated by \citet{Nordhaus1991,nordhaus2008} and adopted in DSGE climate models such as \citet{Heutel2012}, a convex quadratic damage specification is implemented:
\begin{align}
D(M_t) = 1 - \gamma_2 M_t^2 - \gamma_1 M_t - \gamma_0,
\end{align}
where $\gamma_0, \gamma_1, \gamma_2$ govern the level, slope and curvature of climate damages.

Although the pollution stock is determined in equilibrium by aggregate emissions, the representative firm takes \(M_t\) as given when choosing inputs and abatement. The firm therefore accounts for private fossil-energy costs and carbon-policy payments, but not for the effect of its emissions on the aggregate pollution stock and the resulting productivity damages. As a result, the private marginal cost of emissions is lower than the social marginal cost, generating a production-side climate externality in the competitive equilibrium.

Because labor is not modeled as a separate factor, the residual share \(1-\alpha_K-\alpha_E\) captures the contribution of labor and other fixed factors. Residual income in the final-goods sector should therefore be interpreted as payments to the normalized factor rebated to households rather than as pure economic profits.

Energy used in production is a CES composite of renewable and non-renewable inputs, analogous to the household energy aggregator:
\begin{align}
E_{F,t} =\left[\Omega_F^{1/\theta_F} E_{FR,t}^{(\theta_F-1)/\theta_F} + (1-\Omega_F)^{1/\theta_F} E_{FNR,t}^{(\theta_F-1)/\theta_F}\right]^{\theta_F/(\theta_F-1)},
\end{align}
where $\theta_F$ governs the substitutability between both types of energy, while $\Omega_F$ specifies the share parameter associated with renewable energy in the CES aggregator. The CES structure allows the elasticity of substitution between renewable and non-renewable inputs to govern the speed and cost of energy reallocation in production.

Final goods firms also choose the abatement effort ($\mu_t$), which reduces their effective emissions but consumes real resources. The cost of abatement ($CA_t$) is modeled as a fraction of the firm's gross output:
\begin{align}
CA_t = \phi_1 \mu_t^{\phi_2} Y_t.
\end{align}
Unlike standard integrated assessment models that embed abatement as a direct productivity penalty within the production function, this formulation explicitly isolates abatement as a distinct use of gross output. This clarifies the macroeconomic trade-off: mitigating emissions consumes real resources that would otherwise be available for consumption or investment. Consequently, the net output available for distribution is $(1-\phi_1\mu_t^{\phi_2})Y_t$.

The firm chooses capital, abatement effort, and renewable and non-renewable energy inputs to maximize the present discounted value of current and future profits. The firm's revenues come from final-goods production, while its costs include energy purchases, capital rental payments, abatement costs, and tax payments on firm emissions \(Z_{F,t}\).\footnote{Defined in equation~(\ref{firm_emissions_def})} The profit function is then defined as
\begin{flalign}
    \pi_{Y,t} &= Y_t - CA_t - r_{Y,t} K_{t-1} - P_{ER,t}(1-\tau_{ER,t}) E_{FR,t} - P_{ENR,t}(1+\tau_{ENR,t}) E_{FNR,t} - p_{z,t} Z_{F,t} &&
\end{flalign}

Notice that while final goods firms have access to an abatement technology ($\mu_t$), households do not. This asymmetry reflects the technological reality that industrial facilities can adopt end-of-pipe mitigation technologies (e.g., carbon capture and storage), whereas households primarily reduce emissions through demand reduction or fuel substitution.

\subsection{Energy sector}
\label{model_energy_sector}

\subsubsection{Renewable energy}
Renewable energy is produced using sector-specific installed capital according to
\begin{equation}
E_{R,t} = A_{R,t} K_{R,t-1}^{\gamma_{ER}},
\end{equation}
where $\gamma_{ER}=1$ as shown in Table \ref{tab:calibration}. The linear specification reflects the interpretation of renewable capital as installed photovoltaic (PV) capacity. Holding productivity \(A_{R,t}\) fixed, doubling installed renewable capacity doubles renewable energy supply. This differs from a decreasing-returns specification and is appropriate when the model abstracts from site heterogeneity, grid congestion, and intermittency constraints.

Renewable energy firms operate in perfectly competitive markets and take prices as given. Given the installed capital stock \(K_{R,t-1}\), profits are defined as
\begin{align}
\pi_{ER,t} = P_{ER,t}E_{R,t} - r_{R,t}K_{R,t-1}.
\end{align}

\subsubsection{Non-renewable energy and resource dynamics}
The non-renewable energy sector features endogenous resource dynamics that capture the trade-offs between current extraction, future depletion, and exploration. The firm produces non-renewable energy ($E_{NR,t}$) by combining physical capital ($K_{NR,t-1}$) and the extracted resource ($S_t$) using a Cobb-Douglas production function:
\begin{equation}
E_{NR,t} = A_{NR,t} K_{NR,t-1}^{\alpha_{K,NR}} S_t^{\alpha_S},
\end{equation}
with $\alpha_{K,NR} + \alpha_{S} = 1.$

Because production requires the extraction of the underground resource ($S_t$), the stock of deposits ($D_t$) declines as resources are extracted. Resource firms can partially replenish the reserve base through exploration investment ($Exp_{NR,t}$), making the total stock of reserves endogenous and responsive to economic incentives. The resource deposit evolves according to
\begin{equation}
\label{constraint_ENR}
D_t = D_{t-1} - S_t + A_{disc}(Exp_{NR,t})^{\alpha_D},
\end{equation}
where $A_{disc}$ represents the efficiency of discovery and $\alpha_D<1$ implies decreasing returns to exploration. Economically, when the most accessible reserves have already been discovered, additional discoveries require deeper drilling or exploration in less favorable geological locations, generating diminishing marginal returns. This law of motion implies that exploration can partially offset depletion, but only imperfectly because discoveries exhibit decreasing returns.

Extraction costs increase as the underground resource stock declines:
\begin{equation}
AC_{ext,t} = \xi S_t \left(\frac{\bar{D}}{D_{t-1}}\right)^{\psi_D}.
\end{equation}
This mechanism, known as the stock effect, implies that as the resource stock $D_{t-1}$ declines relative to the reference level $\bar{D}$, extraction becomes increasingly costly ($\psi_D>0$). This creates an intertemporal incentive to maintain a positive reserve stock in order to limit future extraction costs.

Exploration expenditure ($Exp_{NR,t}$) enters the firm's profit function at a unit price. One unit of output invested in exploration generates $A_{disc}(Exp_{NR,t})^{\alpha_D}$ units of newly discovered reserves, so a separate price for exploration is unnecessary. 

The non-renewable energy firm chooses capital demand, extraction, and exploration to maximize the present discounted value of profits subject to the resource constraint \eqref{constraint_ENR}. Households own the capital stock and choose investment, while firms rent sector-specific capital services at the rental rate \(r_{NR,t}\). Period profits (\(\pi_{ENR,t}\)) are defined as
\begin{align}
    \pi_{ENR,t} = P_{ENR,t}E_{NR,t} - r_{NR,t}K_{NR,t-1} - Exp_{NR,t} - AC_{ext,t}.
\end{align}

Positive profits in the non-renewable sector reflect scarcity rents associated with the exhaustible resource stock. Even under perfect competition, the finite reserve stock and the shadow value \(\lambda_{D,t}\) generate resource rents that are rebated to households through firm dividends.

\subsection{Environment}
\label{model_environment}
Households and firms generate emissions through the use of non-renewable energy:
\begin{align}
Z_{HH,t} &= E_{HHNR,t}^{\zeta_{HH}}, \\
Z_{F,t}  &= (1-\mu_t) E_{FNR,t}^{\zeta_F}, \label{firm_emissions_def}
\end{align}
\noindent where the emissions intensity parameters \(\zeta_{HH}\) and \(\zeta_F\) are held constant over time. The framework therefore abstracts from endogenous improvements in the carbon efficiency of fossil energy use, cleaner fossil technologies, or emissions-specific productivity shocks that would reduce emissions for a given level of non-renewable energy input.

Here $\mu_t \ge 0$ denotes the abatement intensity chosen by firms. When $0 \le \mu_t \le 1$, abatement reduces emissions per unit of fossil energy used in production. Values of $\mu_t > 1$ correspond to net negative emissions, meaning that abatement technologies remove more carbon than is generated by fossil energy use. Achieving higher abatement intensities becomes increasingly costly because the abatement cost function is convex, implying that marginal abatement costs rise with $\mu_t$ and make large-scale carbon removal progressively more expensive. Households do not have access to a separate abatement technology; instead, their emissions adjust through changes in energy consumption and the gradual replacement of capital.

These emissions accumulate in the atmosphere as a pollution stock $M_t$, which generates an environmental externality by reducing aggregate productivity through the damage function. Pollution accumulates according to
\begin{align}
M_t = (1-\delta_M)M_{t-1} + Z_{HH,t} + Z_{F,t},
\label{law_of_motion_emissions}
\end{align}
where $0 < \delta_M < 1$ is the natural depreciation rate of atmospheric pollution. The low depreciation rate of the pollution stock implies that even temporary changes in emissions can generate highly persistent effects on environmental quality.

The same pollution stock also enters household welfare through environmental quality. This reduced-form channel captures non-market welfare effects of accumulated pollution, such as lower air quality, health damages, and environmental degradation, that are not fully reflected in production damages. Environmental quality is specified as a decreasing function of the pollution stock:
\begin{align}
\label{air_quality_definition}
Q_t = \bar Q - \xi_M M_t,
\end{align}
where \(\bar Q\) denotes environmental quality in the absence of pollution pressure and \(\xi_M>0\) maps accumulated pollution into environmental quality. Because household utility depends on \(Q_t\), reductions in the pollution stock generate a direct welfare gain in addition to their effect on production through the damage function. In the decentralized equilibrium, households take \(Q_t\) as an aggregate environmental state and do not internalize how their own non-renewable energy use affects aggregate emissions, the pollution stock, and environmental quality.

\subsection{Government}
\label{model_government}
The government collects taxes ($\tau_{ENR,t}$) on the use of non-renewable energy by both households and final-goods firms, as well as revenue from the sale of emissions permits to the final-goods firm ($p_{z,t} Z_{F,t}$). The tax and subsidy rates ($\tau_{ER,t}$, $\tau_{ENR,t}$) and the carbon price ($p_{z,t}$) are policy instruments set by the government. In the steady-state policy analysis of Section~\ref{section_welfare_analysis} and in the immediate transition experiments of Sections~\ref{section_transition_50_ren_energy} and~\ref{section_cap_emissions}, these instruments are constant at their target values from $t=1$ onward. In the gradual emissions-cap experiments of Section~\ref{section_cap_emissions}, $p_{z,t}$ instead evolves over time according to a feedback rule that tightens policy in response to deviations of emissions from the long-run target, with $\tau_{ER,t}$ and $\tau_{ENR,t}$ scaled proportionally to $p_{z,t}$.

Government revenues are used to subsidize the price of renewable energy at rate $\tau_{ER,t}$, while any surplus is rebated to households through transfers $Tr_t$. The government budget constraint is given by
\begin{align}
Tr_t + \tau_{ER,t} P_{ER,t}(E_{HHR,t}+E_{FR,t}) = p_{z,t} Z_{F,t} + \tau_{ENR,t} P_{ENR,t}(E_{HHNR,t}+E_{FNR,t}).
\end{align}

\subsection{Market equilibrium and distortions}
\label{model_equilibrium}
Under this setup, the aggregate resource constraint is
\begin{align}
Y_t = C_t + \sum_{j\in\{Y,R,NR\}} I_{j,t}  + \sum_{j\in\{Y,R,NR\}} AC_{j,t}  + CA_t + AC_{ext,t} + Exp_{NR,t}.
\end{align}
Installation adjustment costs ($AC_{j,t}$) are borne by households through their capital accumulation decisions and therefore enter both the household optimization problem and the aggregate resource constraint. Abatement costs ($CA_t$), extraction costs ($AC_{ext,t}$), and exploration expenditures ($Exp_{NR,t}$) also use final output.

Energy markets clear separately for renewable and non-renewable energy. Total supply of each energy type must equal total demand from households and final-goods firms:
\begin{align}
    E_{R,t} &= E_{HHR,t} +E_{FR,t},\\
    E_{NR,t} &= E_{HHNR,t} +E_{FNR,t}.
\end{align}

The benchmark economy features no environmental policy: \(\tau_{ER,t}=\tau_{ENR,t}=p_{z,t}=0\). In the absence of policy, the competitive equilibrium is distorted by a climate pollution externality, because individual agents do not internalize the aggregate damages caused by their emissions. Fossil-resource dynamics do not constitute a separate market failure in the benchmark economy: resource firms internalize how current extraction affects their own future reserve stock, extraction costs, and reserve value. Instead, the reserve block provides a dynamic propagation channel through which climate policy affects fossil supply, exploration, and the shadow value of underground resources.

The policy instruments analyzed in Sections~\ref{section_transition_50_ren_energy} and~\ref{section_cap_emissions} are designed to internalize the climate externality and to study how alternative implementations of the same emissions objective interact with energy-capital reallocation and fossil-resource dynamics.

\subsection{Exogenous shocks}
\label{model_exogenous_shocks}
Three exogenous total factor productivity shocks are modeled, one for each sector:
\begin{align}
    \ln(A_{j,t}) = \rho_{j} \ln(A_{j,t-1}) + \epsilon_{j,t}
\end{align}
where \(\rho_j\) is the persistence of the shock process for sector \(j \in \{Y, ER, ENR \}\). Technological change enters the model through these sector-specific productivity processes. The substitution elasticities and CES weights in household and firm energy bundles are held constant over time. The framework therefore abstracts from endogenous changes in substitution possibilities and from directed technical change between fossil and renewable energy technologies.

\subsection{Calibration}
\label{model_calibration}
The parameterization reported in Table \ref{tab:calibration} is conducted at a quarterly frequency and is organized into three main categories: standard macroeconomic parameters, energy sector parameters calibrated to match steady-state targets, and environmental and resource dynamics parameters derived from the literature and empirical estimates.

\subsubsection{Macroeconomic parameters}
Standard macroeconomic parameters follow conventional values used in the real business cycle literature. The benchmark calibration uses a discount factor of $\beta = 0.99$, consistent with a long-run real interest rate of around 1\% per quarter (about 4\% per year). The coefficient of relative risk aversion is set to $\rho_C = 2.0$, a commonly used value in quantitative macroeconomic models \citep{Stern2008,Weitzman2007}. Final-goods capital depreciates at a quarterly rate of $\delta = 0.025$, which implies annual capital depreciation of approximately 10\%. The utility weight on energy services is set to $\psi_E = 0.05$, ensuring that household energy use accounts for a realistic share of total expenditure in steady state. The consumption weight $\psi_C$ is normalized to unity.

The calibration assigns output elasticities of $\alpha_K = 0.30$ to capital and $\alpha_E = 0.15$ to energy in the final-goods sector. Because labor and other inputs are not explicitly modeled, these factor shares do not sum to one. The chosen values ensure that energy inputs remain quantitatively important for production while remaining smaller than the contribution of capital, consistent with empirical energy cost shares in advanced economies. In the model's steady state, the residual income in the final-goods sector is quantitatively consistent with the unmodeled factor share: with $\alpha_K = 0.30$ and $\alpha_E = 0.15$, the residual share is $1 - \alpha_K - \alpha_E = 0.55$, close to the share of output rebated to households as residual income.

Investment adjustment costs are assumed to be identical across sectors and are set to $h = h_{ER} = h_{ENR} = 10$. These values generate smooth transitional dynamics in response to shocks without preventing capital reallocation across sectors.

\subsubsection{Energy sector and steady-state targets}
Energy-sector parameters are jointly calibrated to match key steady-state moments in the no-policy equilibrium. The calibration targets the steady-state shares of renewable energy in total supply and household energy demand in aggregate energy use. In particular, the CES weights ($\Omega$, $\Omega_F$), substitution elasticities ($\theta$, $\theta_F$), and the energy utility weight ($\psi_E$) are chosen so that the model reproduces aggregate energy shares and the sectoral allocation of energy demand observed in the European Union. The substitution elasticities are set to $\theta = \theta_F = 2.0$, implying moderate substitutability between renewable and non-renewable energy sources. This allows for meaningful long-run energy substitution while still generating significant transition dynamics due to capital adjustment frictions. The corresponding CES weights are $\Omega = \Omega_F = 0.02$.

In the model’s baseline steady state, renewable energy accounts for $30.48\%$ of total primary energy supply, measured as $E_R/(E_R+E_{NR})$. This is close to the EU-wide renewable share of 25.2\% of gross final energy consumption in 2023 \citep{Eurostat2024Renewables}, while allowing for a slightly higher forward-looking benchmark consistent with the continued expansion of renewable capacity in Europe. The model also reproduces the sectoral distribution of energy demand. In steady state, households account for approximately $23.43\%$ of total energy use, computed as $E_{HH}/(E_{HH}+E_F)$. This compares to $27.8\%$ in the EU data for 2023 \citep{EurostatEnergyUse2024}. While somewhat lower than the empirical benchmark, the model captures the dominant role of firms in aggregate energy demand and preserves realistic sectoral magnitudes.

Depreciation rates for energy-specific capital are calibrated to match empirical lifetimes reported by the International Energy Agency (IEA). Capital in the non-renewable sector reflects the long operational lifetimes of fossil-based generation infrastructure, such as coal and gas power plants, which typically operate for 40–50 years. This corresponds to a quarterly depreciation rate of $\delta_{kENR} = 0.006$. In contrast, renewable energy capital, proxied by solar photovoltaic installations, has a shorter operational lifetime of roughly 25–30 years, implying a higher quarterly depreciation rate of $\delta_{kER} = 0.01$ \citep{IEA2022CoalNetZero,IEA2022SolarPV}.

\subsubsection{Environmental damages and resource dynamics}

Environmental parameters are drawn from a combination of empirical estimation and established integrated assessment models. The emission elasticities for households ($\zeta_{HH}$) and firms ($\zeta_F$) are set to $0.69$, based on a log-log regression of aggregate emissions on non-renewable energy use. The estimated elasticity equals $0.688$, which is rounded to two decimal places.

The climate damage function and abatement cost parameters follow the DICE-2007 framework \citep{nordhaus2008}, as adapted to business-cycle environments by \citet{Heutel2012} and \citet{Annicchiarico2015}. The quarterly decay rate of the atmospheric pollution stock is set to $\delta_M = 0.0021$, consistent with the long persistence of atmospheric carbon dioxide \citep{Heutel2012}. The damage function coefficients ($\gamma_0 = 1.3950\times10^{-3}$, $\gamma_1 = -6.6722\times10^{-6}$, $\gamma_2 = 1.4647\times10^{-8}$) are taken directly from \citet{Heutel2012}. The abatement cost curvature parameter ($\phi_2 = 2.80$) originates from the DICE calibration as implemented by \citet{Heutel2012}, while the abatement cost scale parameter ($\phi_1 = 0.1850$) follows \citet{Annicchiarico2015}.

The non-renewable production technology is capital-intensive, with $\alpha_{K,NR} = 0.80$ and $\alpha_S = 0.20$, reflecting the infrastructure-intensive nature of fossil extraction and energy generation. I assume constant returns in non-renewable production, so scarcity rents arise from the reserve block and extraction costs rather than from decreasing returns at the production stage. The scaling parameters for extraction costs ($\xi$) and the reference resource stock ($\bar{D}$) are selected to ensure a stable interior steady state with $D>0$. The calibration implies a positive steady-state shadow value of reserves, so the reserve stock converges to a finite interior level rather than to a corner solution. The scarcity elasticity $\psi_D = 1.0$ implies that marginal extraction costs rise proportionally as the resource stock declines. In steady state, exploration exactly offsets depletion, and the resource Euler equation satisfies a generalized Hotelling condition adjusted for extraction costs.

The calibration targets steady-state cross-sectional and compositional moments rather than dynamic impulse-response statistics. Accordingly, the model is intended to discipline long-run allocations and sectoral shares more tightly than short-run transition dynamics. The quantitative transition effects reported below should therefore be interpreted as model-based rather than directly estimated empirical magnitudes. Transitional dynamics and policy counterfactuals therefore emerge from the structural mechanisms embedded in the model.

Three calibration features are particularly important for the quantitative results. First, the initial renewable share is moderate, implying that fossil energy remains central in the benchmark equilibrium. Second, substitution elasticities between renewable and non-renewable energy are finite, limiting instantaneous reallocation across energy inputs. Third, energy production depends on sector-specific capital stocks that evolve through accumulation equations, so renewable capacity cannot immediately replace a sharp contraction in fossil energy use. Capital adjustment frictions affect the timing of this reallocation, but the counterfactual exercises below show that they do not account for most of the welfare loss. Together, these features generate sizable transitional effects following climate policy shocks.

\begin{table}[H]
\centering
\caption{Calibration Parameters}
\rowcolors{2}{gray!10}{white}
\renewcommand{\arraystretch}{0.9}
\label{tab:calibration}
\begin{tabular}{lll}
\toprule
\textbf{Parameter} & \textbf{Description} & \textbf{Value} \\
\midrule

$\beta$ & Discount factor & 0.990 \\

$\psi_C$ & Weight on consumption utility & 1.000 \\
$\rho_C$ & CRRA coefficient (consumption) & 2.000 \\
$\psi_E$ & Weight on energy utility & 0.050 \\
$\rho_E$ & CRRA coefficient (energy composite) & 2.000 \\

$\theta$ & Elasticity of substitution (household energy CES) & 2.000 \\
$\Omega$ & Renewable share in household energy CES & 0.020 \\

$\theta_F$ & Elasticity of substitution (firm energy CES) & 2.000 \\
$\Omega_F$ & Renewable share in firm energy CES & 0.020 \\

$\alpha_K$ & Capital share in final goods production & 0.300 \\
$\alpha_E$ & Energy share in final goods production & 0.150 \\

$\delta$ & Depreciation rate, final goods capital & 0.025 \\
$\delta_{kER}$ & Depreciation rate, renewable capital & 0.010 \\
$\delta_{kENR}$ & Depreciation rate, non-renewable capital & 0.006 \\

$h$ & Adjustment cost parameter, final goods capital & 10.000 \\
$h_{ER}$ & Adjustment cost parameter, renewable sector & 10.000 \\
$h_{ENR}$ & Adjustment cost parameter, non-renewable sector & 10.000 \\

$\gamma_{ER}$ & Capital elasticity, renewable energy production & 1.000 \\

$\alpha_{K,NR}$ & Capital elasticity, non-renewable production & 0.800 \\
$\alpha_S$ & Resource extraction elasticity, non-renewable & 0.200 \\

$A_{disc}$ & Exploration efficiency parameter & 0.100 \\
$\alpha_D$ & Returns to exploration & 0.600 \\
$\xi$ & Extraction cost scale parameter & 0.500 \\
$\psi_D$ & Stock scarcity elasticity & 1.000 \\
$\bar D$ & Reference resource stock & 10.000 \\

$\tau_{ER}$ & Subsidy on renewable energy (benchmark)& 0.000 \\
$\tau_{ENR}$ & Tax on non-renewable energy (benchmark) & 0.000 \\
$p_z$ & Carbon price (benchmark) & 0.000 \\

$\phi_1$ & Abatement cost scale parameter & 0.1850 \\
$\phi_2$ & Abatement cost curvature parameter & 2.800 \\

$\zeta_F$ & Emission elasticity, firms & 0.690 \\
$\zeta_{HH}$ & Emission elasticity, households & 0.690 \\

$\delta_M$ & Depreciation of pollution stock & 0.0021 \\

$\phi_Q$ & Weight on environmental quality in utility & 0.2000\\
$\rho_Q$ & CRRA coefficient for environmental quality & 2.0000\\
$\bar Q$ & Environmental quality absent pollution pressure & 1.0000\\
$\xi_M$ & Pollution-to-environmental-quality sensitivity & 0.0001\\

$\gamma_0$ & Damage function constant term & 1.395e-3 \\
$\gamma_1$ & Damage function linear term & -6.6722e-6 \\
$\gamma_2$ & Damage function quadratic term & 1.4647e-8 \\

$\rho_{A_Y}$ & Persistence of TFP, final goods & 0.900 \\
$\rho_{A_{ER}}$ & Persistence of TFP, renewable sector & 0.900 \\
$\rho_{A_{ENR}}$ & Persistence of TFP, non-renewable sector & 0.900 \\

\bottomrule
\end{tabular}
\end{table}

\section{Results}
\label{section_results}
\subsection{Unanticipated shock to non-renewable energy productivity}
\label{subsection_shock_AENR}

To illustrate the propagation mechanisms of the model, I consider a one-percent unexpected increase in non-renewable energy productivity, $A_{ENR}$, around the benchmark laissez-faire equilibrium with $p_{z,t} = 0$, $\tau_{ER,t} = 0$, and $\tau_{ENR,t} = 0$. Figures \ref{fig:AENR_shock1} and \ref{fig:AENR_shock2} report the impulse responses. The shock behaves as a conventional positive fossil supply shock: it lowers fossil energy prices, increases output and fossil energy use, weakens abatement incentives, and raises emissions through both higher energy demand and lower mitigation effort. Robustness to alternative parameter values is reported in \ref{appendix_sensitivity_shocks}.\\

\begin{figure}[!ht]
\centering
\begin{subfigure}{\linewidth}
\centering
\includegraphics[width=\linewidth]{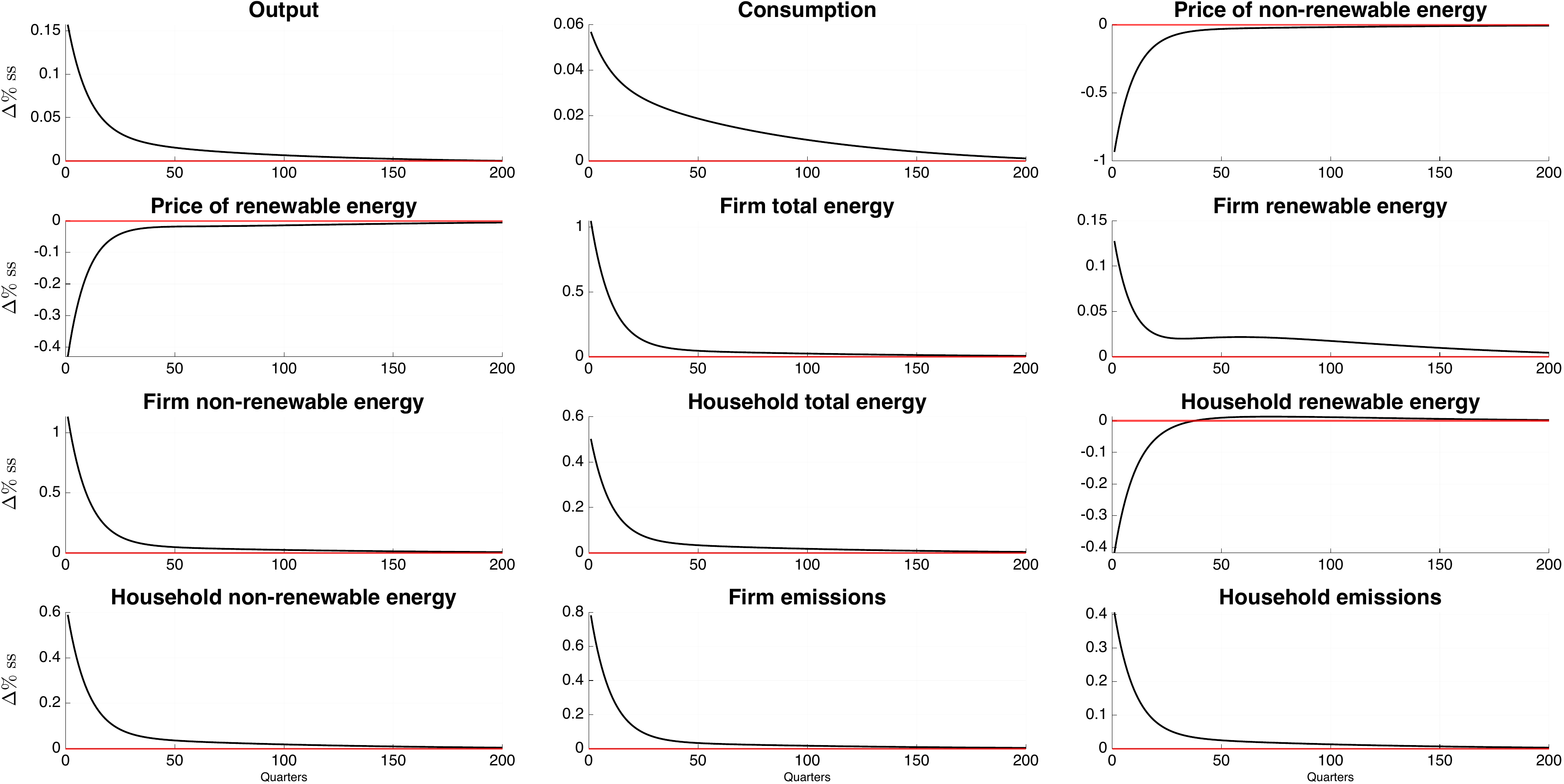}
\caption{Macroeconomic and energy-market impulse responses to a 1\% increase in non-renewable energy productivity ($A_{ENR}$).}
\label{fig:AENR_shock1}
\end{subfigure}
\end{figure}

\begin{figure}[!ht]\ContinuedFloat
\centering
\setcounter{subfigure}{1}
\begin{subfigure}{\linewidth}
\centering
\includegraphics[width=\linewidth]{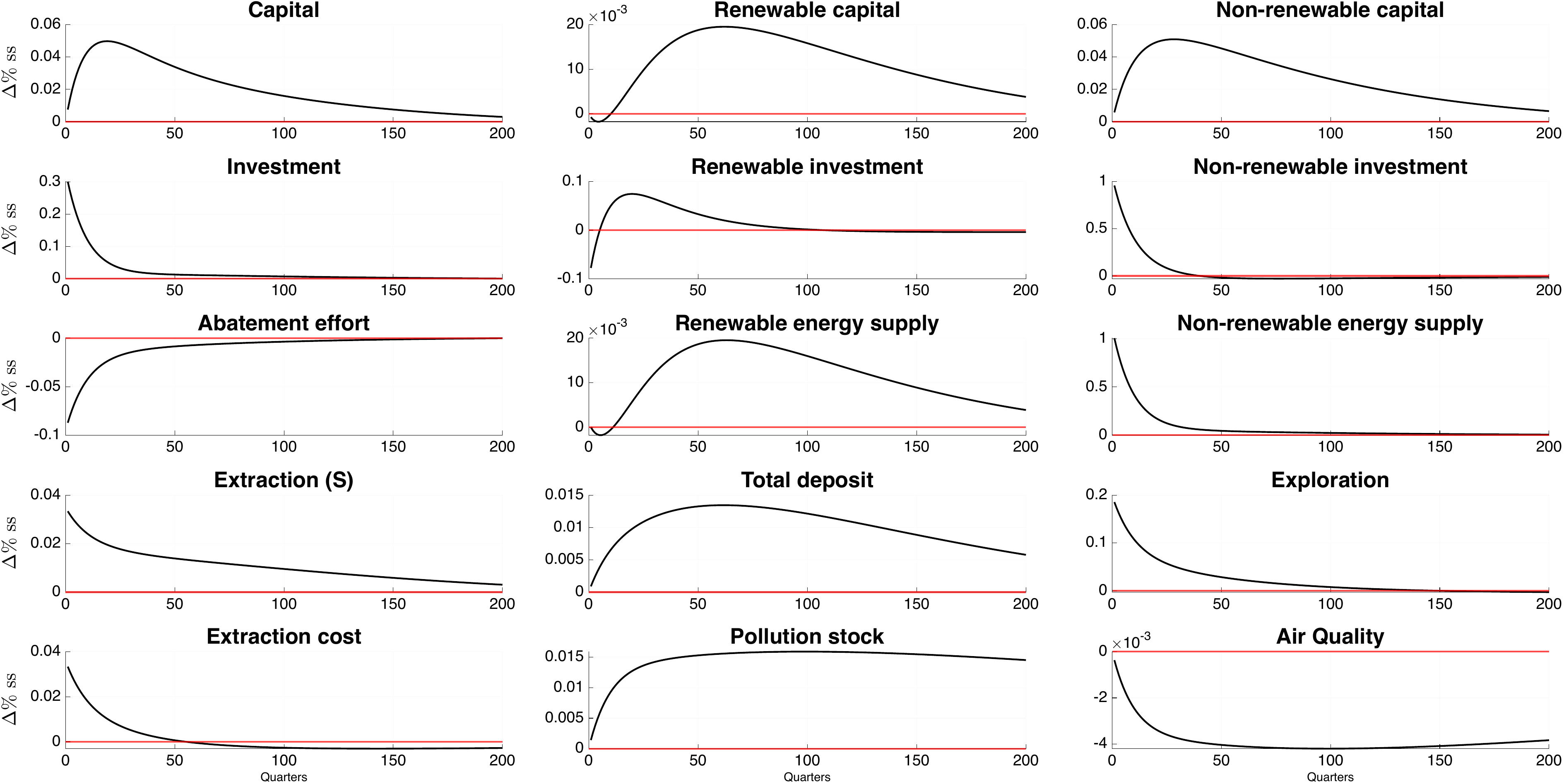}
\caption{Capital accumulation, resource sector, and environmental impulse responses to a 1\% increase in non-renewable energy productivity ($A_{ENR}$).}
\label{fig:AENR_shock2}
\end{subfigure}
\label{fig:AENR_shock_combined}
\end{figure}

\textit{Aggregate and price effects.} Higher fossil productivity lowers the marginal cost of non-renewable energy production, causing the fossil energy price $P_{ENR,t}$ to fall by roughly 0.9\% below its steady state on impact. Because energy enters both utility and production, the decline in energy prices acts as a positive real income shock. Output rises by about 0.15\%, closely reflecting the calibrated energy share in production ($\alpha_E = 0.15$). Consumption increases by roughly 0.06\%, while the marginal utility of wealth $\lambda_t$ declines. Higher returns to final-goods capital also stimulate investment in the final-goods sector.\\

\textit{Energy demand and substitution patterns.} The decline in the relative price $P_{ENR,t}/P_{ER,t}$ induces households to substitute toward fossil energy within the CES energy bundle. Household fossil energy demand $E_{HHNR,t}$ rises by about 0.6\%, while renewable demand $E_{HHR,t}$ falls by approximately 0.4\%. Firms respond through both substitution and scale effects. Lower energy costs increase output and total energy demand, causing fossil energy use $E_{FNR,t}$ to rise by more than 1\%. Renewable energy use in production also increases slightly because the expansion in production dominates the substitution away from renewables. Since no carbon price or energy taxes are present, the lower fossil price translates directly into higher fossil demand.\\

\textit{Resource sector and sectoral spillovers.} As demand shifts toward fossil energy, the renewable energy price \(P_{ER,t}\) falls, reducing renewable-sector profitability and weakening incentives to accumulate renewable capital. Higher fossil productivity also increases the profitability of extraction and exploration: extraction \(S_t\) rises by roughly 0.03\%, while exploration expenditures increase by about 0.18\%. Since exploration exhibits diminishing returns, discoveries initially exceed extraction, causing underground reserves \(D_t\) to rise temporarily in a hump-shaped manner. The shadow value of reserves \(\lambda_{D,t}\) also increases on impact, reflecting the higher marginal productivity of underground resources. Abatement effort declines by about 0.087\% below steady state and gradually returns to baseline, but this response is small in economic terms because laissez-faire abatement is close to zero.\\

\textit{Environmental implications.} The fossil productivity shock increases emissions from both firms and households. Firm emissions \(Z_{F,t}\) rise by about 0.8\%, while household emissions \(Z_{HH,t}\) rise by roughly 0.4\%. The main environmental effect is therefore driven by higher fossil energy use rather than by economically meaningful changes in mitigation. Because pollution depreciates slowly, the rise in emissions generates a persistent increase in the pollution stock and a corresponding decline in air quality. \\

Overall, the experiment shows that productivity improvements in the fossil sector generate short-run macroeconomic expansions while amplifying the climate externality through higher fossil use. The qualitative responses are consistent with DSGE climate models such as \cite{Heutel2012}, although the explicit resource and exploration block amplifies persistence in the pollution dynamics.

\subsection{Unanticipated shock to renewable energy productivity}
\label{subsection_shock_AER}

Figures \ref{fig:AER_shock1} and \ref{fig:AER_shock2} report the impulse responses to a one-percent persistent productivity shock in renewable energy productivity, $A_{ER}$, with persistence $\rho_{A,ER}=0.9$. As in Section \ref{subsection_shock_AENR}, the economy is initially in the benchmark laissez-faire equilibrium with $p_{z,t} = 0$, $\tau_{ER,t} = 0$, and $\tau_{ENR,t} = 0$. Relative to the fossil productivity shock, the aggregate responses are substantially smaller because renewable energy initially represents only a small share of the production energy bundle. Nevertheless, the shock generates environmentally favorable dynamics by shifting energy demand toward renewables and reducing aggregate fossil use. An analogous experiment with an anticipated increase in renewable productivity is reported in \ref{appendix_anticipated_AER_shock}.

\begin{figure}[!ht]
\centering
\begin{subfigure}{\linewidth}
\centering
\includegraphics[width=\linewidth]{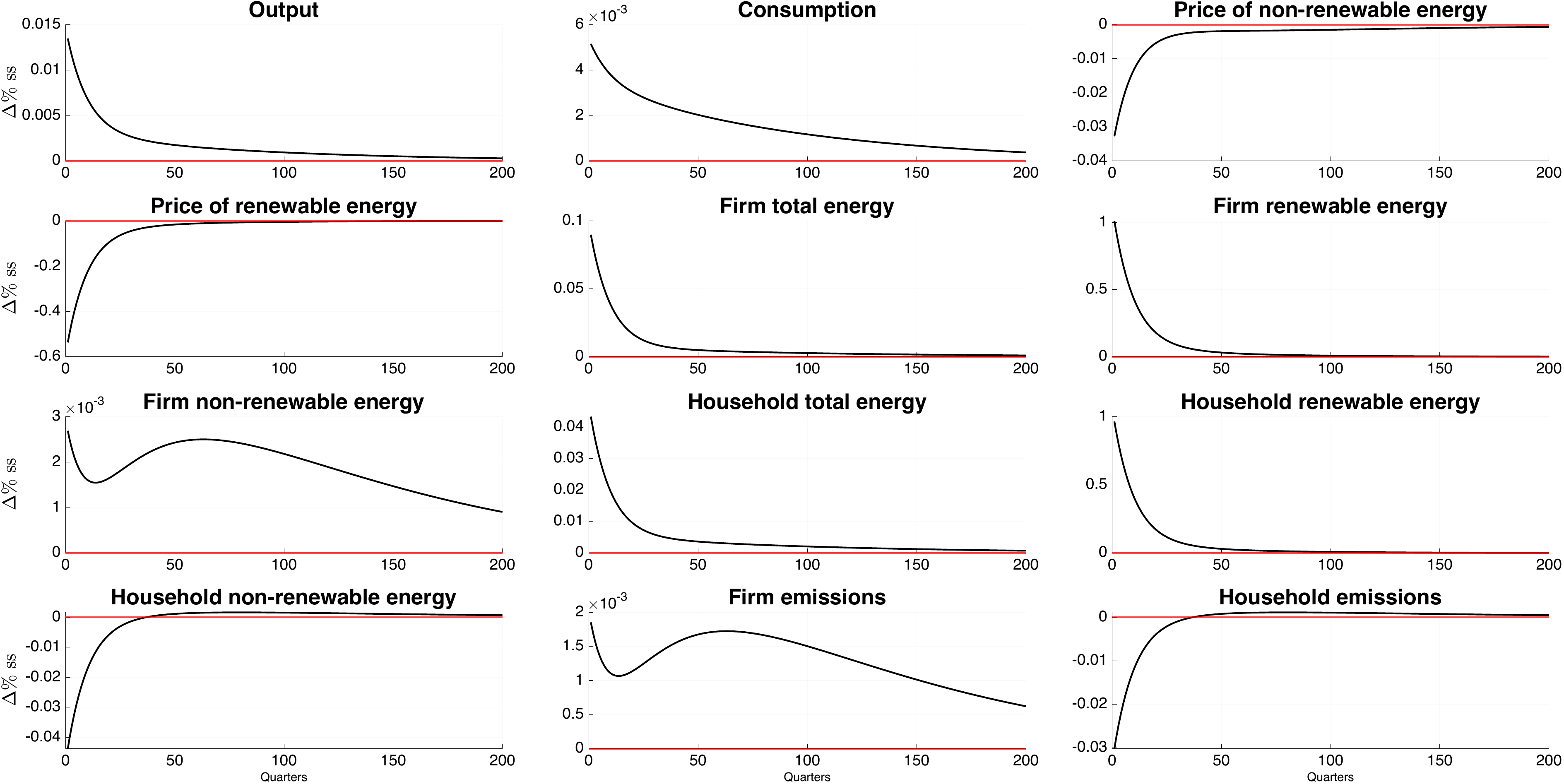}
\caption{Macroeconomic and energy-market impulse responses to a 1\% increase in renewable energy productivity ($A_{ER}$).}
\label{fig:AER_shock1}
\end{subfigure}
\end{figure}

\begin{figure}[!ht]\ContinuedFloat
\centering
\setcounter{subfigure}{1}
\begin{subfigure}{\linewidth}
\centering
\includegraphics[width=\linewidth]{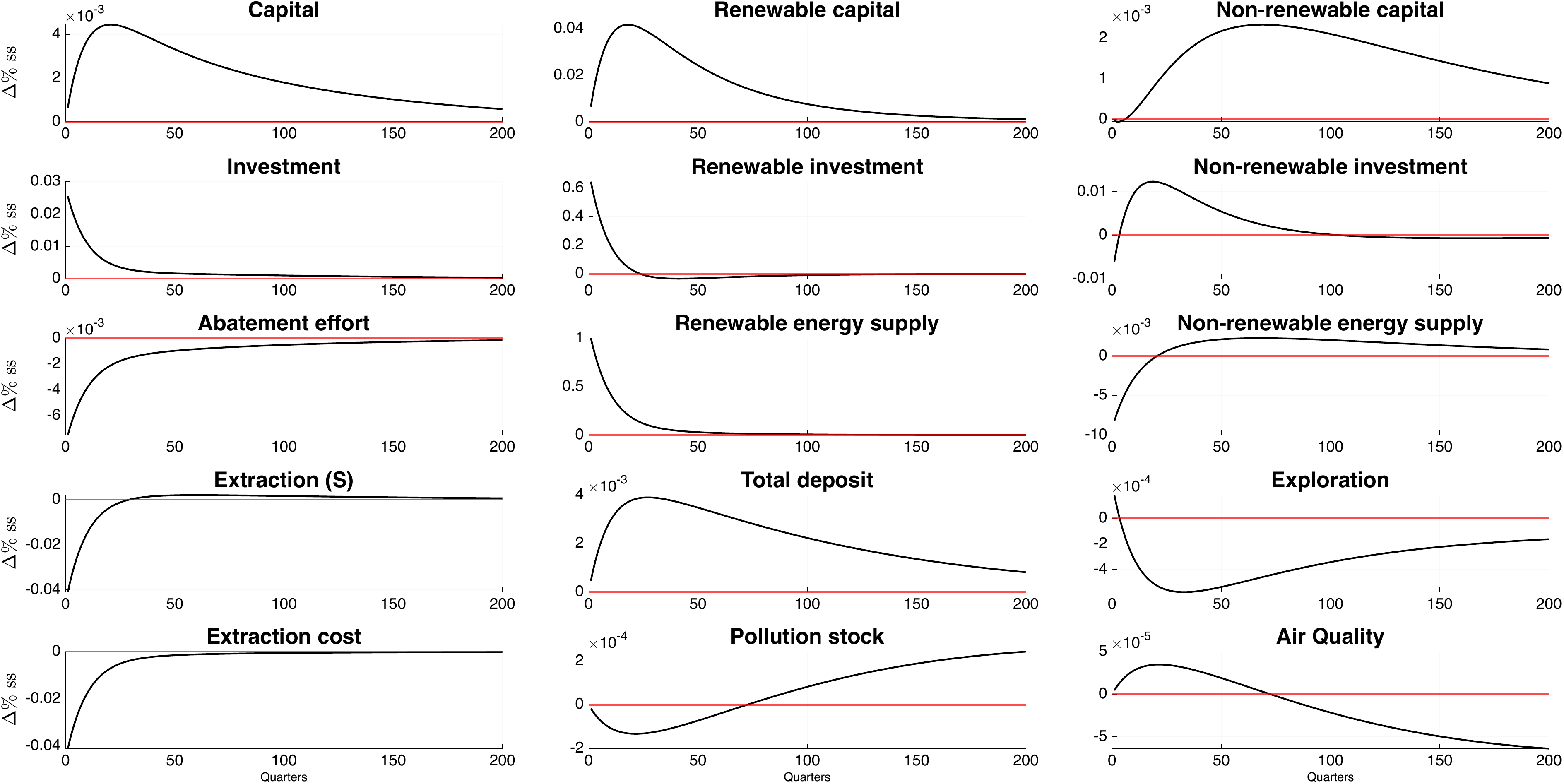}
\caption{Capital accumulation, resource sector, and environmental impulse responses to a 1\% increase in renewable energy productivity ($A_{ER}$).}
\label{fig:AER_shock2}
\end{subfigure}
\label{fig:AER_shock_combined}
\end{figure}

\textit{Aggregate and price effects.} Higher renewable productivity shifts the renewable supply schedule outward. Renewable energy production rises immediately by 1\%, while the renewable energy price $P_{ER,t}$ declines by roughly 0.55\%. Despite the fall in prices, renewable-sector revenues increase because the quantity expansion dominates. As a result, the rental rate of renewable capital rises and renewable investment increases by about 0.6\%, generating gradual accumulation of renewable capital. Due to convex adjustment costs, renewable capital follows a hump-shaped transition path. Nevertheless, the aggregate macroeconomic effects remain modest. Output rises by only about 0.015\%, while consumption increases slightly. This muted response reflects the benchmark energy mix, where production relies predominantly on non-renewable energy and renewables receive a weight of only $\Omega_F = 0.02$. Consequently, even a large increase in renewable productivity generates only a small increase in total energy input and aggregate production.\\

\textit{Energy demand and substitution patterns.} The decline in the relative price of renewable energy induces substitution toward renewables in both household and firm energy demand. Household renewable energy use rises by about 1\%, while household fossil energy demand declines slightly. Firms similarly substitute toward renewable inputs, causing renewable energy use in production to increase by 1\%. However, because fossil energy remains the dominant input in the firm energy bundle, the associated scale effect offsets much of the substitution effect, leaving industrial fossil energy demand nearly unchanged.\\

\textit{Resource sector spillovers.} Although industrial fossil demand changes little, the substitution away from fossil energy by households lowers aggregate fossil demand. Fossil energy prices therefore decline, generating a contractionary effect on the resource sector. Extraction and exploration both fall modestly, reflecting weaker expected profitability of fossil resources. The shadow value of reserves $\lambda_{D,t}$ initially rises slightly because of forward-looking extraction incentives, but subsequently declines below steady state as future fossil rents fall.\\

\textit{Environmental implications.} The renewable productivity shock generates environmentally favorable dynamics. Household emissions decline because households substitute away from fossil energy, while industrial emissions rise only marginally due to the small expansion in aggregate output. Overall emissions therefore fall initially, causing a temporary decline in the pollution stock. Over time, the pollution stock gradually returns close to its steady-state path. The asymmetry relative to the fossil productivity shock highlights that, when renewables represent only a small share of the initial energy mix, clean productivity improvements have limited aggregate macroeconomic effects despite reducing emissions.

\subsection{Environmental policy analysis}
\label{section_welfare_analysis}
Having characterized the propagation of energy productivity shocks in the benchmark economy, the analysis now turns to the normative evaluation of climate policy. To assess the welfare implications of alternative policy regimes, equilibrium allocations are compared across different combinations of climate policy instruments. For each regime \(j\), the available policy variables are chosen to maximize the representative household's lifetime utility subject to the decentralized equilibrium conditions. When multiple policy instruments are available simultaneously, they are chosen jointly.

Unlike private agents in the decentralized equilibrium, the Ramsey planner chooses policy instruments to maximize household lifetime utility subject to the full set of equilibrium conditions, including the laws of motion for the pollution stock and environmental quality. Because \(Q_t\) enters household utility directly and \(M_t\) affects production through the damage function \(D(M_t)\), the planner internalizes the welfare cost of emissions when choosing \(\tau_{ER}\), \(\tau_{ENR}\), and \(p_z\), whereas private agents take \(M_t\) and \(Q_t\) as given. In this section, the Ramsey exercise is restricted to constant policy instruments. For each admissible policy set, I solve for the constant values of \(\tau_{ER}\), \(\tau_{ENR}\), and/or \(p_z\) that maximize household lifetime utility. Welfare comparisons are then conducted relative to the laissez-faire benchmark using the consumption-equivalent variation (CEV), evaluated over the full transition path generated by each policy regime. Let
\[ V^{LF}=\sum_{t=0}^{\infty} \beta^t U\big(C_t^{LF},E_{HH,t}^{LF}, Q_{t}^{LF}\big) \]
denote lifetime utility under laissez-faire and
\[ V^{j}=\sum_{t=0}^{\infty}\beta^t U\big(C_t^{j},E_{HH,t}^{j},Q_{t}^{j}\big) \]
the corresponding lifetime utility under policy regime \(j\).

The consumption-equivalent variation $\lambda_j$ is defined as the constant proportional change in consumption in the laissez-faire allocation that makes the representative household indifferent between the two allocations:
\[
\sum_{t=0}^{\infty}\beta^t U\big((1+\lambda_j)C_t^{LF},E_{HH,t}^{LF},Q_{t}^{LF}\big)=V^j.\]

Given the separable CRRA specification of household preferences, the above expression can be written as
\[(1+\lambda_j)^{1-\rho_C}V_C^{LF}+ V_E^{LF} + V_Q^{LF} = V^j,\]
where
\[ V_C^{LF} = \sum_{t=0}^{\infty} \beta^t \frac{\psi_C (C_t^{LF})^{1-\rho_C}}{1-\rho_C}, \quad V_E^{LF} = \sum_{t=0}^{\infty} \beta^t \frac{\psi_E (E_{HH,t}^{LF})^{1-\rho_E}}{1-\rho_E}, \quad \text{ and} \quad V_Q^{LF}=\sum_{t=0}^{\infty}\beta^t\frac{\phi_Q (Q_t^{LF})^{1-\rho_Q}}{1-\rho_Q}.\]

\noindent Solving for \(\lambda_j\) yields the following expression for the consumption-equivalent variation:
\begin{align}
CEV_j \equiv 100 \lambda_j =100 \left[\left(\frac{V^j - V_E^{LF} - V_Q^{LF}}{V_C^{LF}}\right)^{\frac{1}{1-\rho_C}}-1\right].
\end{align}

With this convention, a positive CEV means that policy \(j\) is preferred to laissez-faire: consumption in the laissez-faire allocation would have to be increased by \(\lambda_j\) to make households indifferent to policy \(j\). A negative CEV means that policy \(j\) is worse than laissez-faire: consumption in the laissez-faire allocation would have to be reduced by \(|\lambda_j|\) to make households indifferent. Expressing welfare in consumption-equivalent units provides a transparent metric that integrates environmental benefits and macroeconomic distortions into a single welfare measure. 

The welfare comparisons in Table \ref{tab:welfare_policy_comparison} evaluate unconstrained optimal policy regimes by computing consumption-equivalent variation over the full transition path from the laissez-faire steady state to the policy steady state. These allocations therefore provide a benchmark for evaluating the constrained transition experiments studied in Sections \ref{section_transition_50_ren_energy} and \ref{section_cap_emissions}. By contrast, the transition experiments solve constrained Ramsey problems in which policy instruments are chosen to maximize welfare subject to a predetermined emissions target, so the resulting welfare differences reflect alternative implementations of the same environmental objective rather than unconstrained optimal policy.

Table \ref{tab:welfare_policy_comparison} reports the optimal constant policy values, steady-state output and emissions relative to laissez-faire, and the dynamic consumption-equivalent welfare change. Output and emissions are normalized by their laissez-faire steady-state levels, denoted \(Y_{LF}\) and \(Z_{LF}\). Thus, \(Y_j/Y_{LF}\) and \(Z_j/Z_{LF}\) reflect steady-state comparisons, while the CEV incorporates both transitional dynamics and long-run allocations. The time subscripts in \(\tau_{ER,t}\), \(\tau_{ENR,t}\), and \(p_{z,t}\) reflect the general model notation; in this table, the reported instruments are constant along the transition.

\begin{table}[H]
\centering
\caption{Policy regimes: steady-state outcomes and dynamic welfare}
\label{tab:welfare_policy_comparison}

\resizebox{\linewidth}{!}{
\begin{tabular}{lcccccc}
\toprule
\textbf{Policy Regime} & $\tau_{ER,t}$ & $\tau_{ENR,t}$ & $p_{z,t}$ & $Y_j/Y_{LF}$ & $Z_j/Z_{LF}$ & \textbf{CEV (\%)} \\
\midrule
Only Renewable Subsidy ($\tau_{ER,t}$) 
& 0.0588 & 0 & 0 & 1.0021 & 0.9972 & 1.1144 \\

Only Fossil Energy Tax ($\tau_{ENR,t}$) 
& 0 & 0.1420 & 0 & 0.9864 & 0.9139 & 1.3079 \\

Only Carbon Price ($p_{z,t}$) 
& 0 & 0 & 0.0271 & 1.0059 & 0.8436 & 1.7073 \\

Fossil Energy Tax + Carbon Price ($\tau_{ENR,t}+p_{z,t}$) 
& 0 & 0.0368 & 0.0232 & 1.0018 & 0.8389 & 1.7159 \\

Renewable Subsidy + Carbon Price ($\tau_{ER,t}+p_{z,t}$) 
& 0.0269 & 0 & 0.0269 & 1.0068 & 0.8431 & 1.7092 \\

Renewable Subsidy + Fossil Energy Tax ($\tau_{ER,t}+\tau_{ENR,t}$) 
& 0.0060 & 0.1406 & 0 & 0.9868 & 0.9140 & 1.3080 \\

All Instruments
& 0.0166 & 0.0354 & 0.0233 & 1.0026 & 0.8387 & 1.7166 \\

\bottomrule

\multicolumn{7}{p{18cm}}{\footnotesize
\textit{Notes:} $\tau_{ER,t}$ denotes the renewable subsidy, $\tau_{ENR,t}$ the fossil energy input tax, and $p_{z,t}$ the carbon price. Instrument values correspond to the Ramsey-optimal steady state conditional on the available policy set. Although written with time subscripts to match the model's notation, the instruments are constant in the regimes reported in this table. $Y_j/Y_{LF}$ denotes steady-state output relative to the laissez-faire steady state and $Z_j/Z_{LF}$ denotes steady-state total emissions relative to laissez-faire. In the benchmark equilibrium, $Y_{LF}=3.4004$ and $Z_{LF}=Z_F+Z_{HH}=4.3776$. CEV denotes the dynamic consumption-equivalent variation computed over the entire transition from the laissez-faire steady state to the policy steady state, including the continuation value at the terminal steady state.}
\end{tabular}
}
\end{table}

Several patterns emerge from the welfare comparison. First, all policy regimes generate positive welfare gains relative to laissez-faire once transition dynamics are taken into account. The dynamic consumption-equivalent variation ranges from 1.11\% under a stand-alone renewable subsidy to about 1.72\% under regimes that combine a carbon price with at least one additional instrument. These gains reflect the fact that environmental quality enters household utility directly, so emissions reductions generate welfare benefits beyond their effect on production damages. The gains arise because the policies reduce the environmental externality while the macroeconomic distortions remain relatively modest in this calibration.

Second, among single-instrument policies, the carbon price produces the largest welfare improvement (CEV $=1.71\%$). By directly pricing emissions, it internalizes the climate externality at its source and delivers the largest emissions reduction among the single-instrument regimes. In contrast, the renewable subsidy and fossil input tax operate through narrower input margins and therefore generate smaller welfare gains (CEV $=1.11\%$ and $1.31\%$, respectively).

Third, policy interactions reveal that combining instruments yields only limited additional gains once the carbon externality is priced. The carbon-price-only regime generates a CEV of 1.7073\%. Adding a renewable subsidy raises this to 1.7092\%, while adding a fossil energy tax raises it to 1.7159\%. The larger gain from adding \(\tau_{ENR}\) rather than \(\tau_{ER}\) reflects the larger role of non-renewable energy in the initial energy mix, which makes the fossil-demand margin quantitatively more important than the renewable-investment margin in this calibration. The full three-instrument regime delivers the highest welfare gain, but only slightly, with a CEV of 1.7166\%. These small differences indicate that, once emissions are directly priced, additional instruments mainly reallocate distortions across margins rather than correcting a separate market failure.

The welfare ranking should be interpreted in light of the model's structure. The economy features flexible prices, a representative household, and no labor supply distortions or endogenous innovation. Moreover, only firms have access to a dedicated abatement technology, while households reduce emissions through energy demand substitution. In this environment, policies that act directly on emissions perform well because they internalize the externality while preserving the allocation of productive capital. Policies operating through indirect input wedges affect production and capital allocation more broadly, which limits their welfare gains. Introducing labor supply distortions, heterogeneous agents, or directed technical change could amplify the welfare differences across instruments. The Ramsey results in Table~\ref{tab:welfare_policy_comparison} therefore establish a benchmark in which directly pricing emissions captures most of the welfare gains generated by the available policy instruments. The constrained transition experiments in Sections \ref{section_transition_50_ren_energy} and \ref{section_cap_emissions}, together with the welfare comparison in Table \ref{tab:dynamic_welfare_comparison}, show how this benchmark ranking interacts with the timing and coverage of policy when the long-run emissions reduction is held fixed but the transition path differs.

\subsection{Transition to a 50\% renewable energy economy under a carbon tax}
\label{section_transition_50_ren_energy}
This section analyzes a price-based transition implemented through a carbon tax calibrated to achieve a 50\% renewable energy share. The carbon price is chosen to decentralize a steady state with $E_R = E_{NR}$, implying a 50\% renewable energy share in total energy use.

An unanticipated and permanent increase in the industrial carbon tax from $p_{z,t} = 0$ to $p_{z,t} = 0.30409$ is simulated. The policy change occurs in the first period and remains fixed thereafter. From that date onward, agents have perfect foresight over the entire transition path. In short, this experiment captures a front-loaded, price-based transition in which a permanently high carbon tax is used to reach a cleaner energy mix, generating large immediate adjustment costs before renewable capital has time to expand. This assumption is useful for isolating the model’s transition mechanism, but it abstracts from uncertainty about policy credibility, technological progress, and climate damages.

The discrete increase in \(p_{z,t}\) immediately raises the effective cost of fossil energy faced by firms, even though the equilibrium producer price of non-renewable energy declines on impact due to the sharp contraction in demand. Because fossil capital depreciates slowly and renewable capacity expands only gradually due to convex installation costs, the reallocation of the energy capital stock unfolds over several decades. The transition therefore isolates the resource costs generated by capital reallocation, abatement, and changes in the sectoral composition following a sudden tightening of climate policy.

This experiment should not be interpreted as the globally optimal dynamic policy path. Instead, it illustrates the macroeconomic adjustment required to reach a cleaner energy mix when the policy instrument is a constant carbon price chosen to satisfy a predetermined renewable share. The resulting steady state features an emissions reduction of approximately 52.26\% relative to the laissez-faire benchmark, providing the reference emissions target for the quantity-based policies studied in Section \ref{section_cap_emissions}. In contrast to a fully optimal time-varying policy, the tax is held fixed so that the transition dynamics reflect the economy's internal adjustment through investment, capital depreciation, and energy substitution.\\

\begin{figure}[!ht]
    \centering
    \includegraphics[width=\linewidth]{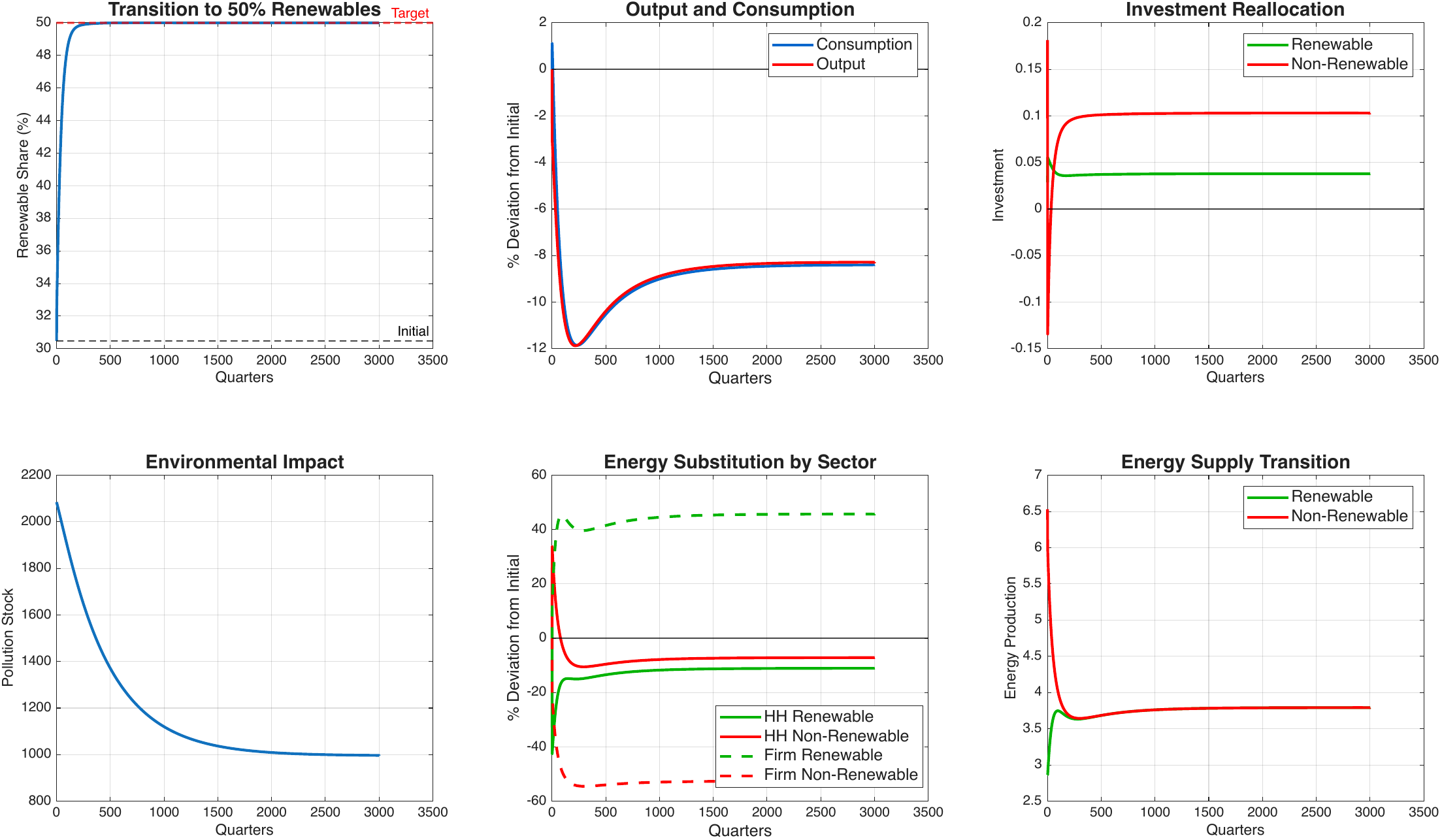}
    \caption{Transition dynamics under a permanent unanticipated increase in the carbon price to $p_{z,t} = 0.30409$, targeting a 50\% renewable energy share.}
    \label{fig:transition_half_renewable}
\end{figure}

\textit{Immediate Macroeconomic Contraction.} The carbon tax sharply increases the effective marginal cost of fossil energy. Firms respond immediately by reducing fossil energy demand and sharply increasing abatement effort on impact, moving from a negligible baseline under laissez-faire to a larger positive level. The large percentage deviation reflects the near-zero initial level rather than an extreme real adjustment. Output and consumption fall immediately following the policy shock. As shown in Figure \ref{fig:transition_half_renewable}, both variables decline by 12\% at their peak and then recover partially, converging to a new steady state approximately 8\% below the initial level. Additional details on other variables are reported in \ref{appendix_transition_to_50_renewable_energy}.

The contraction reflects three mechanisms operating simultaneously: (i) higher effective energy input costs induced by the carbon tax, (ii) increased abatement expenditures that divert resources from consumption and investment, and (iii) a sudden contraction in fossil energy supply before renewable capacity has expanded sufficiently to replace it. Because renewable capital is predetermined and installation costs are convex, the economy temporarily operates with insufficient effective energy capacity.

In this experiment, carbon-tax revenue is rebated to households through lump-sum transfers, which isolates the macroeconomic adjustment generated by the carbon-price wedge itself and holds the revenue-recycling channel neutral. Alternative schemes that direct carbon revenue toward renewable subsidies or other investment support could ease the transition by accelerating renewable capacity accumulation. Section \ref{section_cap_emissions} therefore compares this carbon-price-only transition with broader policy packages that pair the carbon price with renewable subsidies and non-renewable energy taxes; these should be interpreted as showing how complementary fiscal instruments support energy-capital reallocation, rather than as a fully specified earmarking rule. Because the model abstracts from distortionary labor and capital taxation, the standard double-dividend channel emphasized by \citet{BovenbergGoulder1996} and \citet{Parry1997} is absent.

Over time, renewable capital accumulation partially restores productive capacity. Output and consumption recover slightly and converge to levels that remain permanently below the laissez-faire steady state, reflecting the lower long-run fossil energy use under the carbon-price wedge and the persistent resources devoted to abatement and renewable capacity maintenance.\\

\textit{Intersectoral Capital Reallocation.} The transition is driven by large investment reallocation across energy sectors. Fossil investment collapses immediately and becomes strongly negative relative to the initial steady state, implying net disinvestment in that sector. Consequently, fossil capital \(K_{ENR}\) declines persistently and contracts by roughly 43\% in the long run.

In contrast, renewable investment jumps sharply and remains elevated, leading to a steady expansion of renewable capital. The renewable capital stock increases by about 32\% as the economy moves toward the new steady state consistent with a 50\% renewable energy share.

Final goods investment initially falls together with output and converges to a new steady state 11.5\% below the baseline. Because capital accumulates gradually, the decline in investment slows the adjustment of the capital stock, contributing to the persistence of output losses.\\

\textit{Fossil Resource Sector Dynamics.} Exploration expenditures \(Exp_{NR,t}\) fall by about 45\% on impact and remain substantially below their initial steady state throughout the transition, reaching a maximum decline of about 53\%. Extraction \(S_t\) also declines sharply, falling by about 35\% on impact and reaching a trough of about 37\% below the initial steady state. Because discoveries decline more strongly than extraction, the stock of reserves \(D_t\) gradually declines over the transition. The fossil sector therefore contracts along two dimensions: lower current extraction and reduced reserve replenishment.

The contraction of fossil-sector activity is also reflected in the shadow value of reserves, $\lambda_{D,t}$, which declines following the carbon tax shock. This multiplier measures the value of an additional unit of recoverable reserves and can therefore be interpreted as the scarcity rent associated with underground fossil resources. As the long-run demand for fossil energy falls, future extraction becomes less profitable, reducing the value of reserves and weakening incentives for both exploration and extraction.\\

\textit{Structural Energy Substitution.} The carbon tax induces a large and persistent reallocation of energy demand across sectors. Firm fossil energy use declines by approximately 53\% relative to its initial steady state, while renewable energy demand by firms increases by roughly 45\%. Despite this substitution, total firm energy demand falls substantially because renewable capacity cannot immediately replace the sharp contraction in fossil inputs. The economy therefore reaches the emissions target through a combination of substitution and aggregate energy compression.

Household adjustment differs from that of firms. On impact, the sharp decline in the producer price of fossil energy, driven by the collapse in firm demand, induces households to substitute toward fossil energy within their energy bundle. This substitution effect is partly offset over time by the income effect from lower output, which compresses overall household energy demand. Because households are not directly subject to the carbon tax, they respond to the relative price movement. As income declines and the transition unfolds, however, both renewable and non-renewable household energy use fall below steady state, leading to a persistent decline in total household energy demand.

Energy substitution therefore operates through distinct margins across sectors: immediate policy-induced substitution in production, temporary relative-price substitution in household demand, and a broader contraction in aggregate energy use driven by reduced output and capital adjustment.\\

\textit{Environmental Effects.} The structural decline in fossil energy use results in a large decrease in emissions. Both firm and household emissions fall substantially. As illustrated in Figure \ref{fig:transition_half_renewable}, the pollution stock declines monotonically and converges to a new steady state approximately 52\% below its initial level. Given the low atmospheric depreciation rate, improvements in environmental quality occur gradually but accumulate over time as emissions remain persistently lower. \\

\textit{Interpretation.} The transition cost arises from the interaction between immediate fossil energy compression and the gradual adjustment of the broader energy system. A counterfactual transition with near-zero capital adjustment costs raises welfare only modestly, from \(-5.90\%\) to \(-5.73\%\), indicating that installation frictions affect the timing of adjustment but do not account for most of the welfare loss. The main cost instead comes from the abrupt reduction in effective fossil energy input, the associated fall in output and consumption, and the persistent resource reallocation required to sustain the lower-emissions steady state.

The preceding results characterize the transition generated by a carbon-price-only policy that is imposed immediately and calibrated to achieve a 50\% renewable energy share. This provides a useful benchmark against which alternative policy designs can be evaluated. Section~\ref{section_cap_emissions} considers whether the same long-run emissions target can be achieved at lower welfare cost through broader instrument packages, alternative sectoral coverage, or more gradual implementation paths. As the welfare comparison in Table~\ref{tab:dynamic_welfare_comparison} shows, both implementation timing and instrument breadth contribute to the welfare loss: an immediate full policy package, which retains immediate implementation but adds a renewable subsidy and a non-renewable energy tax, reduces the welfare loss by about 2.52 percentage points relative to the carbon-price-only transition. Gradual emissions caps reduce the welfare loss further, especially under firm-only regulation. Both broadening the instrument set and slowing the implementation path therefore reduce transition costs, with gradualism the larger of the two effects. Consistent with this mechanism, the 50\% renewable transition under a constant carbon tax generates the largest welfare loss among the decarbonization policies compared in Table~\ref{tab:dynamic_welfare_comparison}, with a dynamic CEV of \(-5.90\%\) relative to laissez-faire.

% The model therefore emphasizes that rapid decarbonization involves a substantial transition problem even when renewable and non-renewable energy are substitutable. The key challenge is that fossil energy use can contract immediately following policy tightening, whereas renewable capacity expands only gradually through investment and capital accumulation. A more complete set of transition dynamics for most endogenous variables is reported in Figure~\ref{fig:transition_half_renewable_all_vars}. Appendix~\ref{appendix_sensitivity_adj_costs} shows that alternative calibrations of capital adjustment frictions affect the timing and peak magnitude of the output contraction in the carbon-tax transition, while leaving the long-run levels of output, consumption, and energy composition essentially unchanged.

% Compared with the emissions-cap regimes analyzed in Section~\ref{section_cap_emissions}, the tax-based transition operates through an immediate relative-price adjustment rather than a gradual feedback rule. Although capital stocks themselves adjust only gradually, the immediate increase in the carbon price compresses fossil energy use before renewable capacity has expanded sufficiently. The transition cost therefore reflects both the limited instrument set of the carbon-price-only policy and the timing mismatch between rapid fossil energy contraction and slow renewable-capital accumulation.

The model therefore shows that the transition cost arises because fossil energy demand contracts immediately following policy tightening, while the economy initially operates with a renewable-capital stock that is too small to fully replace the lost fossil-energy services. As a result, effective energy input falls during the transition, generating persistent declines in output and consumption. Counterfactual exercises with near-zero capital adjustment costs show that installation frictions affect the timing of reallocation only modestly, indicating that the main cost comes from the reduction in effective energy input and the movement toward a lower-emissions allocation rather than from adjustment frictions themselves. Compared with the emissions-cap regimes analyzed in Section~\ref{section_cap_emissions}, the carbon-price-only transition relies on an immediate relative-price adjustment and a single policy instrument. The resulting welfare loss therefore reflects both the abrupt compression of fossil energy use and the absence of complementary instruments that could support the transition. A more complete set of transition dynamics for most endogenous variables is reported in Figure~\ref{fig:transition_half_renewable_all_vars}, while Appendix~\ref{appendix_sensitivity_adj_costs} shows that alternative calibrations of capital adjustment frictions affect the timing and peak magnitude of the output contraction but leave the long-run levels of output, consumption, and energy composition essentially unchanged.

\subsection{Cap on emissions}
\label{section_cap_emissions}
The policy experiment studied in this section is motivated by current European Union (EU) climate commitments. The European Climate Law commits the EU to climate neutrality by 2050, and in December 2025 the Council of the European Union and the European Parliament reached a provisional agreement to amend the European Climate Law by introducing a binding intermediate target of a 90\% reduction in net greenhouse-gas emissions by 2040 relative to 1990 levels \citep{EU2040ClimateTarget}. These commitments reflect an increasingly stringent trajectory for decarbonization in the European Union, requiring a rapid transformation of the energy system over the coming decades. Yet, the feasibility of this trajectory is increasingly contested. In March 2026, TotalEnergies, one of Europe's largest energy companies, stated in its annual sustainability report that global carbon neutrality by 2050 as envisioned under the Paris Agreement is no longer achievable at the current pace of transition, and that its own net-zero ambitions must therefore be reassessed \citep{ReutersTotal2026}. This divergence between stated policy ambition and observed transition speed underscores the importance of studying not only the long-run properties of deep decarbonization but also the transition dynamics and macroeconomic costs associated with moving toward substantially lower emission levels.

While the EU target refers to a 90\% reduction relative to 1990 emissions, the experiment here focuses on the emissions reduction implied by the 50\% renewable energy transition analyzed in Section~\ref{section_transition_50_ren_energy}, where emissions converge to approximately 52\% below the laissez-faire steady state. To ensure comparability across policy regimes, the emissions cap $\bar{Z}$ is set equal to this long-run emissions level. The objective of this section is therefore not to determine the welfare-maximizing emissions level, but rather to compare alternative ways of implementing a common environmental target. All policy regimes considered below are calibrated to achieve the same long-run emissions outcome, so that differences across experiments reflect policy design rather than target stringency.

This target is imposed exogenously and is not chosen to maximize household welfare. The Ramsey welfare analysis in Section~\ref{section_welfare_analysis}, by contrast, studies the unconstrained policy problem and characterizes the welfare-maximizing combination of policy instruments. The cap experiments should therefore be interpreted as constrained decarbonization exercises rather than as the household-welfare optimum. \ref{appendix_emissions_reduction_90_percent} considers a substantially more stringent 90\% emissions reduction scenario. That experiment highlights that beyond a certain threshold, further decarbonization cannot be achieved through energy reallocation alone and instead requires large-scale negative emissions technologies, which substantially increase the macroeconomic cost of the transition.\\

\textit{Policy design.} The analysis begins with a policy that allows the carbon price to adjust endogenously over time in response to deviations of emissions from the target, so that the transition dynamics arise from gradual policy tightening rather than from an immediate change in relative prices. The endogenous feedback rule is:
$$p_{z,t} = p_{z,t-1} + \phi_Z(Z_{tot,t} - \bar{Z})$$
where $Z_{tot,t} = Z_{F,t} + Z_{HH,t}$. The parameter $\phi_Z$ governs the speed at which the carbon price adjusts when aggregate emissions exceed the target. The calibration $\phi_Z = 0.001$ implies a gradual decline in emissions toward the long-run cap without generating unstable or excessively abrupt price dynamics. Under this calibration, emissions achieve roughly 90\% of the total reduction required to reach the target within approximately 150 quarters (about 35--40 years), producing transition dynamics consistent with the slow adjustment of installed energy capital documented in the empirical energy transition literature. Because the rule adjusts the carbon price only gradually, the economy avoids the sharp, front-loaded contraction that would arise under an immediately binding emissions cap.

The rule also resembles institutional mechanisms used in real-world emissions trading systems. In particular, the EU Emissions Trading System's (ETS) Market Stability Reserve (MSR) adjusts the supply of emission allowances according to predefined thresholds for the total number of allowances in circulation (TNAC) \citep{EU_MSR_2023}. Although the MSR operates through quantity adjustments rather than a direct price rule, both mechanisms introduce a rule-based feedback that tightens policy when the system deviates from its intended emissions trajectory. More generally, feedback rules of this type can be interpreted as reduced-form approximations to optimal carbon pricing policies derived in dynamic integrated assessment models, where the carbon price adjusts endogenously over time in response to deviations between realized emissions and the socially optimal emissions path \citep{Golosov2014, Nordhaus2017}. The integral structure of the rule implies that the transition path of the carbon price is determined endogenously by the equilibrium dynamics until emissions converge to \(Z_{tot}=\bar{Z}\).

The carbon price is complemented by a renewable energy subsidy and a non-renewable energy tax, both set proportional to the prevailing carbon price:
$$\tau_{ER,t} = \frac{p_{z,t}}{p_z^*}\tau_{ER}^*, \qquad \tau_{ENR,t} = \frac{p_{z,t}}{p_z^*}\tau_{ENR}^*$$
where $(p_z^*, \tau_{ER}^*, \tau_{ENR}^*)$ are the Ramsey-optimal steady-state instrument values that decentralize the constrained allocation with $Z_{tot} = \bar{Z}$. These are the result of solving a Ramsey problem subject to the competitive equilibrium conditions and the emissions constraint, yielding $p_z^*~=~0.151522$, $\tau_{ENR}^*~=~0.529825$, and $\tau_{ER}^*~=~0.056924$. By scaling all three instruments proportionally to \(p_{z,t}/p_z^*\), the policy package starts at zero in the laissez-faire steady state and converges smoothly to its Ramsey values as the carbon price approaches \(p_z^*\). Government revenue net of subsidy payments is rebated to households through lump-sum transfers, as in Section~\ref{section_transition_50_ren_energy}; the revenue-treatment caveats discussed there apply here as well. Figures~\ref{fig:emisisons_transition_dynamics_part1} and~\ref{fig:emisisons_transition_dynamics_part2} report the resulting transition dynamics.\\

\begin{figure}[!htb]
\centering
\begin{subfigure}{\linewidth}
\centering
\includegraphics[width=\linewidth]{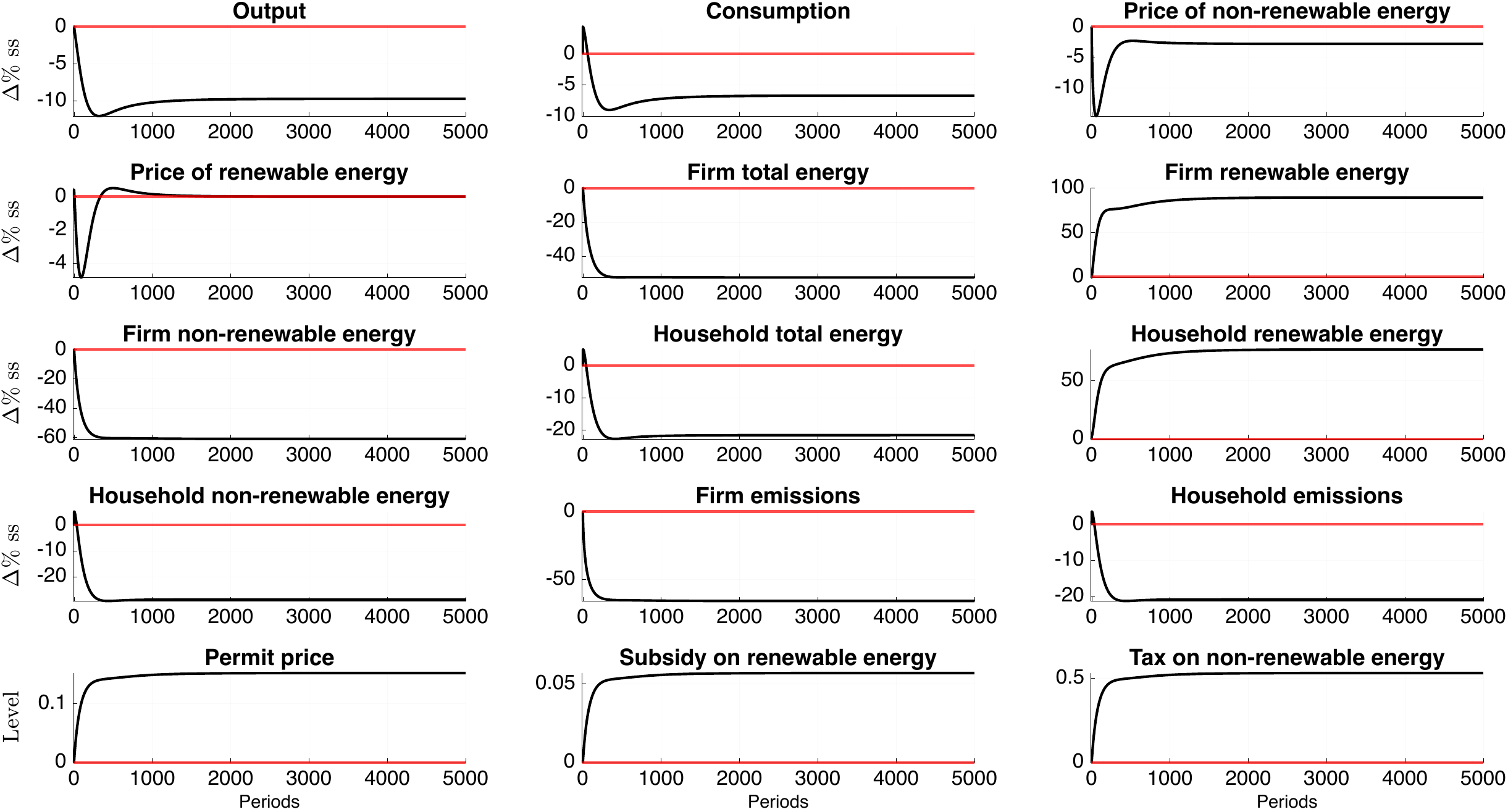}
\caption{Macroeconomic, energy-market, and policy instrument dynamics during the transition to a 52\% emissions reduction under a gradual integral carbon pricing rule. Non-policy variables reflect percentage changes relative to the laissez-faire steady-state benchmark.; policy instruments are reported in levels.}
\label{fig:emisisons_transition_dynamics_part1}
\end{subfigure}
\end{figure}

\begin{figure}[!ht]\ContinuedFloat
\centering
\setcounter{subfigure}{1}
\begin{subfigure}{\linewidth}
\centering
\includegraphics[width=\linewidth]{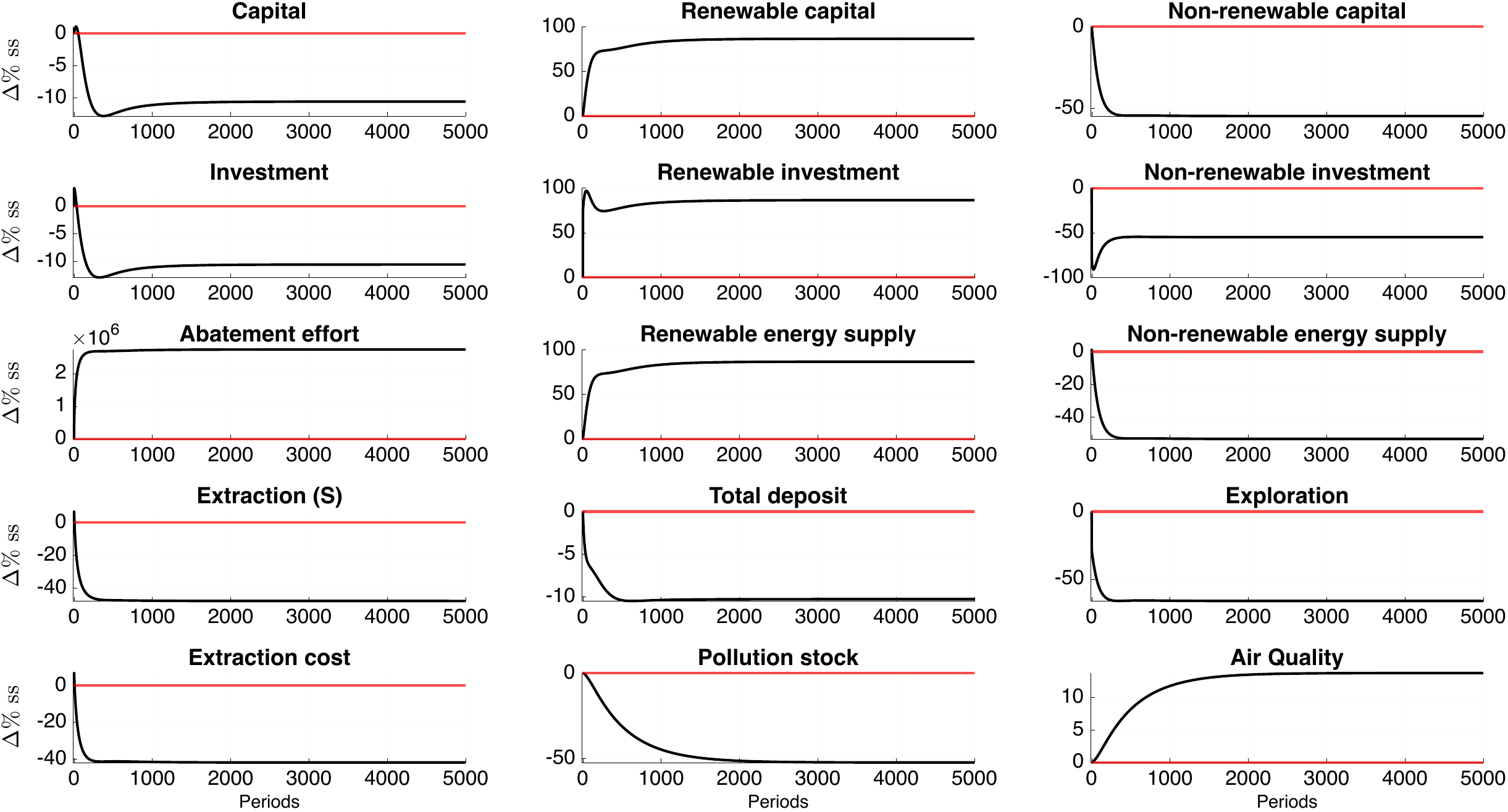}
\caption{Capital reallocation, fossil resource dynamics, abatement, and environmental outcomes during the transition to a 52\% emissions reduction under a gradual integral carbon pricing rule. All variables reflect percentage changes relative to the laissez-faire steady-state benchmark.}
\label{fig:emisisons_transition_dynamics_part2}
\end{subfigure}
\end{figure}

\textit{Gradual tightening and smooth convergence.} Because the carbon price adjusts incrementally, the adjustment burden is distributed over time rather than concentrated at the moment of policy announcement. In the first period, $p_{z,t}$ rises only marginally above zero, so the effective cost of fossil energy increases only slightly. As long as $Z_{tot} > \bar{Z}$, the carbon price continues to rise each period. This generates a smooth and monotone tightening that avoids the abrupt front-loaded contraction observed under immediate implementation. The emissions paths of both households and firms shown in Figure~\ref{fig:emisisons_transition_dynamics_part1} decline monotonically from the laissez-faire level toward the target. Most of the adjustment occurs within the first 150 quarters (approximately 35--40 years), after which emissions gradually converge to the new steady state.\\

\textit{Output and consumption.} Output and consumption both decline persistently, but to different long-run levels. Output converges to a new steady state about 10\% below its laissez-faire value, while consumption settles at roughly 7\% below its laissez-faire level. The contraction is gradual; rather than a sharp on-impact decline, output and consumption fall smoothly as the effective energy input cost rises over time. The renewable energy subsidy and fossil energy tax, rising proportionally with the carbon price, amplify the substitution incentives but do not create discontinuous jumps in production costs.\\

\textit{Energy reallocation.} The key adjustment mechanism is a large and persistent shift from fossil to renewable energy. In the production sector, firm demand for non-renewable energy declines by about 61\% relative to the laissez-faire steady state, while renewable energy demand rises by roughly 90\%. Households also adjust their energy consumption toward renewables: household fossil energy demand declines gradually, while household renewable energy demand increases by about 75\% in the long run. Despite this substitution toward renewable inputs, total firm energy demand contracts because the expansion of renewable energy is insufficient to fully offset the decline in fossil inputs. As a result, the economy reaches the emissions target through a combination of energy substitution and an overall reduction in energy use.\\

\textit{Capital reallocation.} Investment in renewable energy capital rises sharply at the onset of the transition as forward-looking agents anticipate the full future path of the rising carbon price. Renewable capital grows persistently and reaches a new steady state substantially above its laissez-faire level. Non-renewable capital, by contrast, contracts. Fossil investment becomes negative relative to the initial steady state from an early stage of the transition, implying net disinvestment in fossil capacity. The non-renewable capital stock declines by approximately 55\% in the long run. The policy thus generates substantial stranding of fossil capital, concentrated early in the transition as the expected future path of the carbon price renders new fossil investment unprofitable.\\

\textit{Abatement and abatement costs.} The carbon price creates an incentive for firms to increase abatement intensity $\mu_t$. Abatement effort rises persistently throughout the transition, with abatement costs $CA_t$ growing as a share of output. In the new steady state, abatement absorbs a significant fraction of gross output, reflecting the large cost of eliminating residual emissions when the substitution margin toward renewables has already been largely exhausted.\\

\textit{Fossil resource sector and pollution stock.} The gradual policy generates, at most, a quantitatively small Green Paradox response in the resource sector. Because the carbon price rises smoothly and agents have perfect foresight over the policy path, resource owners have an incentive to bring forward extraction before future fossil rents decline. In the numerical transition, this anticipatory response is small relative to the subsequent contraction in fossil activity, so the plotted extraction path is dominated by the persistent decline that follows the tightening of climate policy. The shadow value of reserves \(\lambda_{D,t}\) declines throughout the transition, signaling the progressive stranding of in-ground reserves. Exploration expenditures fall sharply and persistently, generating a managed contraction of the fossil sector in which new discoveries decline below extraction and the total reserve stock gradually falls toward a lower steady state. Despite the gradual policy implementation, the reduction in emissions generates a persistent decline in the pollution stock. The stock declines monotonically, converging to a new steady state approximately 52\% below the initial level, consistent with the lower long-run emission flows.\\

\textit{Policy instruments.} The bottom panels of Figure~\ref{fig:emisisons_transition_dynamics_part1} show the paths (in levels) of $p_{z,t}$, $\tau_{ER,t}$, and $\tau_{ENR,t}$. All three instruments rise smoothly and monotonically, converging to their Ramsey values. The proportional scaling ensures that the relative policy mix is constant throughout the transition, so no individual instrument overshoots its long-run target. The carbon tax reaches approximately half its steady-state value within 56 periods (14 years) and converges to $p_z^* = 0.151522$ in the long run.\\

\textit{Alternative policy designs.} The gradual rule above implements policy along three dimensions at once: a \emph{gradual} adjustment of policy instruments, a \emph{broad} set of instruments combining the carbon price with a renewable subsidy and a non-renewable energy tax, and \emph{comprehensive} sectoral coverage that restricts both firm and household emissions. To isolate the welfare consequences of each dimension, the welfare comparison includes four additional implementations of the same long-run emissions target. The first holds the instrument set and sectoral coverage fixed but switches to immediate implementation: the three instruments are set to their Ramsey-optimal values $(p_z^*, \tau_{ER}^*, \tau_{ENR}^*)$ from the first period onward. This \emph{immediate full policy package} isolates the role of implementation timing while keeping instrument breadth and coverage constant; its transition dynamics are reported in~\ref{appendix_immediate_full_package_irfs}. Two additional regimes vary sectoral coverage by restricting the emissions cap to firm emissions only, leaving household fossil energy use directly unconstrained: a \emph{firm-only gradual cap} and a \emph{firm-only immediate cap}, which together allow the comparison to assess how coverage interacts with implementation timing. The fifth implementation, included for reference, is the narrow carbon-price-only transition already studied in Section~\ref{section_transition_50_ren_energy}, where a constant carbon tax $p_{z,t}=0.30409$ is imposed immediately and calibrated to deliver the same long-run emissions reduction. Comparing these five implementations of the same environmental target separates the welfare consequences of implementation timing, instrument breadth, and sectoral coverage. Table~\ref{tab:dynamic_welfare_comparison} reports the resulting welfare comparison.

\begin{table}[H]
\centering
\caption{Dynamic Welfare Analysis: Transition Costs Under Alternative Decarbonization Policies}
\label{tab:dynamic_welfare_comparison}

\small
\setlength{\tabcolsep}{4.5pt}
\renewcommand{\arraystretch}{1.15}

\begin{threeparttable}
\resizebox{\linewidth}{!}{
\begin{tabular}{lccccc}
\cmidrule(lr){2-6}
& \multicolumn{5}{c}{\textbf{Dynamic CEV comparisons}} \\
\cmidrule(lr){2-6}
& \textbf{Carbon-Price} & \textbf{Firm-Only} & \textbf{Comprehensive} & \textbf{Comprehensive} & \textbf{Firm-Only} \\
& \textbf{Only} & \textbf{Gradual Cap} & \textbf{Gradual Cap} & \textbf{Immediate Cap} & \textbf{Immediate Cap} \\
\midrule
\textbf{Implementation} & Immediate & Gradual & Gradual & Immediate & Immediate \\
\textbf{Instruments}  & $p_{z,t}$ & $p_{z,t},\tau_{ER,t},\tau_{ENR,t}$ & $p_{z,t},\tau_{ER,t},\tau_{ENR,t}$ & $p_{z,t},\tau_{ER,t},\tau_{ENR,t}$ & $p_{z,t},\tau_{ER,t},\tau_{ENR,t}$ \\
\textbf{Relative to LF} & -5.9028 & -0.5955 & -1.2609 & -3.4989 & -5.6677 \\
\textbf{Relative to Carbon-Price Only} & -- & +5.5639 & +4.8668 & +2.5211 & +0.2467 \\
\textbf{Relative to Firm Gradual Cap} & -- & -- & -0.6639 & -2.8975 & -5.0628 \\
\textbf{Relative to Comprehensive Gradual Cap} & -- & -- & -- & -2.2588 & -4.4480 \\
\textbf{Relative to Immediate Full Package} & -- & -- & -- & -- & -2.2389 \\
\bottomrule
\end{tabular}
}

\begin{tablenotes}[flushleft]
\footnotesize
\item \begin{minipage}{\textwidth}
\textit{Notes:} The table compares transition policies calibrated to deliver the same long-run aggregate emissions reduction, \(Z_{tot}=\bar Z\), approximately 52\% below the laissez-faire steady state. Entries are dynamic consumption-equivalent variation (CEV) comparisons, in percentage points, computed over the same perfect-foresight simulation horizon for all policies and including the continuation value at the terminal steady state. Negative values imply welfare losses. Pairwise rows report CEV differences computed directly from the two transition paths over the same horizon; these need not equal arithmetic differences between the CEV values reported relative to laissez-faire. Positive pairwise values indicate that the column policy is preferred to the row comparison policy. The implementation row refers to the speed at which policy instruments are introduced, not to the endogenous speed of adjustment of macroeconomic variables. Energy taxes and renewable subsidies operate through energy markets as specified in the model. The comprehensive immediate policy corresponds to the immediate full policy package reported in \ref{appendix_immediate_full_package_irfs}.
\end{minipage}
\end{tablenotes}

\end{threeparttable}
\end{table}

Several results stand out. First, all five constrained transition policies imply welfare losses relative to laissez-faire once the full adjustment path is taken into account. The losses range from 0.60\% under the firm-only gradual cap to 5.90\% under the immediate carbon-price-only transition. Thus, even when the same long-run emissions reduction is achieved, the path used to reach the target has first-order welfare consequences.

Second, gradual implementation substantially reduces transition costs. Holding the instrument set fixed, switching from immediate to gradual implementation improves welfare by 5.06 percentage points under firm-only regulation and by 2.26 percentage points under comprehensive regulation. The timing of policy implementation is therefore quantitatively important: when policy instruments jump immediately to their long-run values, fossil energy use contracts before renewable capital and the broader energy system have fully adjusted, while gradual tightening allows the energy capital stock to reallocate more evenly over time.

Third, complementary instruments mitigate but do not eliminate the cost of immediate implementation. The immediate full policy package generates a welfare loss of \(-3.50\%\), compared with \(-5.90\%\) under the immediate carbon-price-only transition. Renewable subsidies and non-renewable energy taxes therefore directly support the reallocation from fossil to renewable energy. However, the immediate full package remains substantially more costly than both gradual cap regimes, indicating that complementary instruments cannot fully compensate for the energy-capacity gap created by front-loaded fossil-demand compression. The transition costs under the gradual emissions-cap regime are also not primarily driven by capital adjustment costs. A counterfactual transition with near-zero adjustment costs for renewable, non-renewable, and final-goods capital decreases welfare loss only modestly, from -1.26\% to -1.20\%. This indicates that installation frictions affect the timing of capital reallocation but account for only a small share of the overall welfare loss under gradual implementation. Most of the remaining cost instead reflects the persistent reduction in fossil energy use, the associated decline in effective energy input, and the lower-output allocation required to sustain the emissions target in the long run.

Fourth, under immediate implementation, the firm-only immediate cap is nearly welfare-equivalent to the carbon-price-only transition. The two regimes differ by only about 0.25 percentage points (\(-5.67\%\) vs. \(-5.90\%\)), even though the firm-only immediate cap deploys the full set of three instruments. This shows that broadening the instrument set provides little additional protection when firms are forced to absorb the entire adjustment in the first period: complementary instruments help most when paired with gradual implementation or with comprehensive coverage that spreads the burden onto households.

Finally, the comprehensive gradual cap is more costly than the firm-only gradual cap because coverage changes incidence. Under the comprehensive cap, households and firms both reduce fossil energy use. Since household energy services enter utility directly, this restriction imposes a direct welfare cost in addition to the output loss from the production side. Under the firm-only cap, households remain less exposed to the emissions constraint, while firms undertake a larger share of the aggregate emissions reduction through lower fossil use and higher abatement. This generates internal carbon leakage from firms to households, but it can still raise welfare relative to the comprehensive cap because it protects a utility-relevant household energy margin.

\begin{figure}[ht]
\centering
\includegraphics[width=\linewidth]{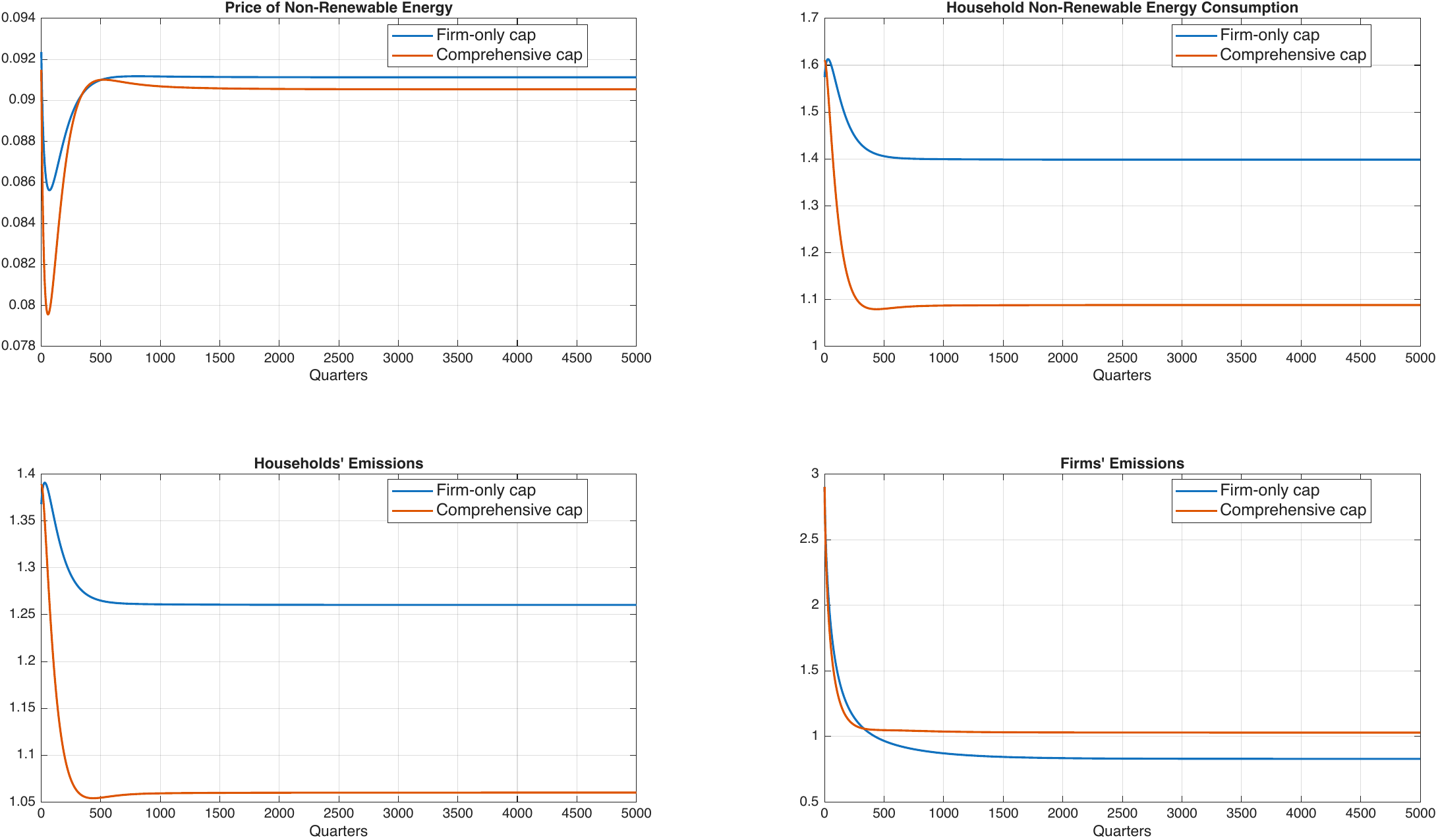}
\caption{Sectoral emissions and fossil energy use under alternative emissions caps. The figure compares the transition dynamics under the firm-only and comprehensive emissions caps calibrated to achieve the same aggregate emissions reduction. Under the firm-only cap, household fossil energy use and household emissions remain higher throughout the transition, while firms undertake a larger share of the aggregate emissions reduction.}
\label{fig:leakage_transition}
\end{figure}

To clarify the incidence mechanism behind the welfare ranking, Table~\ref{tab:leakage_decomposition} reports the terminal allocation of fossil energy price, fossil energy use, and emissions across households and firms under the firm-only and comprehensive cap regimes, while Figure~\ref{fig:leakage_transition} reports the associated transition paths. Relative to the comprehensive cap, the firm-only cap leaves household fossil energy use and emissions higher, while firms account for a larger share of the aggregate emissions reduction. In the terminal allocation, household fossil energy use is about 28.5\% higher and household emissions are about 18.9\% higher under the firm-only regime, whereas firm emissions are about 19.5\% lower. The macroeconomic incidence differs as well: output is higher under the firm-only cap, while consumption is lower, reflecting the different allocation of adjustment costs across production and household energy services. Firms generate lower emissions despite using more fossil energy because they increase abatement effort when they are the regulated sector. Higher abatement intensity lowers emissions per unit of fossil energy and allows the aggregate emissions target to be met while household fossil energy use remains less compressed.

This pattern is consistent with internal carbon leakage within the model economy: under asymmetric regulation, emissions are reallocated from firms to households within the domestic economy. The aggregate target is therefore achieved by placing a larger share of the adjustment burden on firms. The leakage mechanism is visible primarily in the allocation of fossil energy use and emissions across sectors. Under the firm-only cap, households are not directly regulated and therefore reduce fossil energy consumption much less than under the comprehensive cap. As a result, a larger share of the aggregate emissions reduction is achieved by firms through lower emissions and higher abatement effort, while household fossil energy use and emissions remain persistently higher throughout the transition and in the terminal allocation. 

\begin{table}[t]
\centering
\caption{Non-renewable Energy and Emissions Under Alternative Policies}
\label{tab:leakage_decomposition}
\small
\begin{tabular}{lccc}
\toprule
Variable & Firm-only cap & Comprehensive cap & Difference \\
\midrule
% Output $Y$ & 3.1137 & 3.0699 & +0.0439 \\
% Consumption $C$ & 2.1727 & 2.2088 & -0.0362 \\
Fossil energy price $P_{ENR}$ & 0.0911 & 0.0905 & +0.0006 \\
Household fossil energy $E_{HHNR}$ & 1.3986 & 1.0883 & +0.3103 \\
Firm fossil energy $E_{FNR}$ & 2.3297 & 1.9465 & +0.3833 \\
Household emissions $Z_{HH}$ & 1.2604 & 1.0601 & +0.2003 \\
Firm emissions $Z_{F}$ & 0.8294 & 1.0297 & -0.2003 \\
\bottomrule
\end{tabular}
\begin{minipage}{0.75\linewidth}
\footnotesize
\textit{Notes:} Terminal steady-state values under the two emissions-cap regimes used in Table~\ref{tab:dynamic_welfare_comparison}. Under the firm-only cap, households use more fossil energy and generate higher emissions, while firms generate lower emissions through stronger abatement. This sectoral reallocation of emissions is the internal carbon leakage mechanism in the model.
\end{minipage}
\end{table}

\ref{appendix_policy_coefficients} provides additional implementation details for the emissions-cap regimes. It shows that the aggregate-cap Ramsey coefficients decentralize the targeted aggregate-emissions allocation even after removing the explicit cap, whereas the firm-only Ramsey coefficients do not implement the firm-emissions target without the explicit restriction.

% Appendix~\ref{appendix_sensitivity_cev} repeats the welfare comparison under alternative values of the household energy utility weight \(\psi_E\) and the abatement-cost curvature parameter \(\phi_2\). The results show that the qualitative ranking among the constrained cap regimes is robust to the preference and abatement-technology calibrations considered: gradual implementation remains preferred to immediate implementation, firm-only regulation remains preferred under gradual implementation, and comprehensive regulation remains preferred under immediate implementation.

\subsection{Policy implications}
\label{section_policy_implications}
The quantitative results yield several implications for climate policy design. First, the timing of policy implementation is the most important determinant of transition welfare costs in the baseline calibration. Switching from immediate to gradual implementation reduces the welfare loss by 5.06 percentage points under firm-only regulation and by 2.26 percentage points under comprehensive regulation, holding the instrument set fixed. The mechanism is straightforward: gradual tightening allows fossil capital to depreciate and renewable capital to accumulate before the carbon price compresses fossil energy demand, avoiding the energy-capacity gap that drives the front-loaded contraction under immediate implementation. Real-world institutional mechanisms such as the EU ETS MSR, which adjusts allowance supply gradually in response to deviations of emissions from target trajectories, are therefore consistent with the broader principle suggested by the model: rule-based gradual tightening can reduce transition costs relative to front-loaded implementation.

Second, complementary instruments -- renewable subsidies and non-renewable energy taxes -- help reduce the cost of immediate implementation but do not make it welfare-equivalent to gradual adjustment. The immediate full policy package improves welfare by approximately 2.52 percentage points relative to the immediate carbon-price-only transition. However, even with this broad instrument set, the welfare loss under immediate implementation remains more than twice as large as under the corresponding comprehensive gradual cap regime. The implication is that broadening the instrument set is a useful but secondary policy lever, and that governments facing political pressure to implement rapid climate policy should not expect that complementary instruments alone will neutralize the macroeconomic cost of front-loaded tightening.

Third, the welfare ranking of sectoral coverage depends on whether implementation is gradual or immediate. Under gradual implementation, firm-only regulation generates smaller welfare losses than comprehensive regulation. The mechanism is internal carbon leakage: relative-price movements in the fossil energy market reallocate emissions toward unregulated households, while firms compensate through higher abatement effort. Because household energy services enter utility directly, shielding household consumption from direct compression is welfare-improving. Under immediate implementation, however, this advantage reverses. The firm-only immediate cap forces firms to absorb the entire adjustment in a single period, raising abatement costs sharply and generating welfare losses comparable to the carbon-price-only transition. The policy implication is that sectoral coverage cannot be chosen independently of implementation timing. In this calibration, firm-only regulation performs well when adjustment is gradual, while comprehensive coverage performs better under immediate implementation because it spreads the burden across firms and households. This ranking should be interpreted as a model-based incidence result rather than as a general prescription against household regulation.

Fourth, the welfare advantage of firm-only regulation under gradual implementation does not imply that household emissions should be ignored. It implies that extending carbon pricing to households -- for example through policies such as the EU ETS2 introduced under Directive (EU) 2023/959 \citep{EUETS2Directive} -- should be evaluated jointly with the implementation timetable, with compensation and revenue-recycling schemes, and with the availability of household-level substitutes for fossil energy services. In the model, the case for shielding household energy services rests on the assumption that the aggregate emissions target is reached: if extending coverage to households is what makes a more ambitious target politically or administratively feasible, the welfare comparison can flip in favor of comprehensive regulation.

Fifth, these results extend the instrument-choice literature in environmental economics. Classic contributions \citep{GoulderParry2008, BovenbergGoulder1996} emphasize that the welfare effects of carbon taxes depend on their interaction with pre-existing fiscal distortions in second-best static or steady-state environments. The present analysis highlights a complementary dimension of second-best policy design: in dynamic environments with capital adjustment frictions, the welfare effects of climate policy depend on the interaction between policy instruments and the speed of capital reallocation. This suggests that climate policy design should account not only for fiscal interactions but also for the macroeconomic adjustment frictions that govern the transition to a low-carbon energy system.

\section{Conclusion}
\label{section_conclusion}
This paper analyzes how the structure of climate policy and the dynamics of capital reallocation shape the macroeconomic costs of the energy transition. Using a multi-sector macroeconomic framework with fossil and renewable energy sectors and endogenous fossil resource dynamics, the analysis shows that the transition away from fossil energy can generate substantial short-run macroeconomic adjustment costs. When climate policy sharply reduces fossil energy use, fossil capital contracts rapidly while renewable capacity and abatement adjust only gradually. This mismatch creates a temporary shortage of effective energy capacity, leading to sizable declines in output and consumption during the transition. In the baseline calibration, a rapid transition toward a cleaner energy mix reduces output by almost 12\% at its peak and compresses the shadow value of fossil reserves, gradually creating stranded fossil assets.

Comparing alternative policy designs shows that transition costs depend jointly on implementation timing, sectoral coverage, and instrument breadth, with implementation timing emerging as the dominant margin. A narrow immediate carbon-price-only transition generates a large front-loaded contraction because fossil energy use falls before renewable capital has expanded sufficiently. Adding complementary instruments reduces this loss substantially but does not eliminate it: an immediate full policy package generates a welfare loss of \(-3.50\%\), while a gradual cap with the same instrument set generates losses of only \(-1.26\%\) to \(-0.60\%\), depending on sectoral coverage. The welfare ranking of sectoral coverage also depends on timing: firm-only regulation reduces welfare losses under gradual implementation by shielding household energy services, but under immediate implementation it becomes nearly as costly as the carbon-price-only transition because firms must absorb the adjustment at once. The path used to reach the long-run emissions target is therefore central to welfare, and in the baseline calibration it matters more quantitatively than instrument breadth alone.

The results also highlight the role of sectoral incidence in shaping welfare comparisons. Policies that regulate emissions comprehensively across both firms and households reduce emissions more uniformly but compress household energy consumption more strongly during the transition. Because household utility depends directly on energy services, this difference in incidence influences the welfare ranking of alternative policies. Firm-only regulation generates an internal carbon leakage channel by shifting emissions toward households, but it can still reduce welfare losses relative to a comprehensive cap when the aggregate target is fixed. Section~\ref{section_policy_implications} discusses the policy implications of these results, emphasizing that transition costs depend jointly on implementation timing, instrument breadth, sectoral coverage, and fossil-resource dynamics.

Several limitations of the framework point to directions for future research. First, the calibration is designed to match steady-state cross-sectional moments rather than dynamic adjustment paths, so the quantitative transition dynamics should be interpreted as model-based rather than directly estimated empirical magnitudes. Second, the model abstracts from labor-market adjustment by omitting labor as an explicit production input and household choice variable. This keeps the analysis focused on energy-capital reallocation, but it also rules out unemployment dynamics, sectoral labor reallocation, and interactions between climate policy and pre-existing labor-market distortions. Third, technological change is exogenous, whereas environmental policy may influence innovation in renewable technologies and energy efficiency. In particular, the transition experiments abstract from endogenous improvements in renewable productivity that may arise through learning-by-doing or directed innovation. Allowing renewable productivity to evolve endogenously could accelerate the expansion of clean energy and reduce the long-run cost of the transition. Fourth, the analysis focuses on deterministic transition paths under perfect foresight. Agents are assumed to observe the entire future policy path with certainty, which abstracts from potential uncertainty about policy credibility or future tightening of climate regulation. Allowing for stochastic policy paths or imperfect credibility could affect investment timing and fossil-resource extraction decisions. While these extensions could alter the quantitative magnitude of the transition dynamics, the mechanisms highlighted in the present framework provide a tractable benchmark for understanding how capital reallocation, fossil-resource depletion, and policy design interact during the decarbonization process.

The model also abstracts from distortionary labor and capital taxation, so the welfare comparisons do not capture the double-dividend channel emphasized in the environmental tax literature. Revenue recycling operates through lump-sum transfers throughout the analysis. Introducing distortionary taxation would make the use of carbon revenues -- for example, through labor-tax reductions, capital-tax cuts, or targeted investment support -- an additional margin affecting transition costs that operates separately from the energy-instrument channels analyzed here.

Overall, the findings demonstrate that the macroeconomic cost of climate policy depends not only on the stringency of emissions targets but also on the timing and design of the policy package -- in particular, on whether complementary instruments support the reallocation of energy capital and on how the adjustment burden is distributed across firms and households. Accounting for transition dynamics is therefore essential for designing climate policies that achieve environmental objectives while minimizing macroeconomic disruption.

%%%%%%%%%%%%%%%%%%%%%%%%%%%%%%%% APPENDIX %%%%%%%%%%%%%%%%%%%%%%%%%%%%%%%%
\newpage
\appendix

\renewcommand{\thesection}{Appendix \Alph{section}}
\renewcommand{\thesubsection}{\Alph{section}\arabic{subsection}}
\renewcommand{\thesubsubsection}{\Alph{section}\arabic{subsection}.\arabic{subsubsection}}

\section{Equilibrium and FOC derivations}
\setcounter{figure}{0}
\setcounter{table}{0}
\setcounter{equation}{0}
\renewcommand{\thefigure}{\Alph{section}\arabic{figure}}
\renewcommand{\thetable}{\Alph{section}\arabic{table}}
\renewcommand{\theequation}{\Alph{section}\arabic{equation}}

This appendix derives the competitive equilibrium conditions of the model. The optimization problems and first-order conditions for households and firms in each sector are presented, together with the dynamics of fossil resource extraction. In deriving the decentralized first-order conditions, households and firms take aggregate environmental quality \(Q_t\) and the aggregate pollution stock \(M_t\) as given.

%--------------------------------------------------
\subsection{Households}

The representative household chooses consumption, household energy use, and sectoral investment to maximize expected lifetime utility. Preferences are defined over consumption, a composite of household energy services, and environmental quality.

\begin{align}
\max_{\{C_t,E_{HHR,t},E_{HHNR,t},I_{j,t},K_{j,t}\}} 
\mathbb{E}_0 \sum_{t=0}^{\infty} 
\beta^t \left[
\frac{\psi_C C_t^{1-\rho_C}}{1-\rho_C} +
\frac{\psi_E E_{HH,t}^{1-\rho_E}}{1-\rho_E}+
\frac{\phi_Q Q_{t}^{1-\rho_Q}}{1-\rho_Q}
\right]
\end{align}

subject to the intertemporal budget constraint
\begin{align}
C_t &+ P_{ER,t}(1-\tau_{ER,t})E_{HHR,t} + P_{ENR,t}(1+\tau_{ENR,t})E_{HHNR,t} + \sum_{j\in\{Y,R,NR\}} I_{j,t} \nonumber \\
&+ \sum_{j\in\{Y,R,NR\}} \frac{h_j}{2}\left(\frac{I_{j,t}}{K_{j,t-1}}-\delta_j\right)^2 K_{j,t-1} 
= \sum_{j\in\{Y,R,NR\}} r_{j,t} K_{j,t-1} + \Pi_{Y,t} + \Pi_{ER,t} + \Pi_{ENR,t} + T_t .
\end{align}

The law of motion of capital is defined as
\begin{align}
K_{j,t} = (1-\delta_j)K_{j,t-1} + I_{j,t}, \qquad j \in \{Y,R,NR\}.
\end{align}

Household energy services are produced using a CES aggregator of renewable and non-renewable energy:
\begin{align}
E_{HH,t} =\left[ \Omega^{1/\theta}E_{HHR,t}^{(\theta-1)/\theta} + (1-\Omega)^{1/\theta}E_{HHNR,t}^{(\theta-1)/\theta} \right]^{\theta/(\theta-1)}.
\end{align}

\paragraph{Lagrangian}\mbox{}\\[0.5em]

Let $\lambda_t$ denote the multiplier associated with the household's budget constraint and $\Lambda_{j,t}$ the multiplier corresponding to the capital accumulation equation. The household's Lagrangian is

\begin{align*}
\mathcal{L} &= \mathbb{E}_0 \sum_{t=0}^{\infty} \beta^t \Bigg\{
\frac{\psi_C C_t^{1-\rho_C}}{1-\rho_C}
+ \frac{\psi_E E_{HH,t}^{1-\rho_E}}{1-\rho_E} +
\frac{\phi_Q Q_{t}^{1-\rho_Q}}{1-\rho_Q}
+ \lambda_t \Big[ -C_t - P_{ER,t}(1-\tau_{ER,t})E_{HHR,t} \\
&\quad - P_{ENR,t}(1+\tau_{ENR,t})E_{HHNR,t} - \sum_j I_{j,t}
- \sum_j \frac{h_j}{2}\left(\frac{I_{j,t}}{K_{j,t-1}}-\delta_j\right)^2 K_{j,t-1}
+ \sum_j r_{j,t}K_{j,t-1} \\
&\quad + \Pi_{Y,t} + \Pi_{ER,t} + \Pi_{ENR,t} + T_t \Big]
+ \sum_j \Lambda_{j,t}\left[(1-\delta_j)K_{j,t-1}+I_{j,t}-K_{j,t}\right]
\Bigg\}.
\end{align*}

Define Tobin's marginal value of installed capital as
\begin{align}
q_{j,t} = \frac{\Lambda_{j,t}}{\lambda_t}.
\end{align}

\paragraph{First-order conditions}\mbox{}\\[0.5em]

\textit{Consumption}
\begin{align}
\psi_C C_t^{-\rho_C} = \lambda_t .
\end{align}

\textit{Renewable household energy}
\begin{align}
P_{ER,t}(1-\tau_{ER,t}) = \frac{\psi_E}{\lambda_t} E_{HH,t}^{-\rho_E} \Omega^{\frac{1}{\theta}}
\left(\frac{E_{HHR,t}}{E_{HH,t}}\right)^{-\frac{1}{\theta}} .
\end{align}

\textit{Non-renewable household energy}
\begin{align}
P_{ENR,t}(1+\tau_{ENR,t}) = \frac{\psi_E}{\lambda_t} E_{HH,t}^{-\rho_E} (1-\Omega)^{\frac{1}{\theta}}
\left(\frac{E_{HHNR,t}}{E_{HH,t}}\right)^{-\frac{1}{\theta}} .
\end{align}

\textit{Investment}
\begin{align}
q_{j,t} = 1 + h_j\left(\frac{I_{j,t}}{K_{j,t-1}}-\delta_j\right).
\end{align}

\textit{Capital Euler equation}
\begin{align}
\lambda_t q_{j,t} = \beta \mathbb{E}_t \left\{ \lambda_{t+1} \left[ r_{j,t+1} + (1-\delta_j)q_{j,t+1} - \frac{h_j}{2}\left(\frac{I_{j,t+1}}{K_{j,t}}-\delta_j\right)^2 + h_j\left(\frac{I_{j,t+1}}{K_{j,t}}-\delta_j\right)\frac{I_{j,t+1}}{K_{j,t}} \right] \right\}.
\end{align}

%--------------------------------------------------
\subsection{Final goods firm}

The representative final goods firm rents capital and chooses renewable and non-renewable energy inputs together with abatement effort to maximize profits:
\begin{align}
\max_{K_{t-1},E_{FR,t},E_{FNR,t},\mu_t} \Pi_{Y,t} &= Y_t - CA_t - r_{Y,t}K_{t-1} - P_{ER,t}(1-\tau_{ER,t})E_{FR,t} \nonumber \\
&- P_{ENR,t}(1+\tau_{ENR,t})E_{FNR,t} - p_{z,t}(1-\mu_t)E_{FNR,t}^{\zeta_F}.
\end{align}

Output is produced according to
\begin{align}
Y_t = D(M_{t}) A_{Y,t} K_{t-1}^{\alpha_K} E_{F,t}^{\alpha_E},
\end{align}
where the firm energy bundle is
\begin{align}
E_{F,t} = \left[ \Omega_F^{1/\theta_F}E_{FR,t}^{(\theta_F-1)/\theta_F} + (1-\Omega_F)^{1/\theta_F}E_{FNR,t}^{(\theta_F-1)/\theta_F} \right]^{\theta_F/(\theta_F-1)}.
\end{align}

Abatement costs are given by
\begin{align}
CA_t = \phi_1 \mu_t^{\phi_2} Y_t,
\end{align}
and defining $\Phi_t \equiv 1-\phi_1 \mu_t^{\phi_2}$ allows profits to be written as
\begin{align}
\Pi_{Y,t} &= \Phi_t Y_t - r_{Y,t}K_{t-1} - P_{ER,t}(1-\tau_{ER,t})E_{FR,t} - P_{ENR,t}(1+\tau_{ENR,t})E_{FNR,t} \nonumber \\
&- p_{z,t}(1-\mu_t)E_{FNR,t}^{\zeta_F}.
\end{align}

\paragraph{First-order conditions}\mbox{}\\[0.5em]

\textit{Capital demand}
\begin{align}
r_{Y,t} = \alpha_K \Phi_t \frac{Y_t}{K_{t-1}} .
\end{align}

\textit{Renewable energy demand}
\begin{align}
P_{ER,t}(1-\tau_{ER,t}) = \alpha_E \Phi_t \frac{Y_t}{E_{F,t}} \Omega_F^{\frac{1}{\theta_F}}
\left(\frac{E_{FR,t}}{E_{F,t}}\right)^{-\frac{1}{\theta_F}} .
\end{align}

\textit{Non-renewable energy demand}
\begin{align}
P_{ENR,t}(1+\tau_{ENR,t}) + p_{z,t}(1-\mu_t)\zeta_F E_{FNR,t}^{\zeta_F-1}
= \alpha_E \Phi_t \frac{Y_t}{E_{F,t}} (1-\Omega_F)^{\frac{1}{\theta_F}}
\left(\frac{E_{FNR,t}}{E_{F,t}}\right)^{-\frac{1}{\theta_F}} .
\end{align}

\textit{Abatement choice}
\begin{align}
p_{z,t} E_{FNR,t}^{\zeta_F} = \phi_1 \phi_2 \mu_t^{\phi_2-1} Y_t .
\end{align}

%--------------------------------------------------
\subsection{Renewable energy firm}

The renewable energy firm rents installed capital and produces energy according to
\begin{align}
E_{R,t} = A_{ER,t} K_{ER,t-1}^{\gamma_{ER}} .
\end{align}

Its period profit is
\begin{align}
\Pi_{ER,t} = P_{ER,t} E_{R,t} - r_{ER,t} K_{ER,t-1},
\end{align}
which yields the optimality condition
\begin{align}
r_{ER,t} = \gamma_{ER} P_{ER,t} \frac{E_{R,t}}{K_{ER,t-1}} .
\end{align}

%--------------------------------------------------
\subsection{Non-renewable energy firm}

The non-renewable energy firm produces energy using installed capital and extracted resources:
\begin{align}
E_{NR,t} = A_{ENR,t} K_{ENR,t-1}^{\alpha_{K,NR}} S_t^{\alpha_S}.
\end{align}

    The law of motion for reserves is given by
\begin{align}
D_t = D_{t-1} - S_t + A_{disc}(Exp_{NR,t})^{\alpha_D},
\end{align}
and extraction costs are given by
\begin{align}
AC_{extract,t} = \xi S_t \left(\frac{\bar D}{D_{t-1}}\right)^{\psi_D}.
\end{align}

Period profits of the non-renewable energy firm are
\begin{align}
\Pi_{ENR,t} = P_{ENR,t}E_{NR,t} - r_{ENR,t}K_{ENR,t-1} - Exp_{NR,t} - AC_{extract,t}.
\end{align}
Because extraction depletes the reserve stock and affects future extraction costs, the firm solves
\begin{align}
\max_{\{K_{ENR,t-1},S_t,Exp_{NR,t},D_t\}} 
\mathbb{E}_0 \sum_{t=0}^{\infty} \beta^t \frac{\lambda_t}{\lambda_0} \Pi_{ENR,t},
\end{align}
subject to the reserve law of motion, where \(\lambda_{D,t}\) denotes the shadow value of reserves.

\paragraph{First-order conditions}\mbox{}\\[0.5em]

\textit{Capital demand}
\begin{align}
r_{ENR,t} = \alpha_{K,NR} P_{ENR,t} \frac{E_{NR,t}}{K_{ENR,t-1}} .
\end{align}

\textit{Extraction}
\begin{align}
P_{ENR,t} \alpha_S \frac{E_{NR,t}}{S_t} = \xi \left(\frac{\bar D}{D_{t-1}}\right)^{\psi_D} + \lambda_{D,t} .
\end{align}

\textit{Exploration}
\begin{align}
1 = \lambda_{D,t} A_{disc} \alpha_D (Exp_{NR,t})^{\alpha_D-1} .
\end{align}

\textit{Reserve Euler equation (modified Hotelling condition)}
\begin{align}
\lambda_{D,t} = \beta \mathbb{E}_t \left[ \frac{\lambda_{t+1}}{\lambda_t} \left( \lambda_{D,t+1} + \xi \psi_D S_{t+1} \frac{1}{D_t} \left(\frac{\bar D}{D_t}\right)^{\psi_D} \right) \right].
\end{align}

%--------------------------------------------------
\subsubsection{Resource sector parameters}

This subsection summarizes the economic interpretation of the parameters governing exploration and extraction in the fossil resource sector.

\paragraph{Discovery efficiency $A_{disc}$}\mbox{}\\[0.5em]

The parameter $A_{disc}$ measures the productivity of exploration activities. A higher value increases the effectiveness of exploration expenditure in generating new reserves, thereby lowering the shadow value of reserves and expanding fossil supply.

\paragraph{Exploration elasticity $\alpha_D$}\mbox{}\\[0.5em]

The parameter $\alpha_D<1$ captures diminishing returns to exploration. Smaller values imply that additional exploration yields progressively smaller discoveries, reflecting the depletion of easily accessible deposits.

\paragraph{Extraction cost parameters}\mbox{}\\[0.5em]

Extraction costs are given by
\begin{align}
AC_{extract,t} = \xi S_t \left(\frac{\bar D}{D_{t-1}}\right)^{\psi_D}.
\end{align}
The parameter $\xi$ controls the overall scale of extraction costs, while $\psi_D$ governs their sensitivity to the remaining stock of reserves. Larger values of $\psi_D$ generate stronger incentives to smooth extraction over time.

\newpage

%%%%%%%%%%%%%%%%%%%% APPENDIX B %%%%%%%%%%%%%%%%%%%%
\section{Sensitivity of shock propagation to key structural parameters}
\label{appendix_sensitivity_shocks}

This appendix examines the robustness of the model’s propagation mechanism to alternative calibrations of key structural parameters under a one-percent unexpected increase in non-renewable energy productivity, $A_{ENR}$. Unlike the large transition experiments in Sections \ref{section_transition_50_ren_energy} and \ref{section_cap_emissions}, these exercises study local dynamics around the deterministic steady state and therefore isolate how specific parameters affect the short-run transmission of a fossil productivity shock. Throughout, impulse responses are reported as percentage deviations from each calibration’s own steady state. The figures show that the aggregate responses of output and consumption are generally robust, while parameter changes mainly affect the composition of energy demand, the timing of capital adjustment, and the persistence of sectoral responses.

\subsection{Common capital adjustment frictions}
\begin{figure}[H]
\centering
\includegraphics[width=\linewidth]{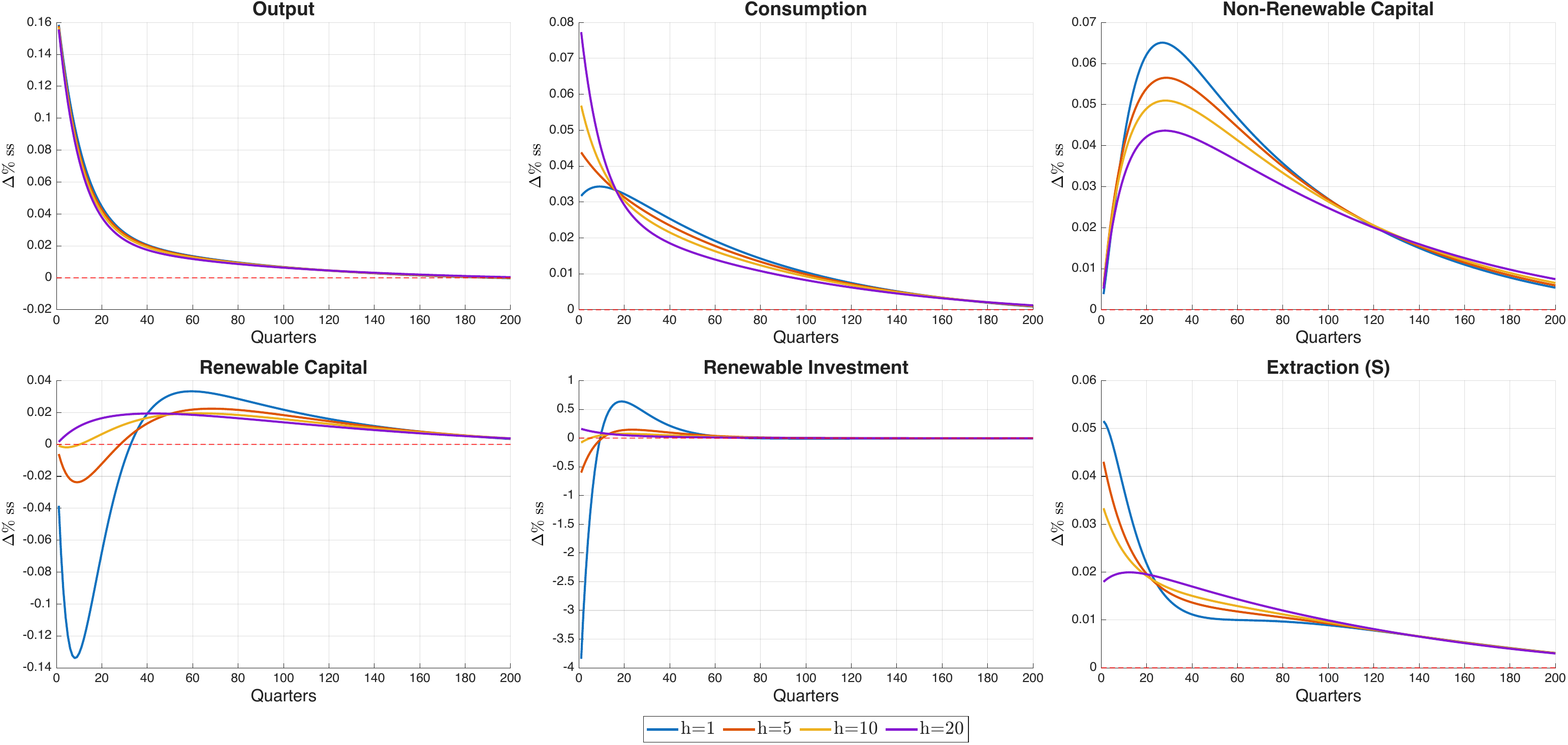}
\caption{Impulse responses to a one-percent increase in $A_{ENR}$ under alternative common capital adjustment frictions.}
\label{fig:sensitivity_same_h}
\end{figure}

Figure~\ref{fig:sensitivity_same_h} varies the common adjustment-cost parameter across all capital stocks, setting $h=h_{ER}=h_{ENR}\in\{1,5,10,20\}$. The figure shows that changing the common adjustment friction has only a limited effect on the impulse responses of output and consumption, which remain very similar across calibrations. The main differences arise in the energy-capital block. Lower adjustment costs generate more front-loaded investment reallocation and somewhat larger short-run movements in sectoral capital stocks, while higher adjustment costs smooth these responses and delay their peak. The extraction response is almost unchanged across cases, indicating that under the $A_{ENR}$ shock the resource sector is considerably less sensitive to the common capital friction than the investment block. Overall, the figure suggests that common adjustment costs mainly affect the speed of capital reallocation rather than the aggregate magnitude of the short-run macroeconomic expansion.

\subsection{Sector-specific capital adjustment frictions}
\begin{figure}[H]
\centering
\includegraphics[width=\linewidth]{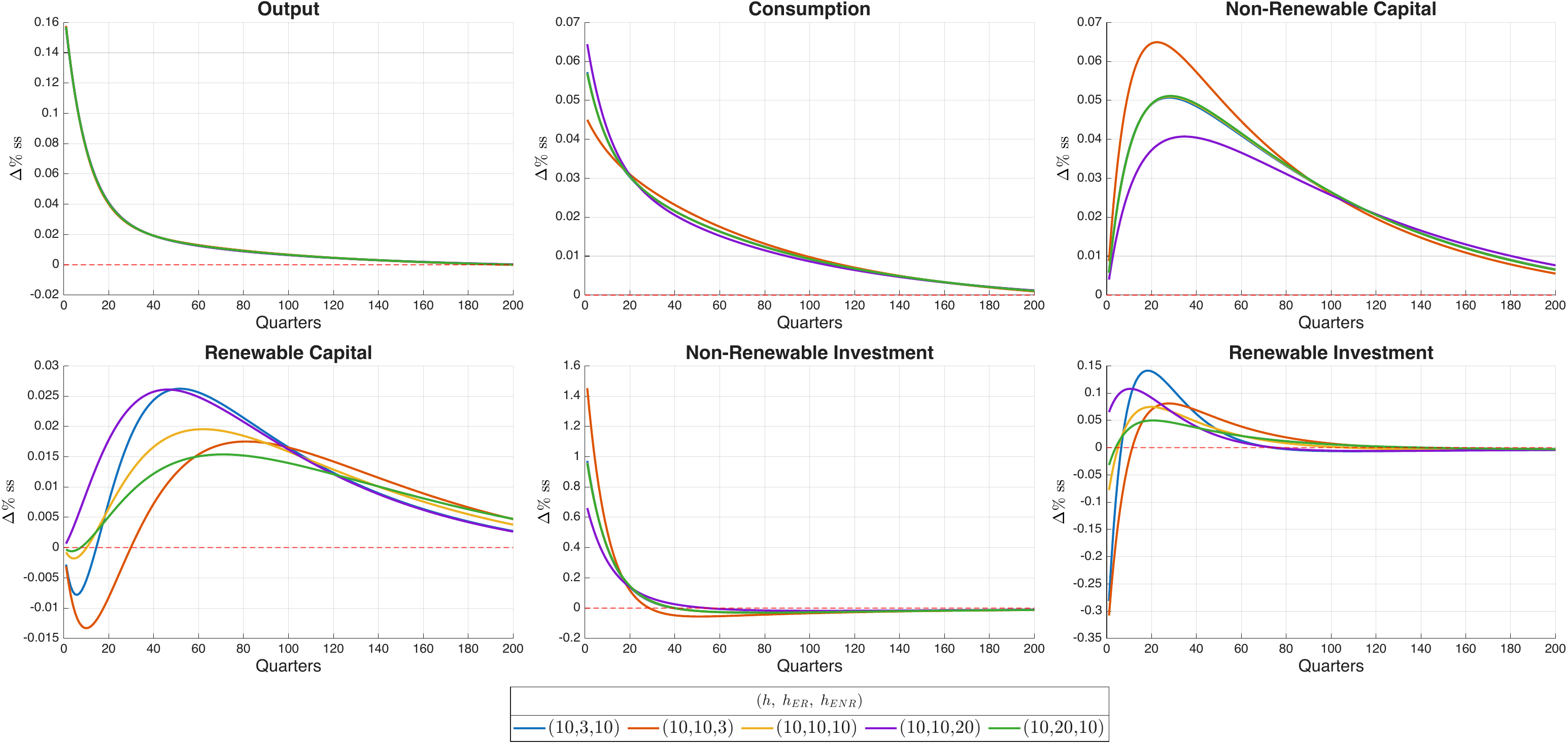}
\caption{Impulse responses to a one-percent increase in $A_{ENR}$ under alternative sector-specific capital adjustment frictions. The legend reports the triplet $(h,h_{ER},h_{ENR})$.}
\label{fig:sensitivity_diff_h}
\end{figure}

Figure~\ref{fig:sensitivity_diff_h} varies the renewable- and fossil-sector adjustment costs separately around the baseline calibration. Aggregate responses remain very similar across cases: the impulse responses of output and consumption differ only slightly. By contrast, the sectoral capital and investment responses are more sensitive. Lower renewable adjustment costs produce a more front-loaded increase in renewable investment and a faster accumulation of renewable capital, whereas lower fossil adjustment costs generate a stronger short-run expansion of non-renewable capital and investment. Higher sector-specific adjustment costs dampen and delay the corresponding capital response. These results confirm that sector-specific frictions matter primarily for the composition and timing of energy-capital adjustment, while leaving the aggregate propagation of the fossil productivity shock largely intact.

\subsection{Energy substitution elasticities}
\begin{figure}[H]
\centering
\includegraphics[width=\linewidth]{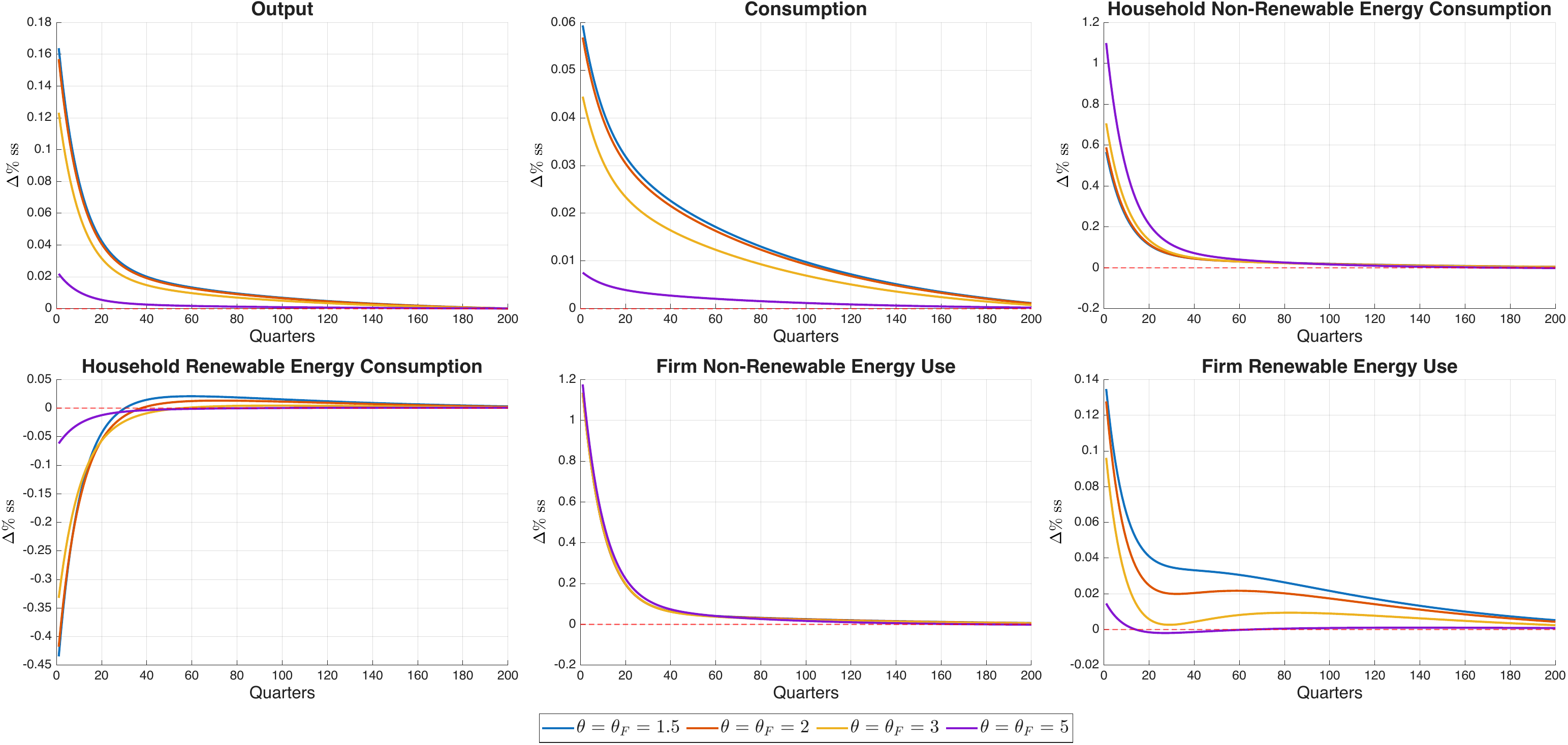}
\caption{Impulse responses to a one-percent increase in $A_{ENR}$ under alternative common substitution elasticities, $\theta=\theta_F$.}
\label{fig:sensitivity_theta}
\end{figure}

Figure~\ref{fig:sensitivity_theta} varies the common elasticity of substitution in household and firm energy bundles, setting $\theta=\theta_F\in\{1.5,2,3,5\}$. The aggregate responses of output and consumption remain relatively stable across calibrations, indicating that the short-run macroeconomic effect of the fossil productivity shock is not highly sensitive to this parameter in the considered range. The main differences appear in the energy-composition responses. Higher substitution elasticities strengthen the within-bundle reallocation between renewable and non-renewable energy, while lower elasticities make the energy mix more inertial. The figure therefore confirms that $\theta$ mainly governs how strongly households and firms reshape the composition of energy demand following the shock, rather than the overall size of the aggregate expansion.

\subsection{Energy-bundle weights}
\begin{figure}[H]
\centering
\includegraphics[width=\linewidth]{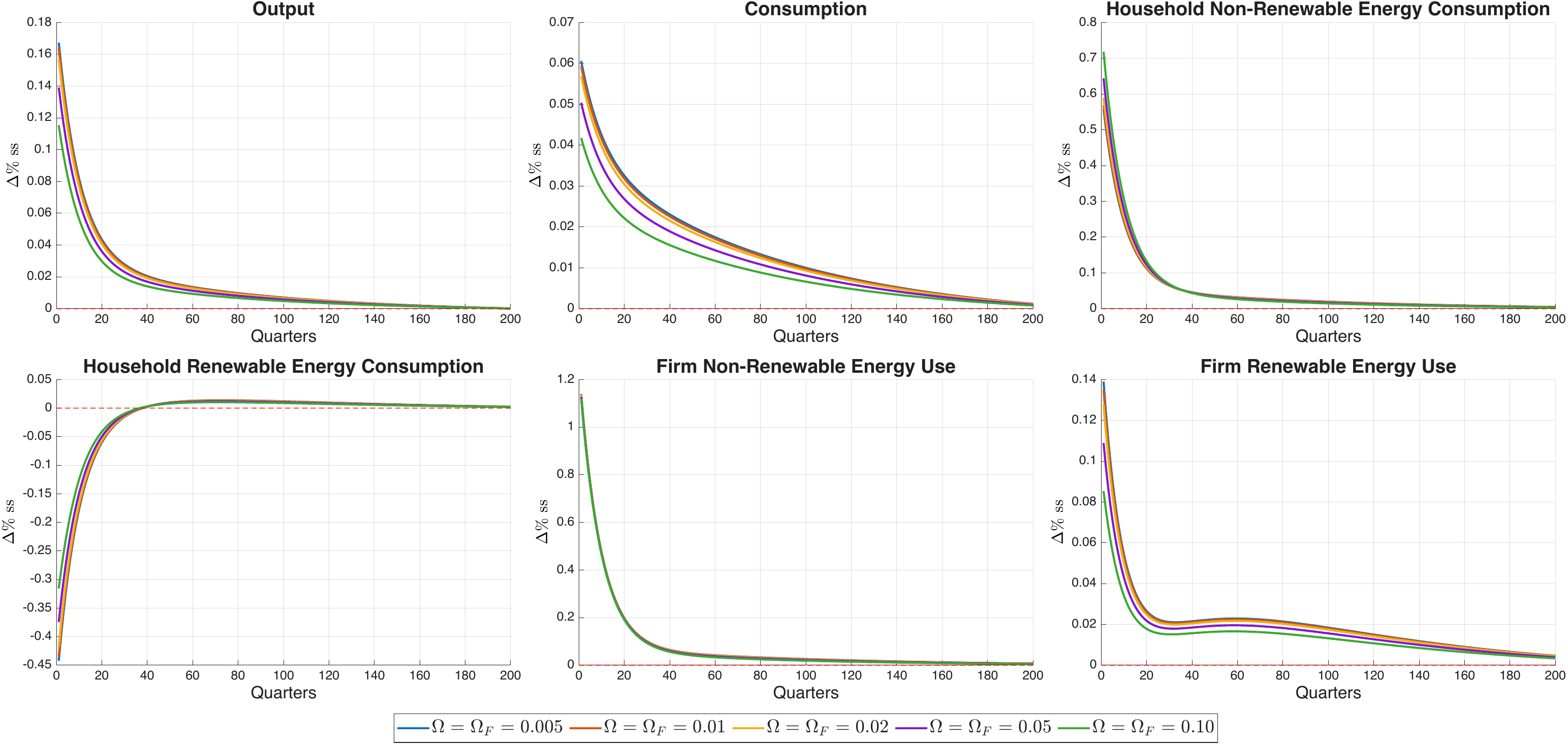}
\caption{Impulse responses to a one-percent increase in $A_{ENR}$ under alternative common renewable weights, $\Omega=\Omega_F$.}
\label{fig:sensitivity_omega}
\end{figure}

Figure~\ref{fig:sensitivity_omega} varies the common CES weight on renewable energy in household and firm energy bundles, setting $\Omega=\Omega_F\in\{0.03,0.05,0.10\}$. The impulse responses remain very similar across calibrations, indicating that moderate changes in the renewable weight have only limited quantitative effects on the propagation of the fossil productivity shock. A higher renewable weight is associated with slightly smaller responses of output and consumption, and also with a modestly smaller increase in firm renewable energy use. The responses of household and firm fossil energy demand are nearly unchanged across calibrations. Overall, the figure suggests that the renewable weight matters only marginally for both aggregate macroeconomic dynamics and the composition of energy demand in this experiment.

\subsection{Household energy utility weight}
\begin{figure}[H]
\centering
\includegraphics[width=\linewidth]{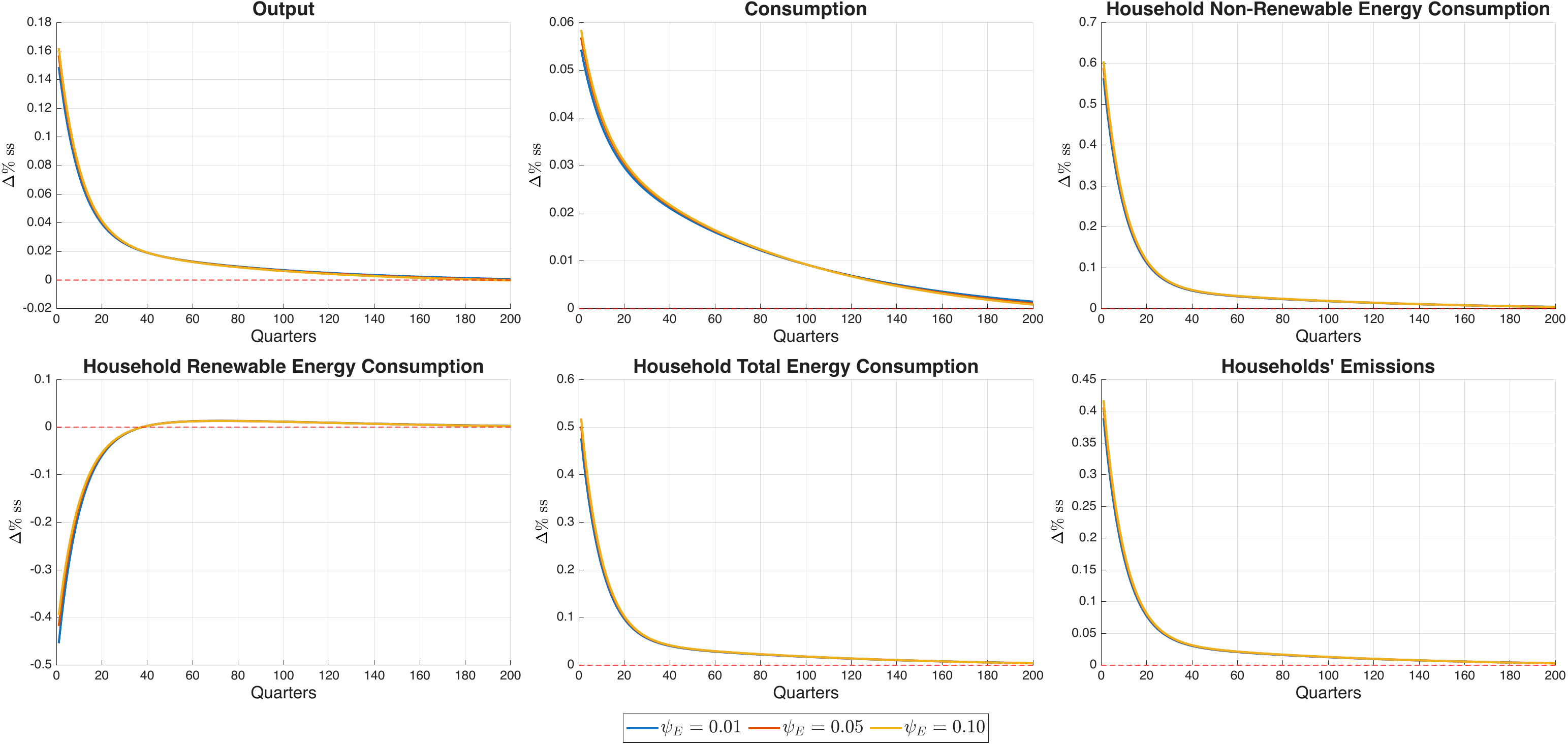}
\caption{Impulse responses to a one-percent increase in $A_{ENR}$ under alternative values of the household energy utility weight, $\psi_E$.}
\label{fig:sensitivity_psiE}
\end{figure}

Figure~\ref{fig:sensitivity_psiE} varies the utility weight on household energy services, $\psi_E\in\{0.01,0.05,0.10\}$. In contrast to the previous exercises, the responses are nearly indistinguishable across the reported range. Output, consumption, household total energy use, and household emissions all exhibit very similar dynamics. This suggests that, for a small positive shock to fossil productivity, the local propagation mechanism is not highly sensitive to moderate changes in the utility weight on energy services. In the present calibration, $\psi_E$ affects the level importance of household energy in utility more than it alters the normalized short-run transmission of the shock.

%%%%%%%%%%%%%%%%%%%% APPENDIX C %%%%%%%%%%%%%%%%%%%%
\newpage
\section{Anticipated renewable productivity shock}
\label{appendix_anticipated_AER_shock}
\setcounter{figure}{0}
\setcounter{table}{0}
\setcounter{equation}{0}
\renewcommand{\thefigure}{\Alph{section}\arabic{figure}}
\renewcommand{\thetable}{\Alph{section}\arabic{table}}
\renewcommand{\theequation}{\Alph{section}\arabic{equation}}

\begin{figure}[H]
\centering
\includegraphics[width=\linewidth]{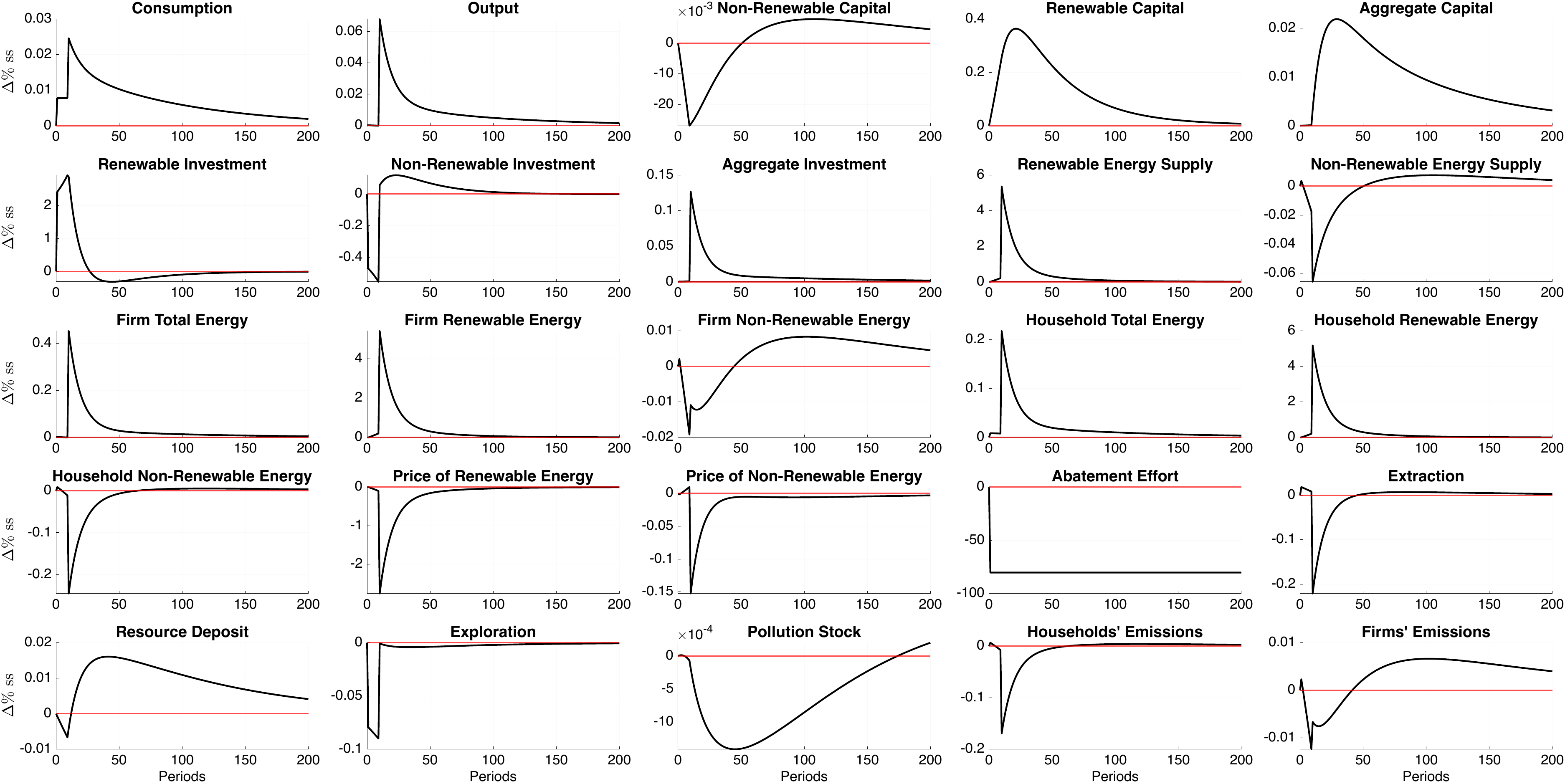}
\caption{Transition dynamics following a pre-announced 5\% permanent increase in renewable energy productivity ($A_{ER}$) realized in period 10. All variables reflect percentage changes relative to the laissez-faire steady-state benchmark.}
\label{fig:preannounced_AER_shock}
\end{figure}

This appendix analyzes the economy’s response to an anticipated improvement in renewable-sector productivity. Agents learn in period 0 that renewable productivity, $A_{ER}$, will increase permanently by 5 percent in period 10. Because agents are forward looking and capital is predetermined, the announcement affects investment and resource allocation decisions immediately, generating an anticipation phase prior to the realization of the shock. Figure \ref{fig:preannounced_AER_shock} reports the resulting transition dynamics.

\textit{Anticipatory capital reallocation.} Renewable investment increases immediately after the announcement as firms anticipate a higher future marginal product of renewable capital. Due to convex installation costs, renewable capital cannot adjust instantaneously and instead accumulates gradually during the anticipation window. Fossil investment declines slightly, leading to a gradual reduction in fossil capital prior to the realization of the shock.\\

\textit{Output and consumption.} Macroeconomic effects are already visible during the anticipation phase. Because the productivity improvement is announced in advance, forward-looking households increase consumption before the shock is realized, reflecting higher expected lifetime income. Once the productivity improvement materializes, output rises modestly as renewable energy becomes more efficient and the renewable capital stock that was installed during the anticipation phase becomes productive. Consumption peaks shortly after the implementation of the shock before gradually approaching its steady-state value as investment and capital stocks adjust.\\

\textit{Energy and resource dynamics.} Energy demand shifts toward renewables in anticipation of the shock. Renewable energy use rises during the anticipation period, while fossil energy demand declines modestly. The announcement creates a weak Green-Paradox-type incentive to bring fossil extraction forward before renewable productivity improves. In the current calibration, this force is quantitatively small and is largely offset by general-equilibrium reductions in fossil demand and fossil investment. As a result, extraction responds only mildly during the anticipation window and exploration remains below its steady-state level.\\

\textit{Emissions and environmental effects.} Despite the temporary increase in extraction, emissions remain close to their steady-state level. The gradual decline in fossil capital offsets the short-run increase in extraction, so effective fossil energy supply changes little during the anticipation window. Consequently, the pollution stock evolves only marginally.\\

\textit{Interpretation.} Overall, the experiment produces a weak intertemporal supply response: anticipation primarily shifts some of the adjustment into the pre-announcement phase without altering the small aggregate impact of renewable productivity improvements in this calibration. Although the announcement induces a modest increase in extraction prior to realization, general equilibrium adjustments in energy demand and capital allocation largely offset the incentive to accelerate fossil production. The results highlight the importance of capital adjustment frictions in shaping the macroeconomic and environmental effects of anticipated technological change.
%%%%%%%%%%%%%%%%%%%% APPENDIX D %%%%%%%%%%%%%%%%%%%%
\newpage
\section{Transition to 50\% renewable energy production}
\label{appendix_transition_to_50_renewable_energy}
\subsection{Transition dynamics of endogenous variables}
\setcounter{figure}{0}
\setcounter{table}{0}
\setcounter{equation}{0}
\renewcommand{\thefigure}{\Alph{section}\arabic{figure}}
\renewcommand{\thetable}{\Alph{section}\arabic{table}}
\renewcommand{\theequation}{\Alph{section}\arabic{equation}}

Figure \ref{fig:transition_half_renewable_all_vars} below depicts the transition dynamics for most of the endogenous variables under a carbon tax increase to 0.30409.
\begin{figure}[H]
    \centering
    \includegraphics[width=\linewidth]{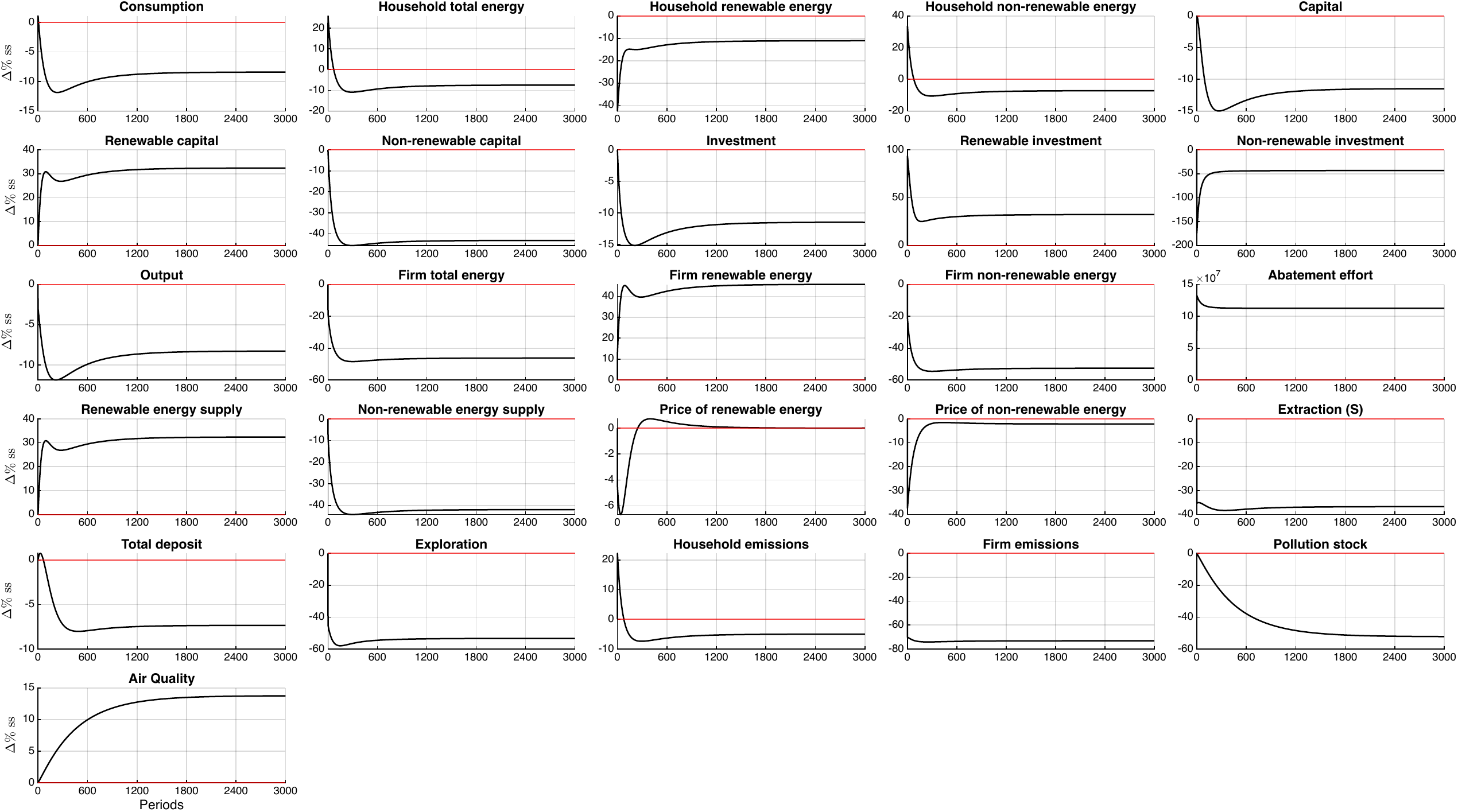}
    \caption{Full set of transition dynamics under a permanent unanticipated increase in the carbon price to $p_{z,t} = 0.30409$. All variables reflect percentage changes relative to the laissez-faire steady-state benchmark.}
    \label{fig:transition_half_renewable_all_vars}
\end{figure}

Figure \ref{fig:transition_half_renewable_all_vars} also illustrates the behavior of the fossil-resource block during the transition to 50\% renewable energy production. The fossil resource sector contracts immediately due to the increase in the carbon price,. Both fossil extraction $S_t$ and exploration expenditures $Exp_{NR,t}$ fall sharply on impact as the carbon tax reduces the profitability of fossil energy production and weakens incentives to discover new reserves.

The shadow value of fossil reserves $\lambda_{D,t}$ also declines, reflecting the lower expected profitability of future extraction following the tightening of climate policy. As a consequence, the stock of deposits $D_t$ gradually decreases along the transition path because discoveries fall while existing reserves continue to be depleted.

Over time, the fossil sector converges toward a smaller steady-state scale consistent with permanently lower fossil demand, while renewable energy production expands and progressively replaces fossil inputs in the energy system.

\subsection{Sensitivity to sector-specific capital adjustment frictions}
\label{appendix_sensitivity_adj_costs}
Tables~\ref{tab:common_h_sensitivity} and \ref{tab:sectoral_h_sensitivity} examine how the transition to 50\% renewable energy depends on the calibration of capital adjustment frictions. Table~\ref{tab:common_h_sensitivity} varies a common adjustment-cost parameter across all capital stocks, while Table~\ref{tab:sectoral_h_sensitivity} changes the renewable and non-renewable energy capital frictions separately around the baseline calibration.

Two patterns emerge. First, larger adjustment frictions smooth the transition but delay adjustment. As the common friction parameter rises, the peak declines in output, consumption, and fossil capital become smaller in magnitude, while the trough occurs later. The peak output decline falls from 12.07\% when $h=h_{ER}=h_{ENR}=1$ to 11.66\% when $h=h_{ER}=h_{ENR}=20$, while the output trough shifts from quarter 200 to quarter 235. Second, lower frictions are associated with much larger front-loaded renewable investment responses. The peak increase in renewable investment falls from 455.54\% under low common frictions to 48.31\% under high common frictions, indicating that renewable investment is the most friction-sensitive margin in the transition.

The sector-specific exercises show that increasing non-renewable capital adjustment frictions delays and smooths the contraction of fossil capital, while lowering them accelerates fossil-capital contraction and modestly increases the short-run output loss. Increasing renewable adjustment frictions mainly dampens the initial renewable investment surge, whereas lowering renewable adjustment frictions amplifies it sharply. Across all calibrations, the terminal levels of output and consumption remain unchanged, reflecting the fact that adjustment frictions affect transition dynamics but not the long-run steady state. Taken together, these results show that the magnitude and timing of transition costs depend materially on the calibration of capital adjustment frictions.

\begin{table}[H]
\centering
\caption{Transition Sensitivity to Common Capital Adjustment Frictions}
\label{tab:common_h_sensitivity}
\small
\setlength{\tabcolsep}{4pt}
\resizebox{\linewidth}{!}{
\begin{tabular}{lcccccccccc}
\toprule
\textbf{$h=h_{ER}=h_{ENR}$} & \textbf{Peak $\Delta Y$} & \textbf{Quarter} & \textbf{Peak $\Delta C$} & \textbf{Quarter} & \textbf{Peak $\Delta K_{ENR}$} & \textbf{Quarter} & \textbf{Peak $\Delta I_{ER}$} & \textbf{Quarter} & \textbf{$\Delta Y_T$} & \textbf{$\Delta C_T$} \\
\midrule
1  & -12.0685 & 200 & -12.0282 & 221 & -45.7909 & 265 & 455.5411 & 1 & -8.2741 & -8.3923 \\
3  & -12.0269 & 204 & -11.9914 & 224 & -45.7521 & 271 & 231.5587 & 1 & -8.2741 & -8.3923 \\
5  & -11.9831 & 207 & -11.9526 & 227 & -45.7122 & 277 & 162.1705 & 1 & -8.2741 & -8.3923 \\
10 & -11.8724 & 217 & -11.8555 & 234 & -45.6102 & 293 & 93.6577  & 1 & -8.2741 & -8.3923 \\
20 & -11.6592 & 235 & -11.6707 & 249 & -45.4004 & 329 & 48.3143  & 1 & -8.2741 & -8.3923 \\
\bottomrule
\end{tabular}}
\begin{minipage}{0.96\linewidth}
\footnotesize
\textit{Notes:} The table reports transition statistics for the immediate permanent carbon-tax transition to a 50\% renewable-energy share under alternative values of a common capital adjustment cost parameter, with $h=h_{ER}=h_{ENR}$. Peak losses and terminal outcomes are measured relative to the initial laissez-faire steady state and reported in percentage terms. $\Delta Y_T$ and $\Delta C_T$ denote the terminal deviations in output and consumption at the end of the simulated transition horizon.
\end{minipage}
\end{table}

\begin{table}[H]
\centering
\caption{Transition Sensitivity to Sector-Specific Capital Adjustment Frictions}
\label{tab:sectoral_h_sensitivity}
\small
\setlength{\tabcolsep}{4pt}
\resizebox{\linewidth}{!}{
\begin{tabular}{lcccccccccc}
\toprule
\begin{tabular}[c]{@{}c@{}}\textbf{Calibration}\\\textbf{$(h,h_{ER},h_{ENR})$}\end{tabular} 
& \textbf{Peak $\Delta Y$} & \textbf{Quarter} 
& \textbf{Peak $\Delta C$} & \textbf{Quarter} 
& \textbf{Peak $\Delta K_{ENR}$} & \textbf{Quarter} 
& \textbf{Peak $\Delta I_{ER}$} & \textbf{Quarter} 
& \textbf{$\Delta Y_T$} & \textbf{$\Delta C_T$} \\
\midrule
$(10,10,10)$ & -11.8724 & 217 & -11.8555 & 234 & -45.6102 & 293 & 93.6577  & 1 & -8.2741 & -8.3923 \\
$(10,10,20)$ & -11.7244 & 230 & -11.7182 & 246 & -45.4357 & 326 & 69.6804  & 1 & -8.2741 & -8.3923 \\
$(10,20,10)$ & -11.8833 & 216 & -11.8672 & 233 & -45.6207 & 292 & 63.2746  & 1 & -8.2741 & -8.3923 \\
$(10,10,3)$  & -11.9679 & 208 & -11.9457 & 226 & -45.7248 & 271 & 126.2566 & 1 & -8.2741 & -8.3923 \\
$(10,3,10)$  & -11.8658 & 217 & -11.8503 & 235 & -45.6016 & 295 & 170.5725 & 1 & -8.2741 & -8.3923 \\
\bottomrule
\end{tabular}}
\begin{minipage}{0.96\linewidth}
\footnotesize
\textit{Notes:} The table reports transition statistics for the immediate permanent carbon-tax transition to a 50\% renewable-energy share under alternative sector-specific capital adjustment frictions. The baseline calibration is $(10,10,10)$. Peak losses and terminal outcomes are measured relative to the initial laissez-faire steady state and reported in percentage terms. $\Delta Y_T$ and $\Delta C_T$ denote the terminal deviations in output and consumption at the end of the simulated transition horizon.
\end{minipage}
\end{table}
%%%%%%%%%%%%%%%%%%%% END OF APPENDIX D %%%%%%%%%%%%%%%%%%%%

%%%%%%%%%%%%%%%%%%%% APPENDIX E %%%%%%%%%%%%%%%%%%%%
\newpage
\section{Transition to a 90\% emissions reduction}
\label{appendix_emissions_reduction_90_percent}
\setcounter{figure}{0}
\setcounter{table}{0}
\setcounter{equation}{0}
\renewcommand{\thefigure}{\Alph{section}\arabic{figure}}
\renewcommand{\thetable}{\Alph{section}\arabic{table}}
\renewcommand{\theequation}{\Alph{section}\arabic{equation}}

This appendix examines the transition dynamics associated with a substantially more stringent climate policy targeting a 90\% reduction in aggregate emissions relative to the laissez-faire equilibrium. Such a reduction is broadly consistent with the European Union’s legally binding objective of achieving a 90\% net reduction in greenhouse gas emissions by 2040 relative to 1990 levels \citep{EU2040ClimateTarget}.

The Ramsey-optimal policy instruments that decentralize this emissions target are a carbon price of $p_z^* = 0.975873$, a tax on non-renewable energy of $\tau_{ENR}^* = 0.808817$, and a subsidy on renewable energy of $\tau_{ER}^* = 0.369387$. Because the distance between the initial and target steady states is large, gradual policy rules can generate substantial transitional oscillations under perfect foresight. Implementing the Ramsey policy as an immediate and permanent policy shock therefore provides a cleaner benchmark for evaluating the structural adjustments required to support deep decarbonization.

\begin{figure}[H]
\centering
\begin{subfigure}{\linewidth}
\centering
\includegraphics[width=\linewidth]{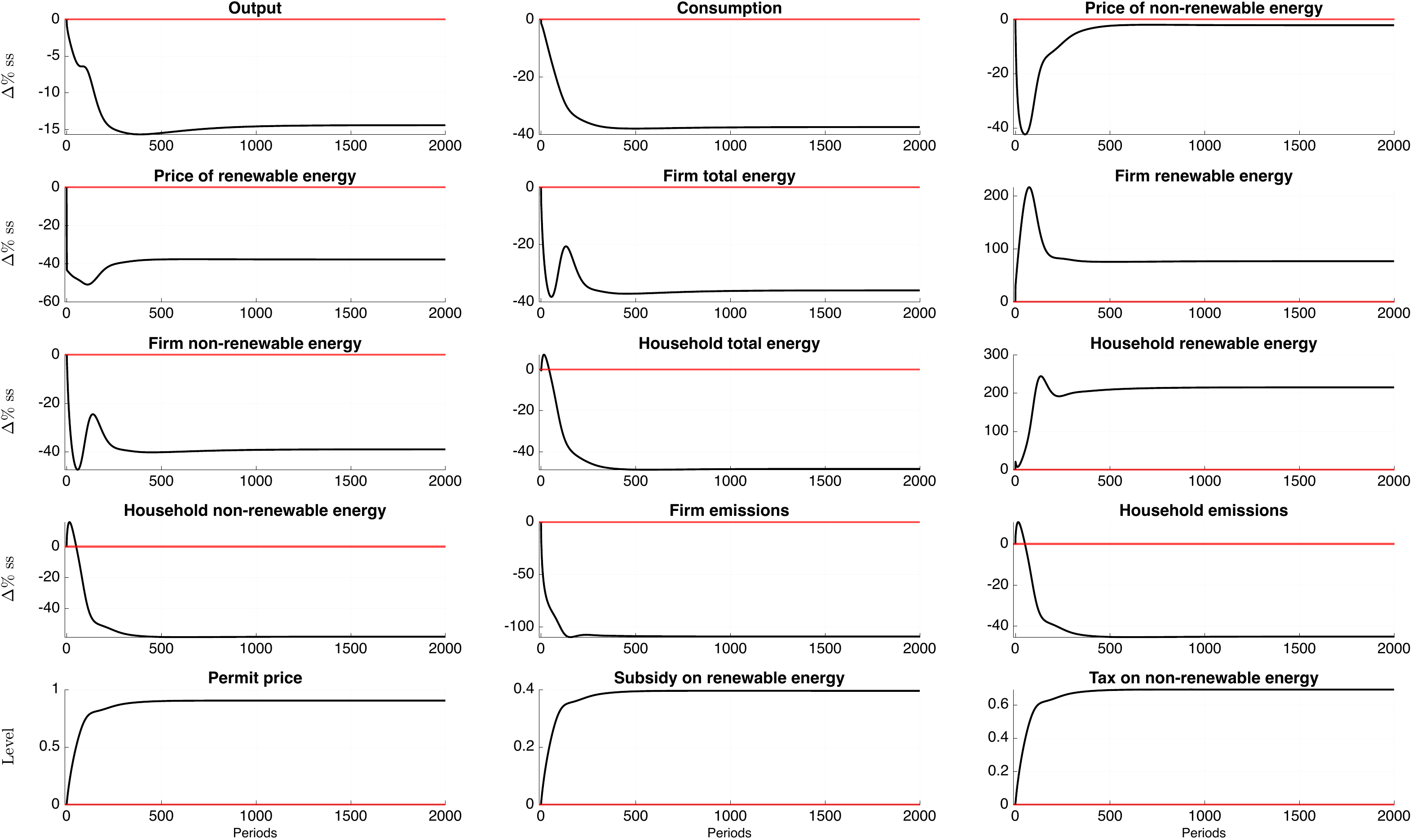}
\caption{Macroeconomic and energy-market dynamics during the transition to a 90\% emissions reduction target. Non-policy variables reflect percentage changes relative to the laissez-faire steady-state benchmark.; policy instruments are reported in levels.}
\label{fig:90_percent_emission_reduction_part1}
\end{subfigure}

\vspace{0.4cm}

\begin{subfigure}{\linewidth}
\centering
\includegraphics[width=\linewidth]{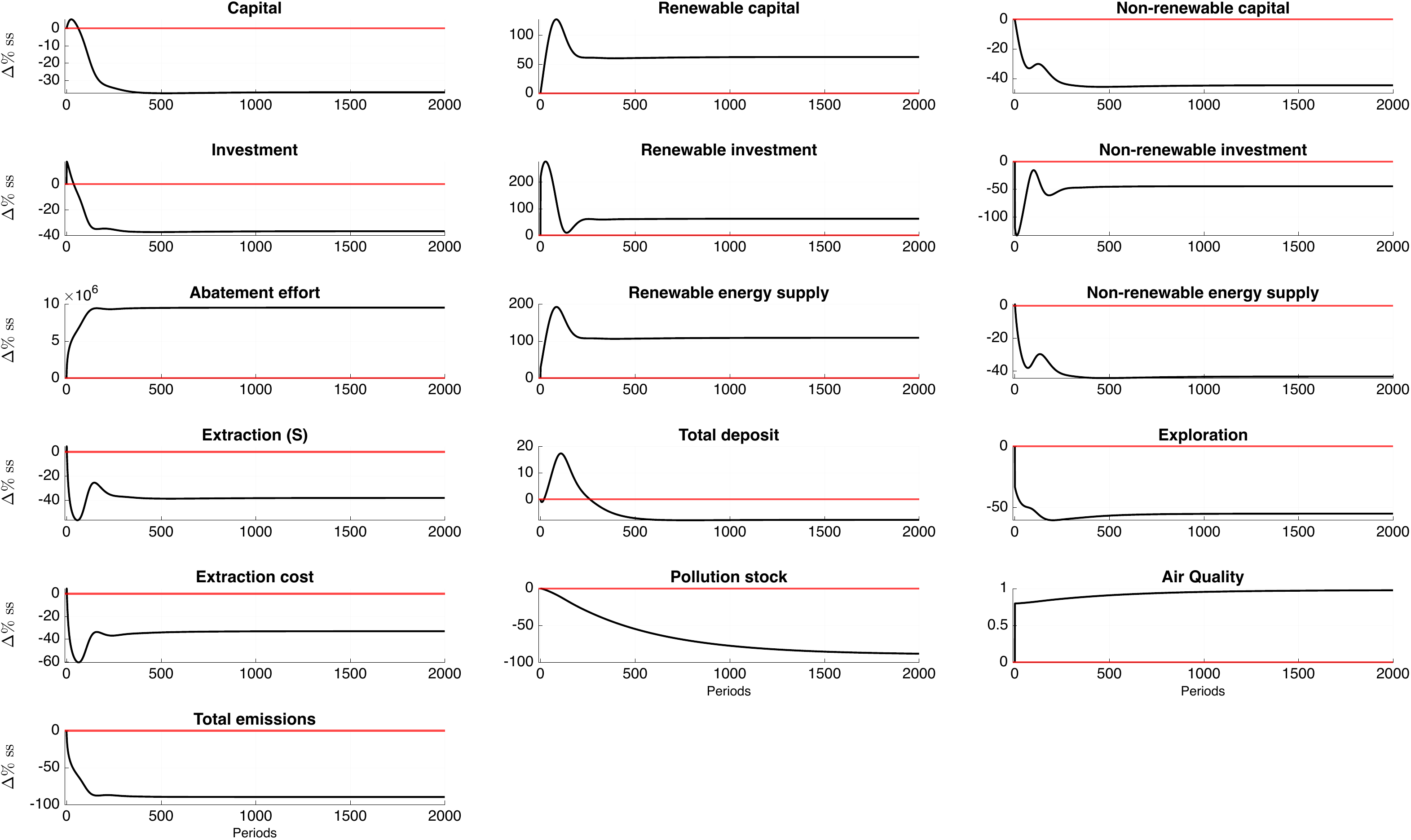}
\caption{Capital reallocation, fossil resource dynamics, abatement activity, and environmental outcomes during the transition to a 90\% emissions reduction target. All variables reflect percentage changes relative to the laissez-faire steady-state benchmark.}
\label{fig:90_percent_emission_reduction_part2}
\end{subfigure}
\end{figure}

The resulting transition dynamics are reported in Figures~\ref{fig:90_percent_emission_reduction_part1} and \ref{fig:90_percent_emission_reduction_part2}. Relative to the baseline emissions-cap experiment in Section~\ref{section_cap_emissions}, the macroeconomic adjustment is considerably larger. Output and consumption decline sharply as the substantial increase in carbon pricing raises the effective cost of fossil energy throughout the economy. Both firms and households substitute rapidly away from fossil fuels toward renewable energy. Firm demand for non-renewable energy collapses, while renewable energy demand rises strongly, reflecting the large-scale reallocation of energy inputs required to satisfy the emissions constraint.

A distinctive feature of this stringent policy is the emergence of large-scale abatement activity. Because the emissions target is extremely ambitious, abatement effort increases substantially and exceeds unity in the new steady state ($\mu = 1.1179$), implying net negative firm emissions. In the model, this outcome reflects the deployment of carbon removal technologies that remove more carbon from the atmosphere than is generated by fossil energy use in production. Firms therefore generate negative emissions that offset the residual fossil energy consumption of households, allowing aggregate emissions to converge to the targeted level. This mechanism is consistent with the growing literature emphasizing the role of carbon dioxide removal technologies such as BECCS (bioenergy with carbon capture and storage) in achieving deep decarbonization targets (e.g., \citep{Fuss2014}). Because the abatement-cost function is extrapolated beyond the conventional range \(0\leq \mu \leq 1\), the 90\% scenario should be interpreted as a diagnostic stress test rather than as a calibrated forecast of carbon-removal deployment.

The model abstracts from several important implementation constraints, including infrastructure requirements, storage availability, permitting, public acceptance, and technological uncertainty. For this reason, the macroeconomic costs reported for this deep-decarbonization scenario should be interpreted cautiously and may represent a lower bound relative to an environment in which large-scale carbon removal is more limited or more costly to deploy.

The transition toward this equilibrium entails substantial structural reallocation across capital and energy sectors. Renewable capital and renewable energy supply expand markedly, while non-renewable capital and fossil energy production decline persistently. Investment shifts toward renewable technologies, whereas investment in the fossil sector contracts sharply. Exploration activity falls rapidly, and the stock of exploitable fossil deposits gradually declines as remaining reserves become progressively stranded.

Environmental outcomes reflect the combined effects of declining fossil energy use and the emergence of carbon removal technologies. Aggregate emissions fall rapidly toward the target level, while the pollution stock declines persistently over time as lower emissions accumulate into a cleaner environmental state.

Overall, the dynamics illustrated in Figures~\ref{fig:90_percent_emission_reduction_part1} and \ref{fig:90_percent_emission_reduction_part2} highlight an important implication of very stringent climate policy. While the baseline experiment in Section~\ref{section_cap_emissions} achieves emissions reductions primarily through energy substitution between fossil and renewable sources, the 90\% reduction scenario additionally requires large-scale negative emissions technologies. Beyond a certain threshold, emissions reductions cannot be achieved solely through energy reallocation and instead necessitate carbon removal technologies that fundamentally alter the structure of production. The reliance on these technologies substantially increases the macroeconomic cost of the transition, as reflected in the pronounced rise in abatement expenditures and the persistent contraction in consumption and output.

%%%%%%%%%%%%%%%%%%%% APPENDIX F %%%%%%%%%%%%%%%%%%%%
\newpage
\section{Immediate Full Policy Package}
\label{appendix_immediate_full_package_irfs}
\setcounter{figure}{0}
\setcounter{table}{0}
\setcounter{equation}{0}
\renewcommand{\thefigure}{\Alph{section}\arabic{figure}}
\renewcommand{\thetable}{\Alph{section}\arabic{table}}
\renewcommand{\theequation}{\Alph{section}\arabic{equation}}

This appendix reports the transition dynamics for the immediate full policy package discussed in Section~\ref{section_cap_emissions}. In this experiment, the three policy instruments are set to their long-run values from the first period onward:
$$p_{z,t} = \bar p_z,\qquad \tau_{ER,t} = \bar\tau_{ER,t}, \qquad \tau_{ENR,t} = \bar\tau_{ENR}. $$

An economic interpretation of the immediate full policy package is that it combines carbon pricing with complementary fiscal instruments that support the reallocation from fossil to renewable energy capital. Since the model rebates carbon-tax revenue lump sum and does not impose an explicit earmarking rule, the experiment should not be read as a literal revenue-recycling scheme. Rather, the comparison with the carbon-price-only transition in Table~\ref{tab:dynamic_welfare_comparison} quantifies the welfare gain from adding renewable subsidies and non-renewable energy taxes to the immediate carbon-price transition. This reduces the welfare loss from \(-5.90\%\) to \(-3.50\%\), a gain of about 2.52 percentage points.

Unlike the gradual emissions-cap experiments, this policy does not use an endogenous carbon-price feedback rule. Instead, it implements the full policy package immediately and keeps the policy instruments fixed along the transition. The purpose of this experiment is diagnostic: it separates the role of the policy instrument set from the role of gradual policy adjustment.

\begin{figure}[H]
    \centering
    \includegraphics[width=.98\linewidth]{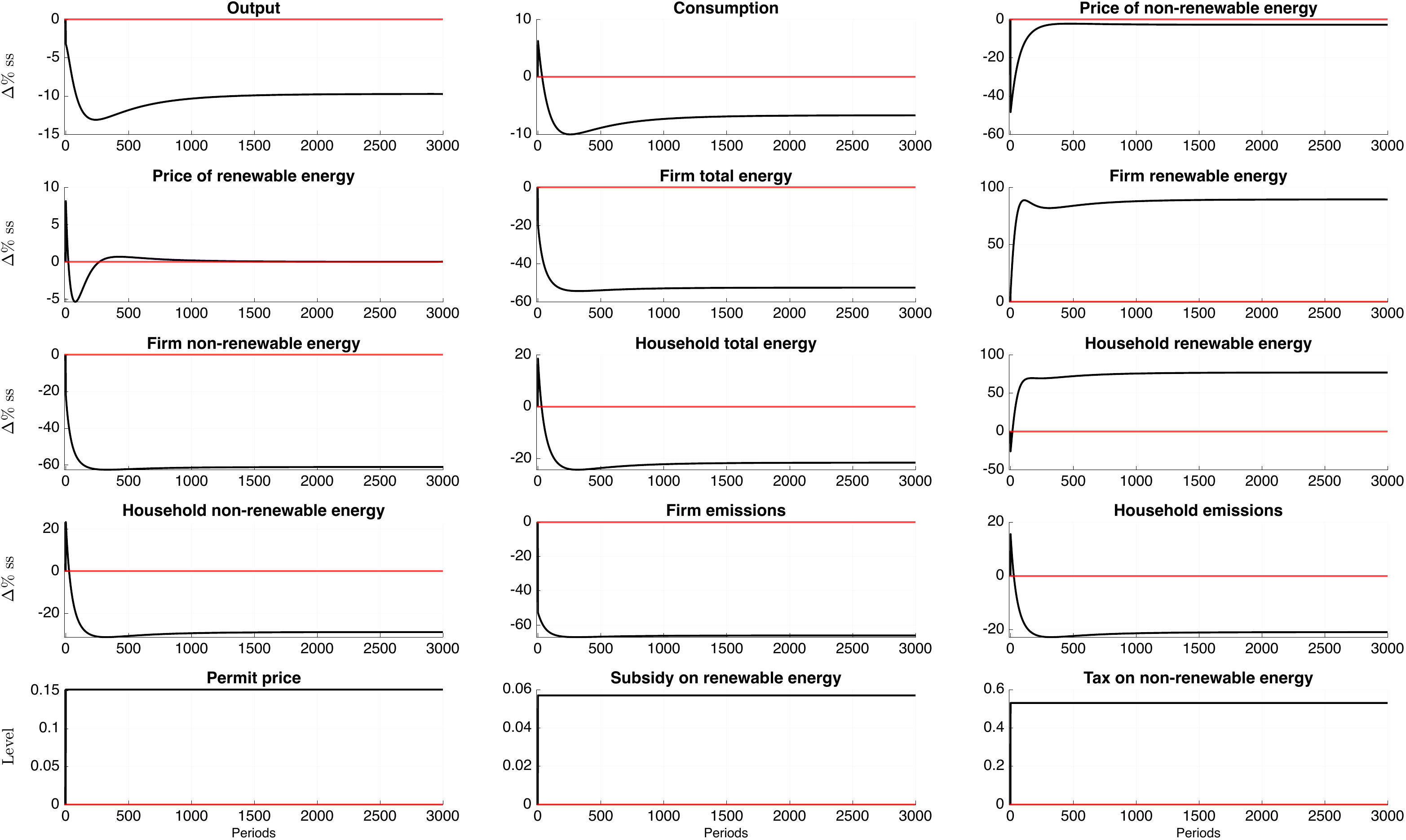}
    \caption{Immediate full policy package: macroeconomic, energy-demand, and policy-instrument dynamics}
    \label{fig:immediate_full_package_irfs_macro_energy}
\end{figure}

\begin{figure}[H]
    \centering
    \includegraphics[width=.98\linewidth]{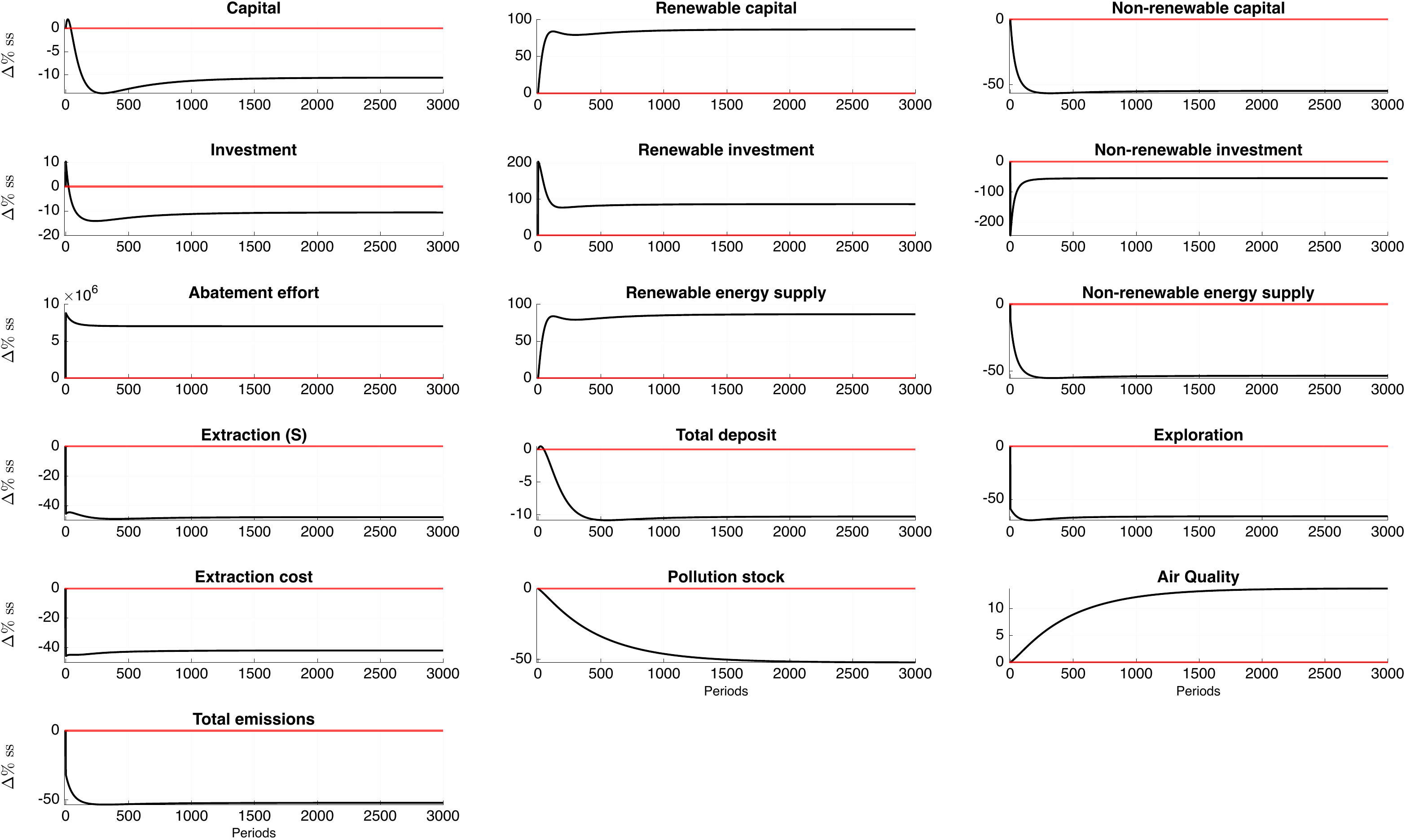}
    \caption{Immediate full policy package: capital reallocation, fossil-resource dynamics, and emissions}
    \label{fig:immediate_full_package_irfs_capital_resource}
\end{figure}

Figures~\ref{fig:immediate_full_package_irfs_macro_energy} and~\ref{fig:immediate_full_package_irfs_capital_resource} report the resulting transition paths. The immediate full package generates a smaller welfare loss than the carbon-price-only transition in Section~\ref{section_transition_50_ren_energy}. This occurs because the renewable subsidy and the non-renewable energy tax directly change the relative price of clean and dirty energy, reducing the burden placed on the carbon price alone. The policy therefore supports energy substitution and renewable-capital accumulation more directly than the carbon-price-only transition. However, because the full policy package is implemented immediately, fossil energy use and fossil-sector investment still contract before the energy capital stock has fully adjusted. The immediate full package therefore attenuates, but does not eliminate, the welfare cost of front-loaded implementation.

The adjustment is front-loaded. Renewable energy use, renewable capital, and renewable investment rise immediately, while non-renewable energy use, fossil capital, extraction, and exploration fall sharply. Total emissions decline rapidly and the pollution stock gradually converges toward its lower long-run level. Because the full set of relative-price incentives is active from the first period, renewable investment and fossil-sector contraction are front-loaded relative to the gradual cap regimes. The welfare results in Table~\ref{tab:dynamic_welfare_comparison} show that the immediate full package performs substantially better than the immediate carbon-price-only transition, but remains more costly than both gradual cap regimes under the baseline calibration. This comparison indicates that both implementation timing and instrument breadth matter for transition welfare, with timing the larger of the two effects in the baseline calibration: switching from a narrow to a broad instrument set under immediate implementation reduces the welfare loss from \(-5.90\%\) to \(-3.50\%\), while switching from immediate to gradual implementation under comprehensive coverage reduces it further to \(-1.26\%\).

This result qualifies a purely instrument-based interpretation of the welfare ranking. The immediate full policy package uses the broad set of instruments, but implements them from the first period onward. The resulting welfare loss shows that complementary instruments reduce the cost of immediate implementation, but do not eliminate the transition cost created by rapid fossil energy contraction and slow renewable-capital adjustment. In the baseline calibration, gradual implementation remains central for lowering welfare losses.

%%%%%%%%%%%%%%%%%%%% APPENDIX H %%%%%%%%%%%%%%%%%%%%
\newpage
\section{Policy coefficients and the implementation of emissions targets}
\label{appendix_policy_coefficients}
\setcounter{figure}{0}
\setcounter{table}{0}
\setcounter{equation}{0}
\renewcommand{\thefigure}{\Alph{section}\arabic{figure}}
\renewcommand{\thetable}{\Alph{section}\arabic{table}}
\renewcommand{\theequation}{\Alph{section}\arabic{equation}}
Table \ref{tab:policy_coefficients_constraints} evaluates whether the policy coefficients obtained from the constrained Ramsey problems are sufficient to implement the targeted emissions allocation once the explicit emissions restriction is removed. For each Ramsey exercise, the planner first chooses the optimal policy coefficients subject to an emissions constraint. The resulting coefficients are then held fixed while the decentralized steady state is recomputed without imposing the corresponding cap.

The table reports steady-state output ($Y$), consumption ($C$), total, firm and household non-renewable energy use ($E_{NR}$, $E_{FNR}$, and $E_{HHNR}$), and sectoral emissions ($Z_F$, $Z_{HH}$, and $Z_{tot}$). Panel A considers the Ramsey problem with an aggregate emissions cap. Removing the explicit aggregate cap while keeping the same policy coefficients fixed leaves the steady state essentially unchanged, indicating that the policy coefficients alone decentralize the targeted aggregate emissions allocation. Panel B considers the Ramsey problem with a firm-only emissions cap. Removing the explicit firm cap leads to a rebound in firm non-renewable energy use and firm emissions, implying that the explicit firm-level restriction remains essential for implementing the targeted allocation.

\begin{table}[H]
\centering
\caption{Policy Coefficients and the Role of Explicit Emissions Constraints}
\label{tab:policy_coefficients_constraints}
\small
\setlength{\tabcolsep}{4pt}
\resizebox{\linewidth}{!}{
\begin{tabular}{lcccccccc}
\toprule
\textbf{Policy system} & $Y$ & $C$ & $E_{NR}$ & $E_{FNR}$ & $E_{HHNR}$ & $Z_F$ & $Z_{HH}$ & $Z_{tot}$ \\
\midrule
\multicolumn{9}{l}{\textit{Panel A: coefficients from Ramsey problem with aggregate emissions cap}} \\
With aggregate cap imposed 
& 3.069876 & 2.208839 & 3.034760 & 1.946471 & 1.088288 & 1.029694 & 1.060116 & 2.089810 \\
Without cap imposed        
& 3.069876 & 2.208839 & 3.034760 & 1.946472 & 1.088288 & 1.029695 & 1.060116 & 2.089811 \\
\midrule
\multicolumn{9}{l}{\textit{Panel B: coefficients from Ramsey problem with firm-only emissions cap}} \\
With firm cap imposed      
& 3.113739 & 2.172663 & 3.728256 & 2.329703 & 1.398554 & 0.829384 & 1.260430 & 2.089814 \\
Without cap imposed        
& 3.314523 & 2.357568 & 4.975315 & 3.681808 & 1.293507 & 2.367986 & 1.194319 & 3.562305 \\
\bottomrule
\end{tabular}}
\begin{minipage}{0.95\linewidth}
\footnotesize
\textit{Notes:} Each panel reports the steady state under a fixed set of policy coefficients obtained from a constrained Ramsey problem. The first row in each panel imposes the emissions restriction used in the Ramsey exercise. The second row removes the explicit emissions constraint and solves for the decentralized steady state under the same policy coefficients. Panel A shows that the aggregate-cap policy coefficients implement the targeted aggregate emissions allocation even after the explicit cap is removed. Panel B shows that the firm-only policy coefficients do not implement the firm-level target on their own: removing the explicit firm cap generates a rebound in firm non-renewable energy use and firm emissions
\end{minipage}
\end{table}

\newpage
\bibliographystyle{apalike} %abbrvnat 
\bibliography{references}
\end{document}